\documentclass[
 reprint,
superscriptaddress,
nofootinbib,
bibnotes,
 amsmath,amssymb,
 aps,
prx,
nobalancelastpage,
floatfix,
]{revtex4-2}
\usepackage{amsmath,amssymb,amsfonts}
\usepackage{braket}
\usepackage{graphicx,color}
\usepackage{enumerate}

\begin{document}

\title{Microwaves in Quantum Computing}
\author{Joseph C. Bardin}
\email{Email: jbardin@engin.umass.edu}
\affiliation{Department of Electrical and Computer Engineering, University of Massachusetts Amherst, Amherst, MA 01003, USA}
\affiliation{Google Inc., Goleta, CA 93117, USA}
\author{Daniel H. Slichter}
\affiliation{Time and Frequency Division, National Institute of Standards and 
Technology, Boulder, CO 80305, USA}
\author{David J. Reilly}
\affiliation{Microsoft Inc., Microsoft Quantum Sydney, The University of Sydney, Sydney, NSW 2050, Australia}
\affiliation{ARC Centre of Excellence for Engineered Quantum Systems (EQuS), School of Physics, The University of Sydney, NSW 2050, Australia}

\begin{abstract}
Quantum information processing systems rely on a broad range of microwave technologies and have spurred development of microwave devices and methods in new operating regimes.  Here we review the use of microwave signals and systems in quantum computing, with specific reference to three leading quantum computing platforms: trapped atomic ion qubits, spin qubits in semiconductors, and superconducting qubits.  We highlight some key results and progress in quantum computing achieved through the use of microwave systems, and discuss how quantum computing applications have pushed the frontiers of microwave technology in some areas.  We also describe open microwave engineering challenges for the construction of large-scale, fault-tolerant quantum computers.  
\end{abstract}

\thispagestyle{empty}
This paper is a preprint (IEEE “accepted” status).  \textcopyright~2020 IEEE.  Personal use of this material is permitted.  Permission from IEEE must be obtained for all other uses, in any current or future media, including reprinting/republishing this material for advertising or promotional purposes, creating new collective works, for resale or redistribution to servers or lists, or reuse of any copyrighted component of this work in other works.

\clearpage
\setcounter{page}{0}
\maketitle
\tableofcontents

\section{Introduction}

Quantum computing and modern microwave engineering share a common ancestor in the pioneering work that led to the development of radar and related technologies in the 1940's~\cite{buderi}. Indeed, many of the fundamental mechanisms underpinning the generation, transmission, absorption, and detection of microwave energy were understood at that time to be governed by quantum mechanics, which describes the light-matter interaction between microwaves and atoms or molecules~\cite{Townes1975}, as well as their constituent charge and spin states \cite{slichter1996principles}. The leveraging of war-time radar technology and methods in the discovery of nuclear magnetic resonance in solids\cite{Purcell1946, Bloch1946} provides an ideal example of the long-standing synergy between microwave engineering and quantum systems\cite{buderi}.

Fundamental to this light-matter interaction is the relation $E = \hbar \omega$, which connects the angular frequency $\omega$ of microwave photons to their energy $E$ ($\hbar$ is the reduced Planck's constant). The quantization of microwave energy describes how microwave photons can interact resonantly with other quantum objects, such as the quantum two-level systems (called \emph{qubits}) that form the building blocks of quantum computers. Today, microwave technology is ubiquitous across many different quantum platforms, enabling the precise control and readout of quantum states.  

\begin{figure}[t!]
\centering
\includegraphics[width=\columnwidth]{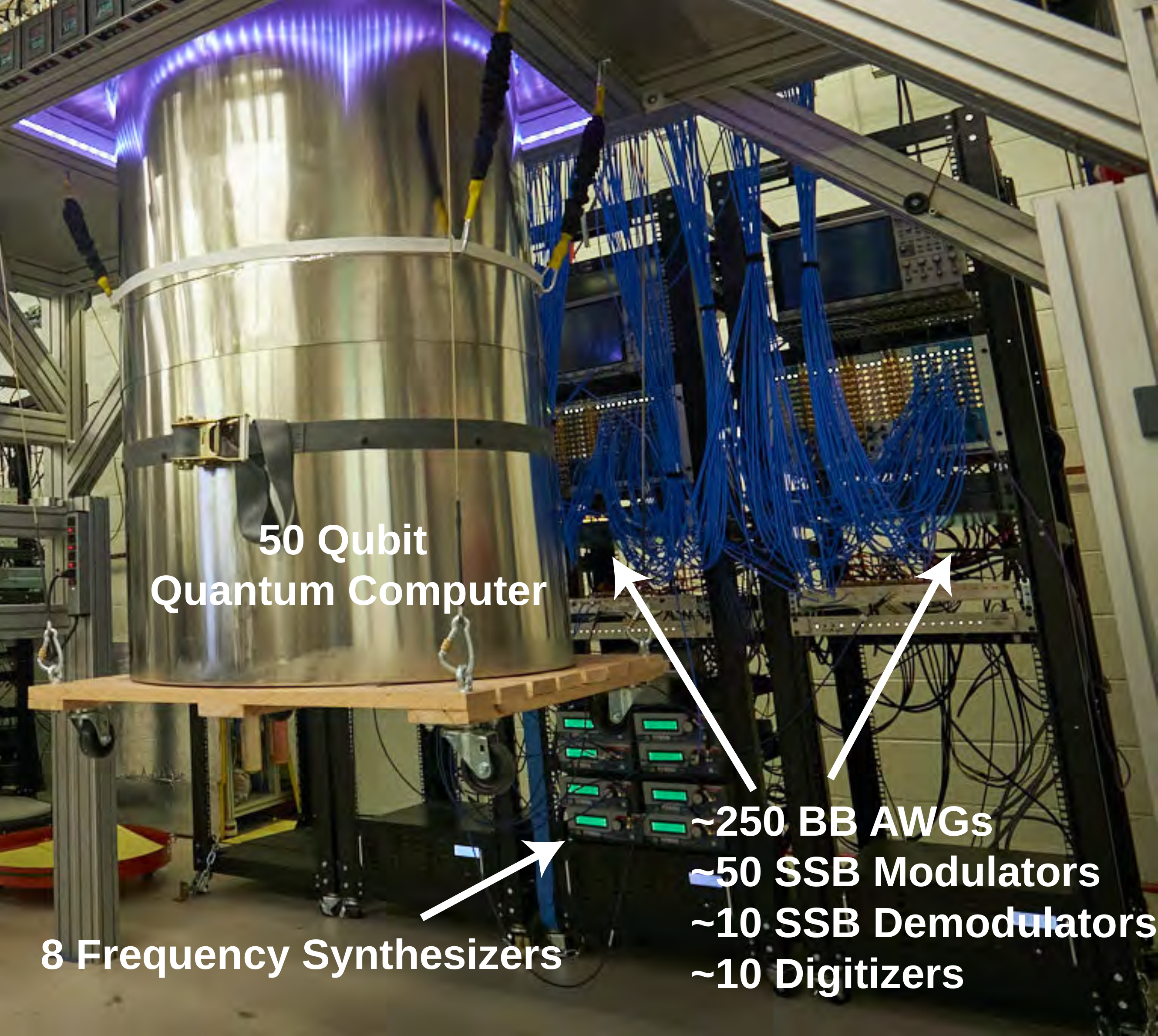}
\caption{Microwave electronics for operating a 50-qubit Google quantum processor.  The system generates and receives signals in the 4-8 GHz band, and was used to execute a demonstration quantum computing algorithm designed to be too complex for simulation by even the largest classical supercomputers~\cite{arute2019quantum}. Four racks of microwave electronics are required to control and measure the quantum processor.  Abbreviations: BB--baseband, AWG--arbitrary waveform generator, SSB--single sideband. Photo credit: R. Ceselin.}
\label{fig:hardware}
\end{figure}

Here, we review the use of microwaves in quantum computing. For the sake of brevity, rather than surveying the complete quantum computing landscape, we focus on three leading qubit technologies: trapped ion qubits, semiconductor spin qubits, and superconducting circuit qubits. We chose these technologies not only because they span the representative qubit technologies and are currently considered among the most promising of all qubit types, but also because they are heavily reliant on microwave technologies (see Fig.~\ref{fig:hardware}). 
The paper is organized as follows:
\begin{enumerate}[I)]
    \setcounter{enumi}{1}
    \item{An overview of qubits and quantum processors is presented; the analogy between a qubit and a microwave resonator is explained, with implications discussed; and physical qubit realizations are described.}
    \item{The coupling of a microwave source to various qubit technologies is described and typical signal and noise levels are compared and contrasted.}
    \item{The quantum gate abstraction is explained,  microwave techniques for implementing single and two qubit quantum gates are described, and typical hardware configurations are presented.}
    \item{Microwave techniques for measuring the state of a qubit are described.}
    \item{Additional microwave techniques required for the operation of trapped-ion qubits are described.}
    \item{Microwave innovations in quantum-limited amplifiers and non-reciprocal devices, driven by quantum computing research, are presented.}
    \item{Outstanding challenges related to microwave engineering that must be overcome to realize the full potential of quantum computing are described.}
\end{enumerate}

\section{Qubits and quantum computing\label{sec:qintro}}

The fundamental information carriers in a quantum computer are quantum bits, or qubits, in analogy to the logical bits used in a classical (non-quantum) computer.  Here we describe the basic characteristics and behavior of qubits, before discussing particular physical implementations of qubits and their connections with microwave technology.  The interested reader can find a much more detailed exposition of this topic, written for microwave engineers, in Ref.~\cite{bardin2020quantum}.

\subsection{Qubit basics\label{sec:qbasics}}
\begin{figure}[tb]
    \centering
    \includegraphics[width=\columnwidth]{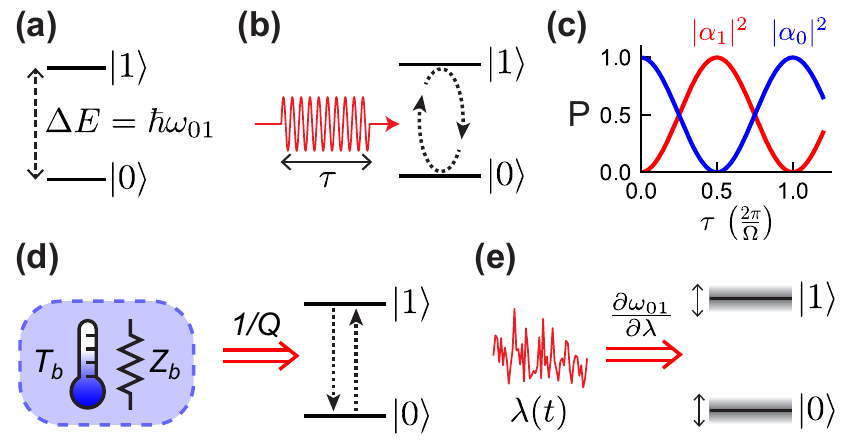}
    \caption{Qubit basics.  (a) A qubit is composed of two quantum states $\ket{0}$ and $\ket{1}$ with an energy difference $\Delta E$.  (b) When excited resonantly at $\omega_{01}=\Delta E/\hbar$, the qubit state can be driven between $\ket{0}$ and $\ket{1}$, including into linear combinations of $\ket{0}$ and $\ket{1}$.  (c) Under cw resonant driving, the qubit state undergoes so-called Rabi oscillations, where the probabilities $|\alpha_0|^2$ and $|\alpha_1|^2$ evolve sinusoidally in time. (d) Coupling to a source of dissipation at frequency $\omega_{01}$ (noise temperature $T_b$, impedance $Z_b$) with a coupling quality factor $Q$ causes qubit state transitions on the timescale $\mathrm{T}_1$. If $k_BT_b\ll\hbar\omega_{01}$, these transitions will be dominated by the $\ket{1}\rightarrow\ket{0}$ process.  (e) If the qubit frequency can be shifted by an environmental parameter $\lambda$ (e.g. magnetic field), fluctuations in $\lambda$ cause fluctuations in $\omega_{01}$, dephasing the qubit state on the timescale $\mathrm{T}_\phi$.}
    \label{fig:qubit_overview}
\end{figure}

A qubit is a quantum mechanical system with two energy eigenstates\footnote{The notation~``$\ket{~}$'' denotes a quantum state.}, which we label as $|0\rangle$ and $|1\rangle$, with corresponding energy eigenvalues $E_\text{0}$ and $E_\text{1}$, taking $E_\text{0}<E_\text{1}$ (see Fig.~\ref{fig:qubit_overview}a).  The ``textbook'' qubit is a spin-1/2 particle (such as an electron) in a magnetic field, but as we shall see, there are many other possible ways to realize a qubit.

The $\ket{0}$ and $\ket{1}$ states are used as computational basis states, analogous to the use of ``0'' and ``1'' in classical computing.  Just as the electromagnetic field can be decomposed into a linear combination of orthogonal modes (such as plane waves, spherical harmonics, or guided modes), the instantaneous state of a qubit $\ket{\psi}$ can be written as a linear combination of the two energy eigenstates, with complex amplitudes $\alpha_0$ and $\alpha_1$:
\begin{equation}
\ket{\psi}=\alpha_0\ket{0}+\alpha_1\ket{1}.
\end{equation}
The customary normalization of $\ket{\psi}$ is that $|\alpha_0|^2+|\alpha_1|^2=1$. 
To meet this criterion, we introduce the suggestive parameterization $\alpha_0=\cos(\theta/2)$ and $\alpha_1=e^{j\phi}\sin(\theta/2)$, where $\phi$ and $\theta$ are real and $j$ is the imaginary unit, for reasons that will become clear later.  For a single qubit, one is free to choose a global phase convention such that $\alpha_0$ is purely real, as seen here, but in general both $\alpha_0$ and $\alpha_1$ are complex.  

The quantum-mechanical nature of the state $\ket{\psi}$ means that the qubit can be in both states $\ket{0}$ and $\ket{1}$ simultaneously\textemdash a phenomenon known as \emph{superposition}\textemdash in contrast to the behavior of classical bits, which can only be in one state at a time.  When the state of the qubit is measured, however, the qubit state is said to ``collapse'' to just one of its eigenstates.  The collapse is probabilistic, with the state being measured to be $\ket{0}$ with probability $P_{|0\rangle}=\left|\alpha_0\right|^2$, or $\ket{1}$ with probability $P_{|1\rangle}=\left|\alpha_1\right|^2$; for this reason, $\alpha_0$ and $\alpha_1$ are called \emph{probability amplitudes}. Because of the collapse process, the post-measurement qubit state no longer contains information about $\alpha_0$ and $\alpha_1$; it has collapsed to either $\ket{0}$ or $\ket{1}$.  The values of $P_{|0\rangle}$ and $P_{|1\rangle}$ can only be determined by many rounds of preparing the same initial qubit state $\ket{\psi}=\alpha_0\ket{0}+\alpha_1\ket{1}$ and measuring it, to build statistics on the measurement collapse probabilities $\left|\alpha_0\right|^2$ and $\left|\alpha_1\right|^2$.  Although beyond the scope of this article, the complex phase $\phi=\arg{\alpha_1}-\arg{\alpha_0}$ can be determined using a procedure known as state tomography~\cite{Nielsen2000}.

Quantum computers require more than just one qubit to perform useful computations.  For $N$ qubits, there are $2^N$ basis states of the system, from $\ket{00...00}$ to $\ket{11...11}$. While $N$ classical bits can only be in one of the $2^N$ basis states at a given time, the phenomenon of superposition means that the quantum state $\ket{\psi}$ of $N$ qubits can be a linear combination of any\textemdash or even all\textemdash of the $2^N$ basis states at the same time, with corresponding complex probability amplitudes $\{\alpha_{00...00},\,\alpha_{00...01},\, ...\,,\, \alpha_{11...11}\}$. Since these amplitudes have a physical interpretation in terms of probabilities of measurement outcomes, they have the normalization condition $\sum_{k=0}^{2^N-1}|\alpha_k|^2=1$, where $k$ indexes the $2^N$ different bitstrings corresponding to the basis states.

While it is possible to access a range of superposition states by  putting each qubit in its own independent superposition state, such an approach can only be used to reach a small fraction of the basis states of the qubit state vector $|\psi\rangle$, since most of the possible linear combinations exhibit correlations between the qubits.  For example, consider the state $\ket{\psi}=\frac{1}{\sqrt{2}}\ket{00...00}+\frac{1}{\sqrt{2}}\ket{11...11}$.  It is not possible to write this state as a product of separate, individual states of the constituent qubits; the state of each qubit is inextricably correlated, or \emph{entangled}, with all the others.  The state $\ket{\psi}$ above is a superposition state, where the process of measurement will cause a collapse to just one basis state.  Let us consider what happens if we measure just one of the $N$ qubits in this entangled state.  If the measured qubit collapses to $\ket{0}$, then this causes all the other qubits to collapse to $\ket{0}$ as well, even though they were not measured directly, and the state becomes $\ket{\psi}=\ket{00...00}$.  Likewise, if the measured qubit happened to collapse to $\ket{1}$, then all the other qubits would to collapse to $\ket{1}$, even though they were not measured directly, giving $\ket{\psi}=\ket{11...11}$.  This phenomenon of \emph{entanglement} is a defining feature of quantum mechanics, and an essential ingredient for quantum computing.  

The $2^N$-dimensional state space of $N$ qubits can hold exponentially more information than that of $N$ classical bits, offering the hope of greatly increased computing power.  However, the phenomenon of measurement collapse means that only $N$ bits of information (a single bitstring representing the state of the $N$ qubits after measurement), selected probabilistically by the measurement process, can be extracted from the quantum state at the end of an algorithm.  In order to realize a speedup over classical algorithms, quantum algorithms generate interference between the $2^N$ complex amplitudes $\{\alpha_{00...00},\,\alpha_{00...01},\, ...\,,\, \alpha_{11...11}\}$ to increase the likelihood of measuring certain output bitstrings (ones which yield the desired outcome of the computation).  The action of an $N$-qubit quantum algorithm is analogous to the scattering matrix for a $2^N$-port passive lossless microwave device, where each quantum basis state (or bitstring) is mapped to a port.  An algorithm with speedup is equivalent to an S-matrix which transforms uniform input excitation at all ports into nonzero output at only a small subset of ports, through constructive and destructive interference, similarly to how a phased array can be used for beamforming, for example.  Since the size of such a theoretical microwave device scales as $2^N$, it would become impossible to realize for sufficiently large $N$.  This highlights the potential for sufficiently large quantum computers to perform calculations which are classically intractable~\cite{Shor1999, arute2019quantum}.  

Quantum algorithms create the desired interference between probability amplitudes by controlling the states of individual qubits, and generating entanglement  between qubits, in the course of the algorithm.  Experimentally, these tasks are usually carried out by microwave signals, or rely critically on microwave techniques.  In some qubit technologies, the measurement process is also carried out by microwave signals.  These topics will be detailed in Sections~\ref{sec:micmethods} and~\ref{sec:measurement}.  

It is useful to characterize the performance of qubit state preparation, control, and measurement operations using a metric known as the \emph{fidelity}.  For each of these three tasks, there exist distinct methods to quantify the fidelity; a detailed discussion is beyond the scope of this paper~\cite{Nielsen2000}.  In general, the fidelity can be thought of as characterizing how close the laboratory implementation of an operation is to its ideal theoretical representation, with a fidelity of 1 representing a perfect implementation and a fidelity of 0 indicating a complete failure.  The presence of noise, drifts, dissipation, or miscalibration can give rise to errors in the implementation and thus cause the fidelity to be less than 1.  The \emph{error rate} characterizes the amount by which the fidelity of an operation is less than 1.  Techniques for efficiently and accurately estimating fidelities and error rates, especially in larger quantum processors, are an active area of research.

\subsection{Qubits as resonators\label{sec:q_as_reson}}

One way to think of a qubit is as a high-quality-factor electromagnetic resonator, with a resonant frequency $\omega_{01}=\left(E_1-E_0\right)/\hbar$ set by the energy difference between the qubit states, and a quality factor $Q\gg 1$. Here $Q=\omega_{01}/\gamma$, where $\gamma$ is the decay rate of the energy in the qubit due to all sources of dissipation.  Good qubits typically have $Q>10^6$, and sometimes considerably higher. 

\begin{figure*}[t!]
    \centering
    \includegraphics[width=2\columnwidth]{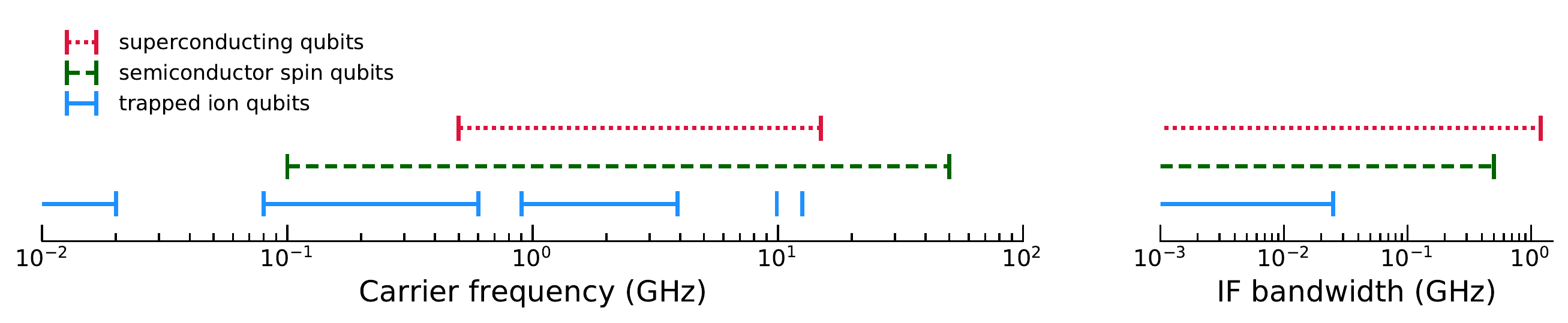}
    \caption{Relevant ranges of the electromagnetic spectrum for quantum computing.  We indicate the regions of the spectrum typically used by different qubit technologies for control and readout.  The carrier frequency for trapped ions can be as low as a few MHz.  The narrow trapped ion carrier frequency bands at $\sim10$ GHz and $\sim12.6$ GHz correspond to $^{133}$Ba$^+$ and $^{171}$Yb$^+$, respectively. Signals at the carrier frequencies are modulated with instantaneous IF/baseband bandwidths shown in the right panel.  These technologies also all use direct baseband control signals with bandwidths as shown in the right panel.}
    \label{fig:spectrum}
\end{figure*}

Unlike ordinary linear resonators, qubits are extremely anharmonic (or nonlinear).  As shown in Fig.~\ref{fig:qubit_overview}b, a resonant cw drive tone can be used to excite the qubit from the ``ground'' $\ket{0}$ state to the ``excited'' $\ket{1}$ state, but further excitation is not possible since there are no higher energy levels resonant with the drive.  Continued application of the drive tone can thus only return the qubit to $\ket{0}$ again.  The principle of superposition allows the qubit state to be driven to arbitrary linear combinations of $|0\rangle$ and $|1\rangle$; accordingly, persistent resonant driving of a qubit causes sinusoidal oscillations of the probabilities $|\alpha_0|^2$ and $|\alpha_1|^2$ in time, as seen in Fig.~\ref{fig:qubit_overview}c.  These oscillations are known as Rabi oscillations, and their angular frequency (called the Rabi frequency $\Omega$) is proportional to the amplitude of the resonant drive. 

As with classical resonators, a qubit's internal quality factor $Q_i$ describes dissipation due to intrinsic loss mechanisms. As we describe in detail in Section~\ref{sec:micmethods}, the introduction of external driving, interaction, and measurement ports---all necessary for quantum computing---will create additional loss channels which further damp the qubit resonance.  Each loss channel can be identified with its own quality factor: $Q_d$ for driving ports (where control signals are applied), $Q_c$ for coupling ports (enabling interactions with other qubits), and $Q_m$ for measurement ports.  The total quality factor of the qubit is then given by the inverse sum:
\begin{equation}
    \frac{1}{Q}=\frac{1}{Q_i}+\frac{1}{Q_d}+\frac{1}{Q_c}+\frac{1}{Q_m}.
\end{equation}
The characteristic time scale $\mathrm{T}_\text{1}$ over which a qubit initially in the $|1\rangle$ state will spontaneously transition to the $|0\rangle$ state\footnote{The direction of the transitions depends on the effective temperature $T_b$ of the loss channel.  If $k_BT_b\ll\hbar\omega_{01}$, the loss channel will cause relaxation from $\ket{1}$ to $\ket{0}$.  However, if $k_BT_b\gtrsim\hbar\omega_{01}$, the loss channel will induce transitions in both directions.} is given by $Q/\omega_{01}$. A qubit's $\mathrm{T}_1$\textemdash analogous to the ring-down time for a high-Q resonator\textemdash is an important metric for measuring qubit performance, and should ideally be much longer than the duration of any algorithm using the qubit.

These loss channels can be thought of as arising from coupling to different sources of dissipation, or ``baths" (for example, the real impedance of a control line), each with some effective noise temperature that is usually (but not always) close to the physical temperature (see Fig. 2(d)).  The qubit will thermalize to these baths on the timescale $\mathrm{T}_\text{1}$.  If $k_BT_b\ll\hbar\omega_{01}$, where $T_b$ is the coupling-weighted average temperature of the baths, then the qubit will ``reset'' thermally to the $\ket{0}$ state by itself. For qubit frequencies of 5 GHz, this corresponds to bath temperatures in the $\lesssim100$ mK range.  Otherwise, the qubit will thermalize to some combination of $\ket{0}$ and $\ket{1}$, and must be actively reset before it is used in a computation.  

In addition to loss or damping, there is a second type of decoherence we must consider.  Even with a completely lossless resonator, the resonance frequency itself can still fluctuate randomly, causing the excitation in the resonator to lose phase coherence over time relative to a stable reference oscillator.  If a qubit exhibits fluctuations in its resonance frequency $\omega_{01}$, the phase information $\phi$ in the complex amplitudes $\alpha_0$ and $\alpha_1$ will be lost on a timescale $\mathrm{T}_\phi$ (assuming infinite $\mathrm{T}_1$).  Real qubits (and real resonators) experience damping too, which also causes loss of phase information.  We can define a characteristic total dephasing time $\mathrm{T}_\text{2}$ over which phase information is lost as $1/\mathrm{T}_\text{2}=1/(2\mathrm{T}_\text{1})+1/\mathrm{T}_\phi$.  The loss of phase information destroys the interference between probability amplitudes $\alpha$ on which quantum algorithms rely, so $\mathrm{T}_\text{2}$ should also be much longer than the duration of any algorithm using the qubit.

Qubit frequency fluctuations occur when $\omega_{01}$ is sensitive to some external parameter $\lambda$. The parameter $\lambda$ could be the local magnetic or electric field, for example (by analogy, for a voltage-controlled oscillator circuit, $\lambda$ could be the tuning voltage).  Small fluctuations $\delta\lambda$ then lead to qubit frequency fluctuations $\delta\omega_{01} = \left(\frac{\partial\omega_{01}}{\partial \lambda}\right)\delta\lambda$, which causes dephasing (Fig. 2(e)).  Experimentally, $\mathrm{T}_\text{2}$ is maximized by reducing both the sensitivity $\frac{\partial\omega_{01}}{\partial \lambda}$ of the qubit frequency to noise and the amount of environmental noise $\langle\delta\lambda(t)^2\rangle$ present.

The qubit properties $\omega_{01}$, $\mathrm{T}_1$, and $\mathrm{T}_2$ can vary widely between different qubit technologies, as we will see in the next section.  We will expand further on the analogy between qubits and high-quality-factor resonators in Section~\ref{sec:interfacing}.

\subsection{Physical realizations of qubits\label{sec:realizations}}

Just as classical bits can have many different physical realizations\textemdash the voltage on the gate of a transistor, the spin orientation of a small magnetic domain on a hard disk, the reflectivity of a small region of an optical storage medium\textemdash qubits can have different physical implementations as well.  In this paper, we focus on three leading physical implementations of qubits: trapped atomic ions, spins in semiconductors, and superconducting circuits.  Below, we briefly explain the fundamentals and properties of these different types of qubits, all of which can have $\omega_{01}$ in the microwave region of the spectrum, as seen in Fig.~\ref{fig:spectrum}.

\subsubsection{Trapped ion qubits\label{sec:ion}}

\begin{figure}[t]
    \centering
    \includegraphics[width=\columnwidth]{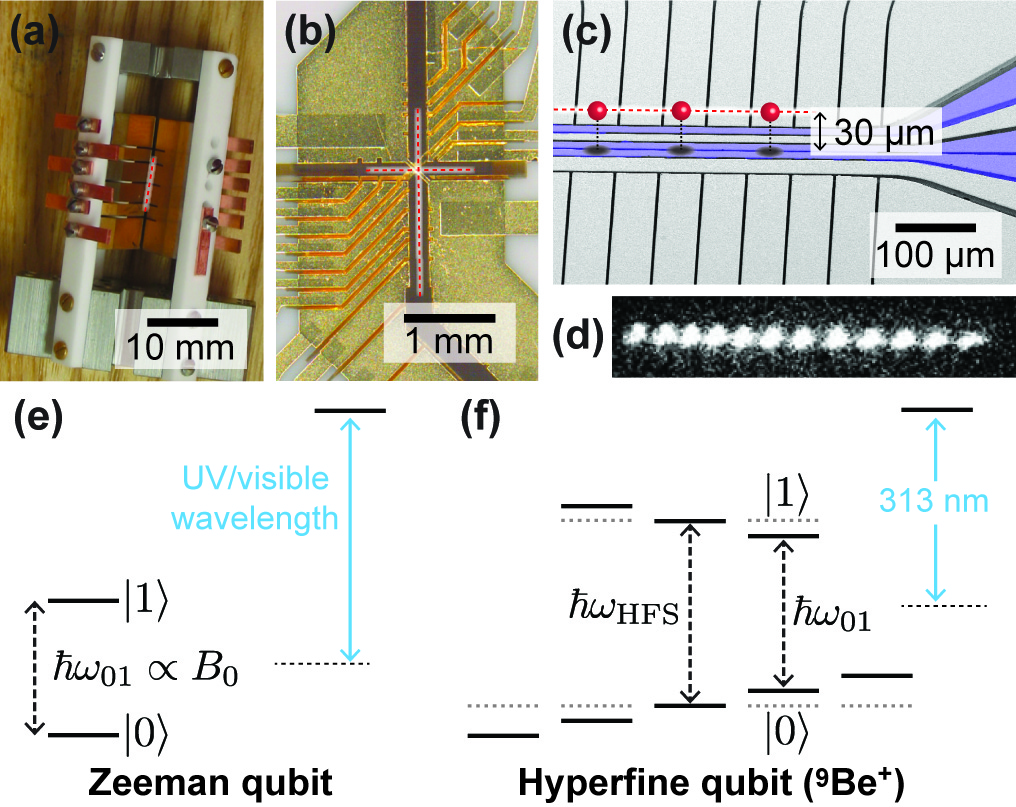}
    \caption{Trapped ion qubits. (a) Machined 3D trap, (b) laser-machined 3D trap, and (c) microfabricated surface-electrode trap with integrated microwave antenna structures for qubit control (blue).  Ions (shown schematically as red spheres in (c), each in a separate trapping potential well) are held at reconfigurable positions along the red dotted lines, between ((a) and (b)) or above (c) the trap electrodes.  (d) Camera image of chain of trapped ions, fluorescing from laser excitation.  Ion-ion spacing is a few $\mu$m.  (e) Energy level diagram of generic trapped ion Zeeman qubit. (f) Energy level diagram of $^{9}$Be$^+$ hyperfine qubit with $B_0>0$ (the levels are degenerate at $B_0=0$, shown as grey dotted lines), and one particular choice of qubit levels.  In $^9$Be$^+$, $\omega_\mathrm{HFS}/2\pi=1.25$ GHz.  In (e) and (f), the optical transitions to excited electronic states (blue arrows, not to scale) are used for laser cooling, qubit state preparation, and qubit readout. Photo credits: Ion Storage Group/NIST}
    \label{fig:iontrap}
\end{figure}

Qubits realized in the quantum states of atomic ions trapped in ultra-high vacuum are one of the most mature and high-fidelity quantum technologies~\cite{Wineland1998, Leibfried2003, Bruzewicz2019}.  Because ions have a net charge, they can be readily trapped and held in isolation in vacuum using electromagnetic fields.  For quantum computing applications, the ions are usually trapped using so-called linear rf Paul traps~\cite{Leibfried2003}, which confine charged particles\textemdash for hours to months, depending on parameters\textemdash using a combination of static electric fields and oscillating radio-frequency (typically between 20 MHz and 150 MHz) electric fields.  These fields are generated by applying dc and/or rf potentials to sets of trapping electrodes; some example ion traps are shown in Fig.~\ref{fig:iontrap}. The largest traps have centimeter-scale electrodes made in a machine shop, usually held together with insulating ceramic parts in a three-dimensional geometry (Fig.~\ref{fig:iontrap}(a)).  Intermediate-scale three-dimensional traps, with electrode dimensions down to hundreds of microns, can be made by depositing patterned metal films on laser-cut or etched insulating substrates (Fig.~\ref{fig:iontrap}(b)).  Two-dimensional traps, known as surface-electrode ion traps~\cite{Seidelin2006a}, are made on planar substrates using microfabrication techniques, and have typical electrode dimensions from $\sim100\,\mu$m down to a few $\mu$m (Fig.~\ref{fig:iontrap}(c)).  Numerous groups, including commercial quantum computing entities, are pursuing surface-electrode traps as a path toward large-scale trapped-ion quantum computing, due to the ability to make complex trap designs with many different trapping regions to hold large numbers of ions~\cite{Wright2019, Pino2020}.  

Ions are typically loaded into traps by electron impact ionization or resonant photoionization of a flux of neutral atoms from an in-vacuum thermal oven or laser ablation target~\cite{Bruzewicz2019}.  When more than one ion is held in a trap, the mutual Coulomb repulsion between ions keeps them spatially separated.  When laser-cooled~\cite{Eschner2003, Wineland1998}, the ions form a static Coulomb crystal with ion-ion spacings on the order of a few $\mu$m, as seen in Fig.~\ref{fig:iontrap}(d).  The motion of the ions in this configuration is strongly coupled by the Coulomb force, and is described as a set collective normal modes of ion motion, whose resonant frequencies are typically between 500 kHz and 10 MHz.  Traps can have multiple spatially separated potential wells, each containing such a Coulomb crystal.  

Trapped ions used in quantum computing applications typically have a single valence electron, with all other electrons in closed shells.  Among the ion species most commonly used for quantum computing applications are $^9$Be$^+$, $^{25}$Mg$^+$, $^{40}$Ca$^+$, $^{43}$Ca$^+$, $^{88}$Sr$^+$, $^{133}$Ba$^+$, $^{138}$Ba$^+$, and $^{171}$Yb$^+$, although numerous others have been employed as well.  Below, we describe several possible choices of qubit levels from among the many quantum states of these ions.    

When the ion species has no nuclear spin, a qubit can be realized using the two spin states of the ground state valence electron in the presence of an external magnetic field.  This type of trapped ion qubit is known as a Zeeman qubit, and is shown in Fig.~\ref{fig:iontrap}(e).  The qubit resonance frequency is proportional to the external magnetic field $B_0$ according to $\omega_{01}/2\pi=(\gamma_e/2\pi)|B_0|$, where $\gamma_e/2\pi\approx28$ GHz/T is the electron gyromagnetic ratio.  Trapped ion Zeeman qubits are typically operated in magnetic fields of less than 1 mT, giving qubit frequencies of $\sim10$ MHz.  For ion species with nonzero nuclear spin, such as $^9$Be$^+$ or $^{171}$Yb$^+$, the hyperfine interaction between the nuclear spin and the valence electron spin gives rise to two manifolds of hyperfine states in the ground electronic state, as shown in Fig.~\ref{fig:iontrap}(f).
At low magnetic fields ($B_0\lesssim50$ mT), these manifolds are separated by the hyperfine splitting $\omega_{\mathrm{HFS}}/2\pi$, which ranges from 1.25~GHz for $^9$Be$^+$ to 12.6~GHz for $^{171}$Yb$^+$, with higher-mass ions having larger splittings. A small magnetic field $B_0$ (typically $<1$ mT, but sometimes up to $\sim$~tens of mT) is applied such that each hyperfine state has a unique energy.  A trapped ion hyperfine qubit consists of two such states, usually chosen to be in separate hyperfine manifolds.  The qubit resonance frequency $\omega_{01}$ may differ by up to several hundred MHz from $\omega_{\mathrm{HFS}}$ for the range of magnetic fields listed above, depending on the choice of qubit states.  Zeeman and hyperfine qubits can be manipulated directly using rf or microwave magnetic fields, either launched in free space by distant antennas or horns, or from local antenna structures fabricated in the trap (as seen in Fig.~\ref{fig:iontrap}(c)).

Trapped ions in ultra-high vacuum are isolated from the nearest surfaces and bulk materials by tens to hundreds of $\mu$m.  As a result, electric and magnetic field noise at the ion are orders of magnitude lower than typically seen inside or on the surface of solids; combined with the relatively weak coupling of the quantum states of the ion to external fields, this means that trapped ion qubits do not thermalize to the environment rapidly.  Typical $\mathrm{T}_1$ values for Zeeman and hyperfine qubits are years.  As a result, optical pumping is used to initialize the internal states of the ions, and laser cooling is used to bring the ion motion near its quantum mechanical ground state~\cite{Wineland1998, Bruzewicz2019}.  This very slow thermalization also means that the temperature of the trap electrodes and the vacuum chamber need not satisfy $k_BT\ll\hbar\omega_{01}$ to achieve quantum behavior.  However, cryogenic operation of ion traps (in the 4 K to 10 K range) can be useful for increasing ion lifetime in the trap by cryopumping background gas, and for reducing electric field noise that heats and decoheres the ion motion~\cite{Bruzewicz2019}.  

Trapped ion qubits can be dephased by magnetic field fluctuations.  Zeeman qubits are directly sensitive to magnetic field fluctuations, with $\mathrm{T}_\text{2}\sim$ tens of ms, but $\mathrm{T}_\text{2}$ values of up to 300 ms have been achieved with appropriate magnetic field shielding~\cite{Ruster2016}.  For hyperfine qubits, it is possible to choose $B_0$ such that a particular hyperfine transition is insensitive to magnetic field noise to first order.  Such a qubit is known as a ``clock" qubit, so named because field-insensitive transitions generally have very long dephasing times and are therefore ideal for realizing microwave-frequency atomic clocks.  Bare clock qubit $\mathrm{T}_\text{2}$ values are usually $\gtrsim1$ s, but values as high as 50 s have been reported~\cite{Harty2014a}; performing a type of qubit ``chopping'' (called \emph{dynamical decoupling}) to counteract $1/f$ magnetic field noise can yield $\mathrm{T}_2$ in excess of an hour~\cite{Wang2020}.  Measurements of qubit coherence on these timescales are generally limited by the frequency stability and drift of the microwave reference oscillator to which the qubit is compared~\cite{Ball2016, Sepiol2019, Wang2020}.  

The interested reader is referred to 
Ref.~\cite{Bruzewicz2019} for further details on trapped ion quantum computing.

\subsubsection{Semiconductor spin qubits}

The spin degree of freedom in solids provides another potential platform for scalable quantum computing systems. Nuclear spins in silicon, for instance, can exhibit hours-long $\mathrm{T}_\text{1}$ times~\cite{Kane1998}. In contrast to the vacuum of a trapped ion, spin qubits are embedded in solids and surrounded by other atoms, many of which may interact with the spin qubit in uncontrolled ways\cite{slichter1996principles}.  Fortunately, these interactions are relatively weak in materials that contain few nuclear spins such as silicon\cite{RMPSilicon}, silicon-germanium\cite{SiGe}, and carbon materials such as diamond\cite{diamond}.

Highly isolated qubits that are only weakly interacting with their environment are also, in general, weakly coupled to any means of control. Weak coupling to control fields results in slow quantum gate times, potentially cancelling any advantage afforded by the long coherence times. For this reason many different `flavors' of spin qubit have been devised, with trade-offs between controllability, device complexity, and sensitivity to charge or voltage noise.  In this review, we will limit our discussion to spin qubits based on confined electrons or holes in semiconductors.

\begin{figure}
    \centering
    \includegraphics[width=.66\columnwidth]{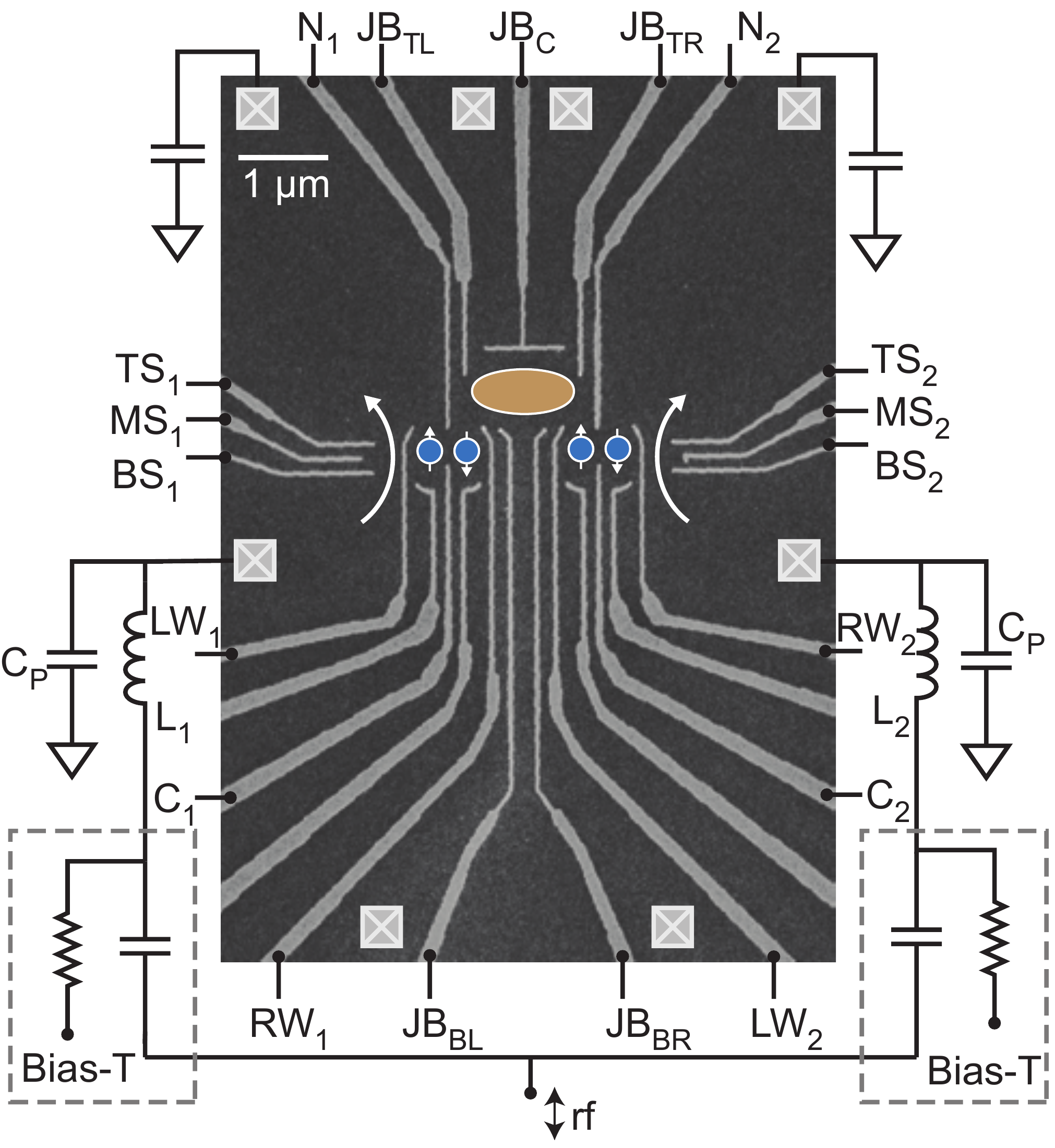}
    \caption{A five quantum dot device in GaAs with additional quantum dot charge sensors either side. White arrows indicate microwave currents used in readout. Blue and orange ovals are electrons.The pairs of blue ovals represent two S-T qubits.}
    \label{fig:spinqubit}
\end{figure}

Modern nanofabrication makes it possible to confine and detect single electron spins in `zero-dimensional' nanostructures referred to as quantum dots (QDs)\cite{Kouwenhoven1997}. The potential that confines the electron (or hole) is produced electrostatically via gate electrodes on the surface of a semiconductor, enabling the number of electrons on the dot and their coupling to the neighboring dots and reservoirs to be tuned by varying gate voltages. The ability to confine, manipulate, and detect single spin states on quantum dots is largely a consequence of the Coulomb blockade of charge, an electrostatic phenomenon arising when the energy to charge a capacitor $C$ by a single electron charge $e$, $E= e^2/2C$, is larger than thermal energy $k_BT$. For sub-micron devices with self-capacitance in the attofarad range, the energy scale for Coulomb blockade requires temperatures below a few kelvin. This necessitates the use of dilution refrigerators for operating spin qubit systems, although work is underway to operate at elevated temperatures\cite{yang2020operation,petit2020universal}.

The initial proposal by Loss and DiVincenzo \cite{Loss1998} for a spin-based quantum computer assumed arrays of coupled quantum dots, each hosting a single electron spin. A large external magnetic field $B_0$ then sets the energy difference between the two spin states aligned or anti-aligned with $B_0$, much as with a trapped ion Zeeman qubit. The qubit resonance frequency is given by $\omega_{01}/2\pi=(\gamma_e/2\pi)|B_0|$. Single electron spin qubits are typically operated in tesla-scale magnetic fields in order to ensure $\hbar\omega_{01}\gg k_B T$, with $\omega_{01}/2\pi\sim1-50$ GHz.

Kane proposed exploiting the exceedingly long coherence of phosphorous donors in isotopically purified $^{28}$Si by coupling their nuclei to localized electron spins for single qubit addressing, two-qubit coupling, and readout~\cite{Kane1998}. Any such qubit architecture requires methods for the precise placement of single atomic donors in a solid. Despite significant experimental progress since Kane's original proposal~\cite{Morello2010}, the MHz (rather than GHz) resonance frequencies of nuclei present a major challenge, leading to kHz clock rates for a quantum computer. The potential for nuclear spins to be used as quantum memories, however, appears more promising~\cite{Morello2010}.

The challenge of requiring GHz-frequency magnetic fields for single spin manipulation can be overcome at the expense of requiring a two-electron system for a single qubit. Here, double quantum dots are used to host two tunnel-coupled electrons, and the qubit is created by the energy splitting of the spin singlet state $\ket{S}$ and one of the three spin triplet\:\footnote{For two spins, each with basis states $\ket{\uparrow}$ and $\ket{\downarrow}$, these states are ${\ket{S}=\frac{1}{\sqrt{2}}\left(\ket{\downarrow\uparrow}-\ket{\uparrow\downarrow}\right )}$, ${\ket{T_+}=\ket{\uparrow\uparrow}}$, ${\ket{T_0}=\frac{1}{\sqrt{2}}\left(\ket{\downarrow\uparrow}+\ket{\uparrow\downarrow}\right)}$, ${\ket{T_-}=\ket{\downarrow\downarrow}}$.} configurations ($\ket{T_+}, \ket{T_0}, \ket{T_-}$)~\cite{Levy2002}. Coupling singlet-triplet (S-T) spin qubits is challenging, and work is underway to devise various architectures to facilitate two-qubit gates. A pair of S-T qubits, each requiring a double quantum dot, is shown in Fig. 5. A multi-electron dot (shown in orange) provides a means of coupling the qubits \cite{jellybean}.

One can extend the idea of using multiple spins to define a qubit by implementing three exchange-coupled electrons. This approach creates a qubit via the relative orientation of spins, rather than the alignment of spins to an external magnetic field. Requiring three quantum dots as host platform, the exchange-only (E-O) qubit~\cite{DiVincenzo2000a} (and its ac variant, the resonant exchange qubit~\cite{Medford2013}) can be controlled entirely with baseband gate voltages rather than microwave magnetic fields. 

The coherence times of semiconductor spin qubits are $\mathrm{T}_\text{1}\sim0.1 - 1$ ms and $\mathrm{T}_\text{2}$ of tens to hundreds of $\mu$s for all of the variants described above.  For a detailed review of spin qubit technology,  we refer the reader to \cite{HansonRMP}.

\subsubsection{Superconducting qubits\label{sec:super}}

Unlike the qubit variants described above, whose degrees of freedom are those of single electrons and atomic nuclei, superconducting qubits are macroscopic devices that are defined at the circuit level and implemented using nominally lossless capacitors, inductors, and Josephson junctions (JJs).  When operated at a low enough physical temperature---typically in the low tens of millikelvins---these circuits display coherent quantum mechanical behavior~\cite{martinis1985energy}, as necessary for use in a quantum processor. Since they are constructed using lumped and/or distributed circuit elements, the properties of superconducting qubits can be engineered similarly to classical circuit structures. As such, a rich family of quantum devices can be realized in this technology platform. Owing to the engineerable nature of these monolithically fabricated devices, the field of superconducting qubit technology has attracted considerable attention since the first device was realized by Nakamura {\it et al.} a little over two decades ago~\cite{nakamura1999coherent}. Here, we focus on one particular type of superconducting qubit, the transmon~\cite{koch2007charge}, currently used in commercial quantum computing efforts~\cite{arute2019quantum}. For a detailed review of superconducting qubit technology, we refer the reader to~\cite{krantz2019quantum}.

\begin{figure}
    \centering
    \includegraphics[width=0.7\columnwidth]{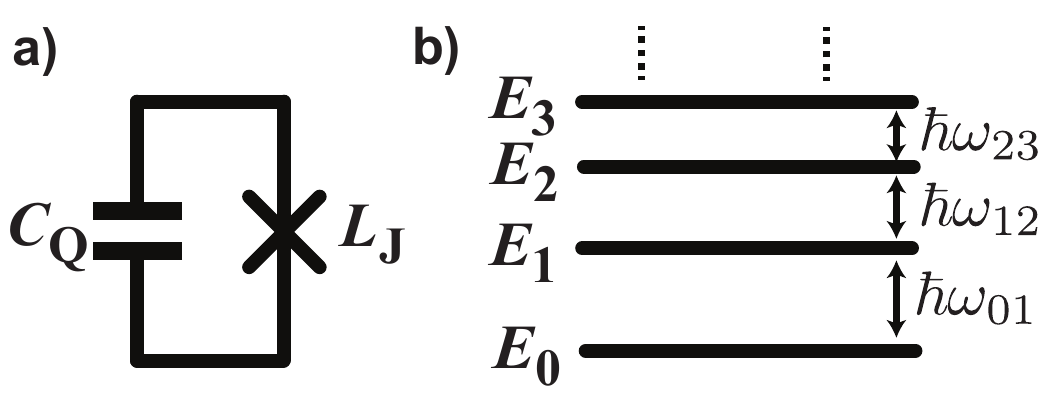}
    \caption{The transmon qubit. (a) Schematic diagram. The ``X'' symbol represents a Josephson junction. (b) Energy diagram. The unequal energy spacings are due to the qubit nonlinearity.}
    \label{fig:transmon}
\end{figure}

A transmon qubit is a 
nonlinear microwave \emph{LC} resonator, constructed by shunting a JJ with a capacitance $C_\text{Q}$, as shown schematically in Fig.~\ref{fig:transmon}(a). The nonlinearity arises from the JJ, which behaves as a current-dependent inductance $L_\text{J}=L_\text{J0}/\sqrt{1-I_\text{J}^2/I_\text{C}^2}$, where $L_\text{J0}=\Phi_0/2\pi{}I_\text{C}$ is the zero-bias inductance of the JJ, $\Phi_0=\pi\hbar/e$ is the magnetic flux quantum, $e$ is the elementary charge, $I_\text{J}$ is the current through the JJ, and $I_\text{C}$ is the critical current of the JJ. The value of $I_\text{C}$ is determined by the JJ geometry and is typically about 40\,nA for transmons, corresponding to $L_\text{J0}\approx{}8\,$nH~\cite{Jimmythesis}.

The nonlinearity causes the transmon's resonant frequency to decrease in proportion to the energy stored in the resonator, leading to an energy level diagram of the form shown in Fig.~\ref{fig:transmon}(b). The non-uniform nature of the energy spacings is referred to as \emph{anharmonicity} and, while quantum engineers typically refer to the transmon qubit as weakly anharmonic, it is actually very non-linear in comparison to typical microwave components.  The resonant frequency shifts by an amount $\eta=-e^2/2\hbar{}C_\text{Q}$ for each microwave photon added to the energy in the transmon, such that all transitions between neighboring transmon energy levels are at different frequencies. For typical component values, $\omega_{01}/2\pi$ and $|\eta/2\pi|$ are in the range of 4--8\,GHz and 150--300\,MHz, respectively. To take an example, for $\omega_{01}/2\pi=6\,$GHz and $|\eta/2\pi|=200\,$MHz, the addition of a single microwave photon of energy $4\times{10}^{-24}\,$J causes a drop in the transmon's resonant frequency of over 3 \%. 

Transitioning from energy $E_\text{i}$ to energy $E_\text{j}$ and vice-versa requires coupling energy into the transmon at frequency $\omega_{ij}$, so one can treat the transmon as an ideal two-level qubit if caution is taken never to excite it in a way which results in leaving the $\left(|0\rangle,|1\rangle\right)$ manifold---that is, one must not drive the device at $\omega_{12}$. Notably, the transmon anharmonicity can be engineered through the choice of $C_\text{Q}$ (while selecting the appropriate JJ sizing to obtain a desired $\omega_{01}$). However, smaller $C_\text{Q}$ results in an increased sensitivity to $1/f$ charge noise, resulting in a practical upper limit on $|\eta|$. As we will see, this places a constraint on the spectral content of the signals used to drive the $\omega_{01}$ transition.

The relaxation time constant $\mathrm{T}_1$ of a transmon is limited by materials losses; recent work has demonstrated values of $\mathrm{T}_1$ as high as 300\,$\mu$s for isolated planar qubits fabricated on a sapphire substrate~\cite{place2020new}. Unfortunately, qubits used in a quantum processor tend to be coupled to additional loss channels, and the best values of $\mathrm{T}_1$ reported for a $\ge25$ qubit quantum processor are about a factor of three lower~\cite{jurcevic2020demonstration}. These losses can occur through couplings to local defects that behave as two level systems (TLSs), among other mechanisms. The transition frequencies of such TLSs can be time dependent, causing $\mathrm{T}_1$ to vary with time \cite{klimov2018fluctuations}. Mitigating such effects is an active area of research.  

The dephasing time constant $\mathrm{T}_\phi$ of single-JJ transmons is typically limited in part by fluctuations in $I_\text{C}$, resulting in time-dependent variations in $L_\text{J0}$. Transmons used in some of today's $\gtrsim$25 qubit quantum processors have $\mathrm{T}_2$ as high as 100\,$\mu$s~\cite{jurcevic2020demonstration}. However, many contemporary quantum processor architectures leverage frequency tunable transmons (described below), and these devices have considerably lower $\mathrm{T}_2$, due to coupling to magnetic flux noise.

\section{Interfacing a microwave source to a qubit\label{sec:interfacing}}

Transitioning between the $|0\rangle$ and $|1\rangle$ states of trapped ion, single-electron spin, and transmon qubits requires exciting the qubit on resonance, which means that we need some mechanism to couple microwave energy to the device. This can be done either electrically or magnetically, and since a qubit can be thought of as a microwave resonator, we can also think of coupling to it just as one would couple to any other microwave resonator. 

In the next sections, we describe some of the considerations related to microwave drive and deterministic state control of qubits in a quantum processor.
\subsection{Drive coupling quality factor, $Q_d$}\label{sec:DC}
The qubit-drive coupling can be quantified in terms of a drive coupling quality factor $Q_d$, defined as the contribution to the loaded quality factor of the qubit resonance due to dissipation in the impedance of the drive source. The value of $Q_d$ sets an upper limit on the qubit's relaxation time constant,
\begin{equation}
    \mathrm{T}_\text{1}\le{}\frac{Q_d}{\omega_{01}}=\mathrm{T}_{1,d}.
\label{T1d}
\end{equation}
Qubits are very under-coupled to the drive source so that $\mathrm{T}_{1,d}\gg{\mathrm{T}_\text{1}}$. However, the degree to which the drive is under-coupled to the qubit varies drastically from technology to technology. 

\begin{figure}
    \centering
    \includegraphics[width=\columnwidth]{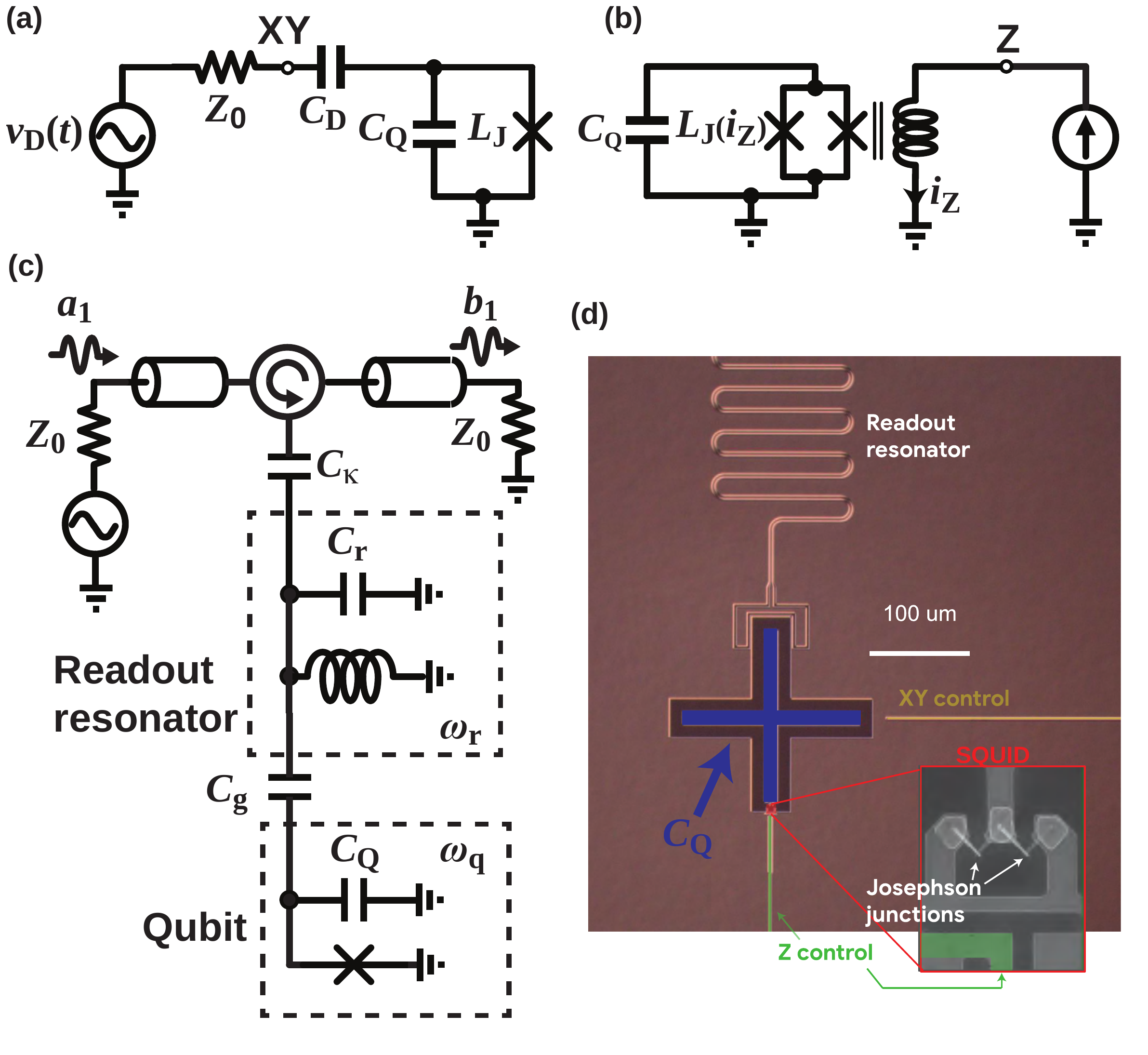}
    \caption{Engineering control and measurement ports into the transmon circuit. (a) An \emph{XY} control port that is capacitively coupled to the qubit permits microwave drive of the $\ket{0}\leftrightarrow\ket{1}$ transition. (b)~Replacing the single JJ with a flux-biased SQUID permits control of $\omega_{01}$ via a current bias. {{(c) A linear resonator is coupled to the qubit for dispersive qubit state readout. (d)~Photograph showing each of these circuit techniques being simultaneously employed.}} Capacitances $C_\text{D}$ and $C_\text{g}$ are those between the coplanar waveguide lines marked ``XY control'' and ``readout resonator'' and the qubit, respectively. In addition, the inductive coupling between the \emph{Z} control line and the qubit is visible in the expanded view.}
    \label{fig:XMONctrl}
\end{figure}

In superconducting qubits, where the qubit is engineered into a circuit environment, the coupling quality factor associated with the microwave drive can be engineered just like in any passive planar circuit. For example, with the capacitive coupling of Fig.~\ref{fig:XMONctrl}(a), $Q_d\approx{}C_\text{Q}/\left(C_\text{D}^2Z_\text{0}\omega_{01}\right)$; thus, $Q_d$ can be set by choosing the coupling capacitor $C_\text{D}$. Materials properties limit the internal quality factor $Q_{i}$ of today's state-of-the-art transmons to about $4\times10^6$~\cite{place2020new}, so $Q_d$ is typically designed to be about an order of magnitude larger such that it does not limit the qubit $\mathrm{T}_\text{1}$. For the typical case of capacitive coupling (Fig.~\ref{fig:XMONctrl}(a)), $Q_d\approx4\times10^7$ corresponds to a coupling capacitance of 30\,aF~\cite{Jimmythesis}. Fortunately, this coupling capacitor---which limits the $\mathrm{T}_1$ of a 6\,GHz qubit to just over 1\,ms---can be readily designed using modern electromagnetic design tools.

The coupling of microwave fields to semiconducting spin qubits, and to trapped ion hyperfine and Zeeman qubits, can be thought of as an inductive coupling: the microwave magnetic field couples to the electron spin, whose magnetic moment is fixed by nature and is ``atom-sized''.  By contrast, superconducting qubits use an electric field coupling to the qubit circuit, where the effective electric dipole moment can be engineered to be much larger than ``atom-sized'' by increasing the dimensions of the qubit circuit.  As an analogy, one can think of the drive as coupling either to an extremely small loop antenna or to a large dipole antenna.  

This difference means that if one puts the different types of qubits at the same distance from a propagating electromagnetic wave on a drive line, the $Q_d$ would be roughly $10^8$ times higher for semiconductor spin qubits and trapped ion hyperfine and Zeeman qubits than for superconducting qubits, given typical superconducting qubit parameters.  The $Q_d$ can also be made larger or smaller by increasing or decreasing (respectively) the distance between the qubit and the drive line, because of the fall-off of drive field strengths.

For semiconductor spin qubits, values of $Q_d$ in the $10^{13}$ to $10^{15}$ range have been reported~\cite{Koppens2006, VanDijk2019}; these qubits were located within $\sim100$ nm of the driving transmission line.  For trapped ions, the smallest reported experimental values of $Q_d$ are ${\sim10^{19}}$~\cite{Ospelkaus2011}, using surface-electrode traps with integrated near-field antenna structures where the ions are $\sim30\,\mu$m from the antenna (see Fig.~\ref{fig:iontrap}(c)).  Many experiments use large three-dimensional traps with microwave horns outside the vacuum chamber, giving substantially weaker coupling and $Q_d$ values up to ${\sim10^{24}}$~\cite{Wang2020}.

\subsection{Rabi oscillation frequency}

Once the value of $Q_d$ is known, we can readily determine the Rabi oscillation frequency as a function of average available power during the pulse $P_\text{av}$, referenced to the qubit drive port 

\begin{equation}
\Omega=2\sqrt{\frac{P_\text{av}}{\hbar{}Q_d}}.
\label{eq:rabi}
\end{equation}

The Rabi frequency is an important metric, in that $\pi/\Omega$ is the time required to flip between the $\ket{0}$ and $\ket{1}$ states. From (\ref{eq:rabi}) we see that the required available power referenced to the drive terminal to achieve a given Rabi frequency is directly proportional to the coupling quality factor. It is also important to emphasize that the Rabi frequency is proportional to signal amplitude (as opposed to power) since the qubit is a coherent device. 

Since the qubit to drive line coupling is strongest for superconducting qubits, the available power required to achieve a given Rabi frequency is lowest. For instance, achieving a Rabi frequency of 50\,MHz with a superconducting qubit having $Q_d$ of $4\times10^{7}$  requires an available power of about $-70$\,dBm. Compare this to a trapped ion qubit with $Q_d\approx10^{19}$, where roughly $+40$\,dBm is required to achieve Rabi frequencies in this range (the highest reported Rabi frequency for a trapped ion qubit using microwaves is 26 MHz~\cite{Ospelkaus2011}).  Most microwave single-qubit gates for trapped ions are implemented with Rabi frequencies below $\sim100$~kHz. For semiconductor spin qubits, achieving a Rabi frequency of 50\,MHz requires an available power in the range of $-16$\,dBm to $+4$\,dBm~\cite{Koppens2006, VanDijk2019}, referenced to the drive line, with the exact level depending upon the mode of inductive coupling to the qubit. This increased sensitivity compared to trapped ions is simply due to the closer proximity of the feed structure to the qubit.   

\subsection{Effect of noise coupled through microwave drive line}
One must also consider the effect of thermal noise coupled to the qubit through the microwave drive line. Since this port is used to drive transitions between $|0\rangle$ and $|1\rangle$, noise at $\omega_{01}$ injected to the qubit through this channel will also induce transitions, leading to a transition rate~\cite{clerk2010introduction}:
\begin{equation}
R_{\uparrow\downarrow,d}=\frac{S_\text{av}}{\hbar{}Q_d},
\label{Rud}
\end{equation}
where $S_\text{av}$ is the spectral density of the noise power at the qubit frequency that is available at the qubit drive port. For a fixed value of $S_\text{av}$, the transition rate is inversely proportional to the coupling quality factor. 

In general, we do not want noise on the drive line to limit coherence, so it is useful to put the required noise levels in context for each of the technologies under consideration. To begin, let us consider the limit in which relaxation through the drive line 
produces the same amount of decoherence as noise emitted from the drive line---that is, $R_{\uparrow\downarrow,d}=1/\mathrm{T}_{\text{1},d}$. Remarkably, keeping the rate of decoherence for these two mechanisms equal requires the effective noise temperature presented by the drive line ($T_{e,d}$) be kept at the single photon level ($T_{e,d}=\hbar\omega_{01}/k_\text{B}$), independent of the coupling strength from the drive line to the qubit.

For the case of superconducting qubits, the rate of relaxation through the drive line is typically within an order of magnitude of the qubit's intrinsic relaxation rate, and it is essential that noise on the drive line not further decohere the qubit. Therefore, the limit described above is relevant, which means that it is essential to attenuate the thermal noise floor well below the single-photon noise temperature ($\approx300\,\mathrm{mK}$ at 6\,GHz). As such, the 
microwave control lines typically feature heavy attenuation, with the final 20--30\,dB of loss thermalized to the lowest temperature stage.

Since semiconductor spin qubits are coupled less strongly to the microwave drive source in comparison to superconducting qubits, they are also considerably less sensitive to noise. For the range of $Q_d$ described in Section~\ref{sec:DC}, the value of $\mathrm{T}_{1,d}$ for $\omega_{01}/2\pi=5$\,GHz is between 5\,minutes and 9\,hours. Since the coherence of semiconductor spin qubits is limited by other mechanisms to timescales considerably shorter than this, the single photon limit described above is not applicable. Limiting the rate of qubit transitions due to noise on the microwave drive line to one per second requires limiting the effective noise temperature of the drive line to between 75 K and 7,500\,K, depending upon the qubit coupling to the drive line. 
\begin{figure*}
    \centering
    \includegraphics[width=2\columnwidth]{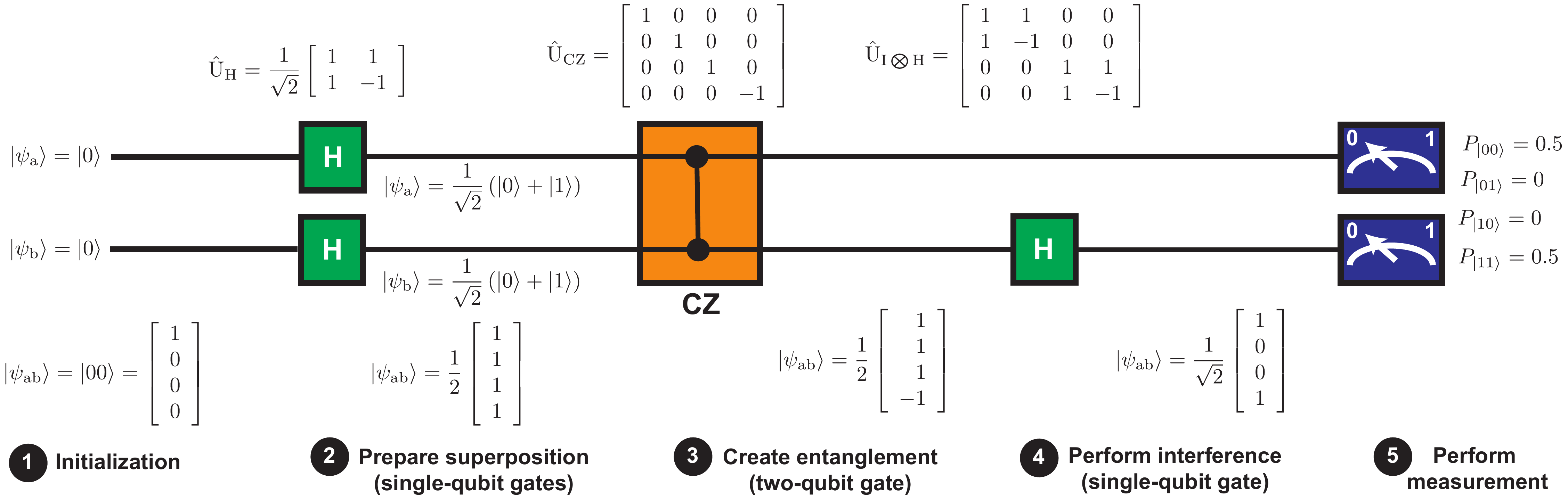}
    \caption{Diagram of a simple quantum algorithm.  The horizontal axis represents time, positive towards the right.  Each horizontal black line represents a qubit (labeled with subscripts as ``a'' and ``b'').  Each green box represents a control operation on a single qubit, while the orange box represents a control operation that entangles two qubits.  The blue boxes at far right indicate measurement of the qubits at the end of the algorithm.  By preparing the identical initial state, running the algorithm, measuring, and then repeating this cycle many times, one can build up statistics about $P_{\ket{00}}$ and the probabilities of the other three possible outcomes.  These probabilities represent the result of the algorithm.}
    \label{fig:q_algo}
\end{figure*}

Trapped ion qubits are so weakly coupled to the microwave drive that a considerable amount of noise (in absolute terms) can be tolerated on the drive line. For instance, in a worst-case scenario using the lowest reported values of $Q_d\approx10^{19}$ and requiring the highest reported Rabi frequency of $\approx25$ MHz, a typical effective drive line noise would be $S_\mathrm{av}=-120$ dBm/Hz, or a noise temperature of $\approx7\times10^7$ K.  This would limit the qubit $\mathrm{T}_1$ to approximately 1 second.  However, since values of $Q_d$ are typically orders of magnitude higher, and substantially lower Rabi frequencies are used, in practice the limit on $\mathrm{T}_1$ due to the drive line noise would be minutes or hours.  

The noise requirements are drastically different for each of the technologies, so it is worth asking whether there is any common ground between the signal and noise requirements. It turns out that it is in fact possible to relate the signal-to-noise ratio on the drive line to the Rabi frequency and transition rates:
\begin{equation}
    \frac{P_\text{av}}{S_\text{av}}=\frac{\Omega^2}{4R_{\uparrow\downarrow}}.
\end{equation}
Thus, once a desired Rabi frequency and Rabi frequency to transition rate ratio (effectively the average number of transitions which can be coherently driven before a noise-induced transition occurs) are determined, the required signal to noise per unit bandwidth is easily calculated using this universal relationship. Shaped pulses have a peak to average ratio which is greater than unity, so additional margin is required if using such control waveforms.

\section{Coherent control of quantum processors using microwave techniques\label{sec:micmethods}}

Thus far, we have described how one might drive a single qubit between the $\ket{0}$ and $\ket{1}$ states, but to perform quantum computing, we need coherent control of the full multi-qubit complex state vector.  This section describes the role of microwaves in this process. 

To get a sense of the type of control needed to implement a quantum algorithm, we will begin by considering the simple quantum algorithm shown in Fig.~\ref{fig:q_algo}, which is used to generate an entangled state known as a Bell state: $\ket{\psi}=\frac{1}{\sqrt{2}}\left(\ket{00}+\ket{11}\right)$.  The quantum state can be written as a four-element vector representing the four complex amplitudes $\alpha_{00}$, $\alpha_{01}$, $\alpha_{10}$, and $\alpha_{11}$.  Control operations, called gates, can be represented by matrices that act on the state vector, as shown in Fig.~\ref{fig:q_algo}.

The algorithm begins by resetting both qubits (labeled ``a'' and ``b'') to the $\ket{0}$ state. Then a series of quantum gate operations are carried out. First, each qubit is placed in a superposition state by applying a so-called Hadamard gate H to each of the qubits. After this step, the two qubits are in an equal superposition of each of the four possible basis states. Next, the qubits are entangled via a controlled \emph{Z} (CZ) gate, which inverts the sign of $\alpha_{11}$ while leaving the other amplitudes unaltered. While it is not obvious from the measurement statistics (which are unaltered by the application of the CZ gate), this is an entangled state, since it is no longer possible to describe the joint state $\ket{\psi_\text{ab}}$ as a product of single qubit states. A Hadamard gate is then applied to the second qubit, which produces constructive and destructive interference between $\alpha_{00}$ and $\alpha_{01}$, as well as $\alpha_{11}$ and $\alpha_{10}$, resulting in the production of the desired entangled Bell state. Finally, a measurement is carried out on both qubits. 

While the algorithm described above only involves a pair of qubits, it turns out that a library of gate operations giving full state control of a single qubit (described by $2\times2$ unitary matrices), combined with a single two-qubit entangling gate (described by a $4\times4$ unitary matrix), is sufficient to implement a universal quantum algorithm---that is, one can decompose any arbitrary $2^N\times2^N$ unitary operator into a sequence of these basic operations, each of which is applied to either a single qubit or a pair of qubits~\cite{Nielsen2000}. This so-called \emph{universal gate set} can be thought of in analogy to how all digital logic operations in a classical computer can be constructed from NAND gates, for example.  Just as NAND gates are only one of many possible choices of a universal gate for classical computing, there are many choices for the universal gate set used in quantum computing; the particular choice of the universal gate set varies from technology to technology, since each technology has its own particularly convenient set of ``native'' gate operations. However, in contrast with classical computing, where a gate is thought of as physical object implemented with transistors to which bits are brought to carry out logical operations, a quantum gate is an operation applied directly to a qubit or a pair of qubits \emph{in situ}.  Quantum gate operations are often carried out using microwave techniques.

When thinking about qubit control and measurement (to be discussed in Section~\ref{sec:measurement}), a natural question to ask is: how good must our control and measurement be?  To what extent can we tolerate errors in either?  The answer depends on a variety of factors, but in general, the lower the errors, the larger the algorithm that can be run successfully, so striving for lower error is important.  State-of-the-art error rates for control and measurement are currently in the range of $10^{-2}$ to $10^{-6}$ per operation, depending on the type of operation and the physical qubit implementation.  These are much higher than typical error rates in classical computing hardware.  The primary reason for this difference is that analog fluctuations on a digital signal in classical computing must have a certain minimum amplitude before the signal crosses a digital logic level and gives rise to a logical error, affording high intrinsic noise immunity.  By contrast, any amount of noise or miscalibration in the operations of a quantum computer can affect the continuously-variable amplitudes $\alpha$ of the computational basis states, potentially altering the result of the computation\footnote{We stress that quantum computing is not analog computing, in the sense of circuit-based analog classical computers.  A quantum computer uses a discrete set of operations performed on a discrete set of basis states, and gives a digital output; control and measurement errors can be thought of as digital errors appearing with some probability, a fact which underlies the ability to perform quantum error correction~\cite{Nielsen2000}.}.

Since it is never possible to eliminate errors completely, error correction techniques must be used for large-scale computations.  Quantum error correction (QEC) is a large field of active research, and we will not attempt to give details in this paper, directing the reader instead to~\cite{Nielsen2000, Brun2019, roffe2019quantum} and references therein.  However, a general rule of thumb is that control errors below a threshold of roughly $10^{-4}$ per gate are low enough for most QEC protocols~\cite{Knill2005}, and some can tolerate errors as high as roughly 1 \% per gate~\cite{Fowler2012}. Another general rule of thumb is that the larger the gate error rate, the more resources (in terms of qubits, gate operations, and measurements) are required to implement the QEC protocol, an overhead that can become very cumbersome as error rates near the threshold. A practical target for average control error rates is $10^{-4}$ per gate. Since control errors are an aggregation of many different physical error mechanisms, including decoherence mechanisms intrinsic to the qubits themselves, meeting this goal means that each error contribution should be significantly lower; here we consider the goal of a maximum error contribution of $10^{-5}$ for each of the control error mechanisms. Readout errors should generally be at similar levels, although certain error correcting codes can tolerate readout errors as high as $\sim1$ \%.

\subsection{Single qubit gates}

\begin{figure}[bt!]
\centering
\includegraphics[width=0.4\columnwidth]{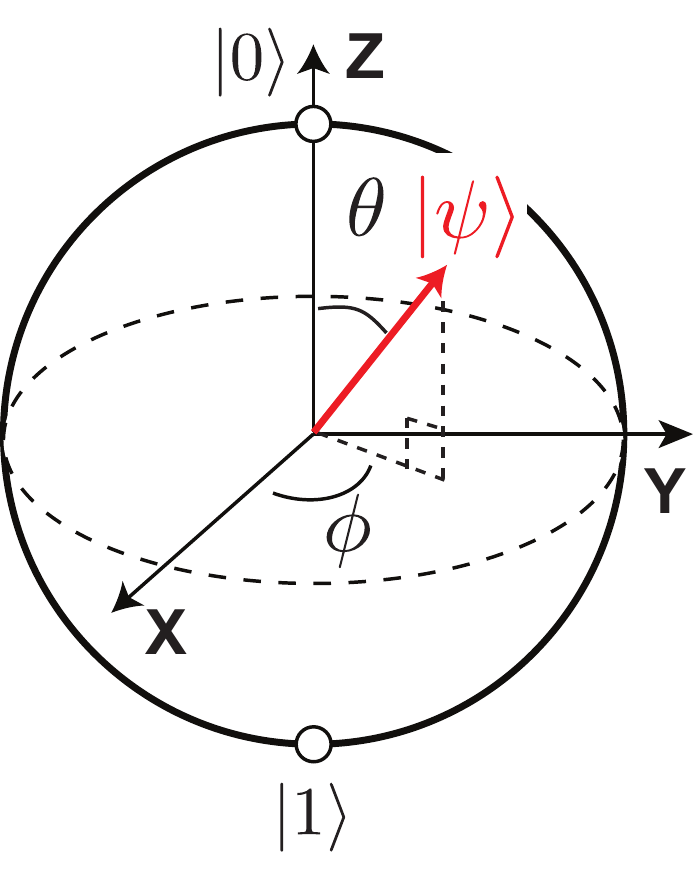}
\caption{Bloch-sphere representation of a single-qubit state.} 
\label{fig:BS}
\end{figure}

While Rabi oscillations describe the response of a qubit to the amplitude of a resonant drive signal, a qubit is a coherent device, and just like the IQ outputs of a direct conversion receiver depend upon the relative phase relationship between the LO and RF signals, the response of a qubit also depends upon the phase of the drive signal. To help understand this relationship, we consider the single-qubit state vector, $|\psi\rangle=\cos\left(\theta/2\right)|0\rangle+e^{j\phi}\sin\left(\theta/2\right)|1\rangle$. 
Since the qubit state is completely described by angles $\theta$ and $\phi$, it can be thought of as a vector terminating on the surface of a unit sphere, which is referred to as the Bloch sphere (Fig.~\ref{fig:BS}).\footnote{While the Bloch sphere representation of a single-qubit state is a valuable tool for gaining intuition into single-qubit gate operations, once two qubits are entangled, we can no longer separate their states and the Bloch sphere picture fails to be meaningful.} In this representation, the $|0\rangle$ and $|1\rangle$ states map to the north and south poles, whereas all other points correspond to unique superposition states. In the Bloch sphere picture, single-qubit quantum gate operations can be thought of as rotations of the qubit state vector. To construct a universal single-qubit gate set, we must be able to perform deterministic rotations of the qubit state about the \emph{X}, \emph{Y}, and \emph{Z} axes. 

\subsubsection{XY gates}
Single-qubit gate operations are typically classified into two types of gates, \emph{XY} and \emph{Z}, which each typically have their own physical implementation. As the name suggests, \emph{XY} gates produce rotations about an axis in the \emph{XY} plane of the Bloch sphere. Since resonant microwave excitation of the qubit produces oscillation between the $\ket{0}$ and $\ket{1}$ states, microwave pulses can be used to mediate \emph{XY} gates. As the Rabi frequency is proportional to the drive amplitude, we can rotate the state by a deterministic amount simply by controlling the integrated envelope amplitude of the microwave pulse.  This can be accomplished by controlling the envelope amplitude and pulse duration. The axis of rotation is set by the microwave carrier phase, so adding control of this degree of freedom allows access to an arbitrary set of \emph{XY} gates. Additionally, detuning the drive from $\omega_{01}$ causes the axis of rotation to tilt away from the \emph{XY} plane, so drive frequency can be used as an additional degree of freedom. Achieving a 180$^\circ$ rotation (also called a `$\pi$ pulse') in a duration $\tau_\text{g}$ nanoseconds while limiting the impact of each of the error mechanisms to below $10^{-5}$ requires control of the integrated envelope amplitude, carrier phase, and carrier frequency offset to better than 0.25 \%, 0.22$^\circ$, and (2/$\tau_\text{g})$\,MHz, respectively. 

In general, it is desirable that gate operations be carried out as quickly as possible to limit the impact of decoherence on the achieved gate fidelities. However, the spectral width of a control pulse is inversely proportional to pulse duration, so it is essential to design these pulses properly to avoid driving undesired off-resonant transitions. This is especially true when working with transmon qubits, where the typical separation of $\omega_{01}$ and $\omega_{12}$ is about $2\pi\times200$\,MHz, on the order of $\Omega$. The leakage rate to an undesired transition $\omega_{ij}$ can be estimated as the relative energy at $\omega_{ij}$ to that at $\omega_{01}$~\cite{steffen2003accurate}, so pulse shaping techniques are typically employed when driving transmons to optimize the tradeoff between pulse duration and $\omega_{12}$ drive. Common envelope waveforms include Gaussian and raised cosine shapes, which have much reduced frequency-domain sidelobes in comparison to a rectangular envelope. These simple envelopes are sufficient to achieve pulse durations as short as about 20\,ns, but reaching shorter gate durations with transmon qubits requires further waveform optimization. First, to further suppress drive of $\omega_{12}$, one can employ the derivative removal by adiabatic gate (DRAG) technique, in which a notch is generated at $\omega_{12}$ by adding a quadrature derivative term to the baseband envelope~\cite{motzoi2009simple,chow2010optimized}. While this takes care of the $\omega_{12}$ leakage term, moving to shorter gate durations requires larger amplitudes and, due to the ac Stark effect, the effective value of $\omega_{01}$ becomes amplitude-dependent. As such, a time-varying detuning also must be applied to the microwave carrier signal~\cite{chen2016measuring}. 

In trapped ion qubits, \emph{XY} gate fidelity is typically limited by errors in the microwave pulse parameters as described above, rather than qubit decoherence during the gate, especially for clock qubits.  For this reason, high-fidelity microwave gate pulses are generally of longer duration, $\sim1\,\mu$s to $\sim100\,\mu$s, which allows more fine-grained control of pulse durations and thus integrated pulse amplitude.  The highest fidelity single-qubit \emph{XY} gates reported to date in trapped ion qubits have infidelities of $1.0(3)\times10^{-6}$ per gate, with a $\pi$-pulse duration of 24 $\mu$s~\cite{Harty2014a}.  Because of their longer duration, the spectral content of these control pulses is fairly narrow.  Off-resonant coupling to other states is generally negligible, given typical separations to the nearest neighboring hyperfine transitions of $\gtrsim5$ MHz, and sometimes $\sim100$ MHz, depending on the ion species and the magnetic field.  Zeeman qubits have no additional neighboring levels and so  off-resonant excitation of other levels is not a concern.

For simplicity, pulses are generally rectangular, without shaped rise and fall times, because even the broader spectral content from the sharp pulse edges is far detuned from other transitions.  Pulses with durations much longer than $\sim100\,\mu$s can become problematic for field-sensitive qubits, if the fluctuations of the qubit frequency due to environmental magnetic field noise start to become comparable to the Rabi frequency $\Omega/2\pi$. 

Unlike with superconducting qubits, where each qubit has a dedicated drive line, microwave control fields are relatively uniform over an array of trapped ions, providing only global control.  This arises because the ion-ion spacing is much smaller than either the distance to the driving antenna/horn or the free-space wavelength of the control fields.  In surface-electrode traps, ``regional'' control can be achieved through ``beamforming'' from multiple antennas spaced a suitable distance apart~\cite{AudeCraik2017}.  Individually addressed \emph{XY} gates for closely-spaced ions can be realized in one of several ways.  For magnetic-field-sensitive qubits, an applied magnetic field gradient along the ion string causes each ion to have a separate qubit frequency, so that frequency domain addressing is possible~\cite{Piltz2014}.  Alternatively, focused laser beams, or microwave field gradients from near-field electrodes, can be used for individual addressing by creating differential Rabi frequencies or differential qubit frequencies on multiple ions, among other techniques~\cite{Leibfried1999, Staanum2002, Warring2013, Srinivas2020}.  These latter methods can be used on both field-sensitive and field-insensitive (clock) qubits.  The full literature for trapped ion qubit individual addressing techniques, including individual addressing using laser beams, is extensive and is not referenced here.  

Implementing \emph{XY} control of single semiconductor spin qubits requires microwave magnetic fields $B_1$, applied orthogonal to the direction of $B_0$, and typically generated from dedicated on-chip antenna structures near each qubit. An alternate approach uses a global cw microwave magnetic field on all qubits in the array~\cite{Kane1998}. Individual qubits are then tuned in and out of resonance with the global field via local gate electrode voltages, effectively pulling or pushing the electron wavefunction towards an interface to modify the $g$-factor \cite{Kane1998}, or away from the nuclear spin of a donor atom to vary the hyperfine coupling \cite{Kane1998, Morello2010}. 

For S-T qubits, static magnetic field gradients between the two dots that make up the qubit can drive \emph{XY} rotations. For control of these rotations, nanosecond rectangular pulses are used to separate the two electrons for a time such that they experience different magnetic fields. These gradients are either produced naturally by hyperfine magnetic fields from neighboring nuclear spins, or engineered using micro-magnets on the surface of the semiconductor (the latter is better controlled and gives reduced gate error). Resonant microwave driving at the frequency corresponding to the exchange energy ($\lesssim1$~GHz) can also be used for manipulation of S-T qubits, with the advantage that the drive frequency is much lower than for single-spin qubits~\cite{Takeda2020}.

The need for magnetic fields can be alleviated altogether using the three-electron E-O spin qubit. Here, qubit \emph{XY} control is implemented using modulation of the exchange energy between two of the electrons, proportional to the overlap of their wavefunctions and controllable using time-dependent voltages applied to surface gates, similar to S-T qubits.  Exchange between the right-most spin pair drives qubit rotations about the \emph{Z} axis, whereas  exchange between the left-most pair drives rotations about an axis tilted by 120 degrees from the \emph{Z} axis. Concatenating up to four pulses produces single qubit rotations around any axis. 
Again, these pulses are typically baseband square pulses with nanosecond rise and fall times. The use of the exchange interaction for qubit \emph{XY} control reduces microwave complexity, but does so at the cost of increased sensitivity to voltage noise on the control line and charge noise in the material system. 

\subsubsection{Z gates}

In addition to $XY$ gates, a universal gate set requires rotations about the $Z$ axis, which are referred to as $Z$ or phase gates. Referring to Fig. \ref{fig:BS}, \emph{Z} gates only affect the qubit phase $\phi$. One can perform these gates either virtually, by applying a phase jump to the RF carrier used for subsequent $XY$ gates~\cite{mckay2017efficient}, or physically, either by applying a sequence of two $XY$ gates whose combined phase values yield a prescribed $Z$ rotation or by detuning the qubit frequency by $\delta\omega_{01}$ for a controlled duration $\tau$, so as to accumulate phase (similar to how phase is adjusted in a phased locked loop during acquisition). Virtual $Z$ gates are attractive since they can be instantaneously applied to baseband IQ envelopes and their accuracy just requires a stable system clock, which is necessary to generate high fidelity $XY$ gates in the first place. However, for systems with more than one qubit, control of the relative phases of qubits is often necessary, which requires physical \emph{Z} gates.  

Minimizing error rates requires carrying out physical \emph{Z} gates quickly, so it is often preferable that these gates be mediated via a frequency tuning mechanism rather than via multiple \emph{XY} gates. For superconducting transmon qubits, control of $\omega_{01}$ can be enabled by replacing the single JJ of Fig.~\ref{fig:transmon}(a) with a two-JJ loop, known as a superconducting quantum interference device (SQUID) (Fig.~\ref{fig:XMONctrl}(b)). In this context, the SQUID behaves as a magnetic-flux-tunable nonlinear inductor, so the qubit frequency can be externally controlled via a current bias, analogous to how a voltage controlled oscillator is tuned via a control voltage. Coupling is kept very weak to minimize frequency fluctuations due to noise on the flux-bias tuning line, with typical mutual inductances in the range of just 2\,pH. This weak coupling leads to control currents of a hundred microamps or more, and nanoamp-level resolution is required to minimize errors. Digital-to-analog converters (DACs) with at least 14 bits of resolution are typically used to generate these currents.

For gate-defined semiconductor spin qubits, the specific method of producing  \emph{Z} rotations depends on the flavor of qubit, but generally amounts to applying a rectangular pulse to modulate the energy detuning or the exchange interaction between electrons. For single spins, square-shaped pulses can be applied to vary the position of the electron wavefunction, modulating the Zeeman energy for the duration of the pulse. For S-T qubits, 
turning on the exchange coupling swaps the two spins in the presence of the magnetic gradient, effectively performing a rotation about the \emph{Z} axis. Similarly, for the three electron E-O qubit, rotations about \emph{X,Y,} or \emph{Z} are produced via application of concatenated rectangular pulses to modulate the exchange interaction between left and middle electrons, or right and middle electrons. 

Physical \emph{Z} rotations in trapped ion Zeeman qubits and magnetic-field-sensitive hyperfine qubits (where $\frac{\partial\omega_{01}}{\partial{}B_0}\neq 0$) could be implemented by changing the magnetic field $B_0$ to shift the qubit frequency by $\delta\omega_{01}$ for a specified duration $\tau$, yielding a \emph{Z} rotation by $\delta\omega_{01}\tau$.  However, the static magnetic fields in trapped ion experiments are usually generated by highly stabilized current sources or permanent magnets, which are not amenable to rapid field shifting and would provide only global \emph{Z} rotations.  Therefore, physical \emph{Z} rotations are typically implemented by compiling them to \emph{XY} gates, applying ac Stark shifts to the ions using detuned laser beams, or applying ac Zeeman shifts to the ions using detuned microwave or rf magnetic fields.  Individually addressed \emph{Z} rotations can be applied using focused laser beams to shift only specific ions, or by using a gradient of an rf/microwave magnetic field~\cite{Warring2013, Srinivas2019}.  These techniques have the added benefit that they can be applied to hyperfine clock qubits as well.

\subsection{Multi-qubit gates}
\begin{figure}
    \centering
    \includegraphics[width=\columnwidth]{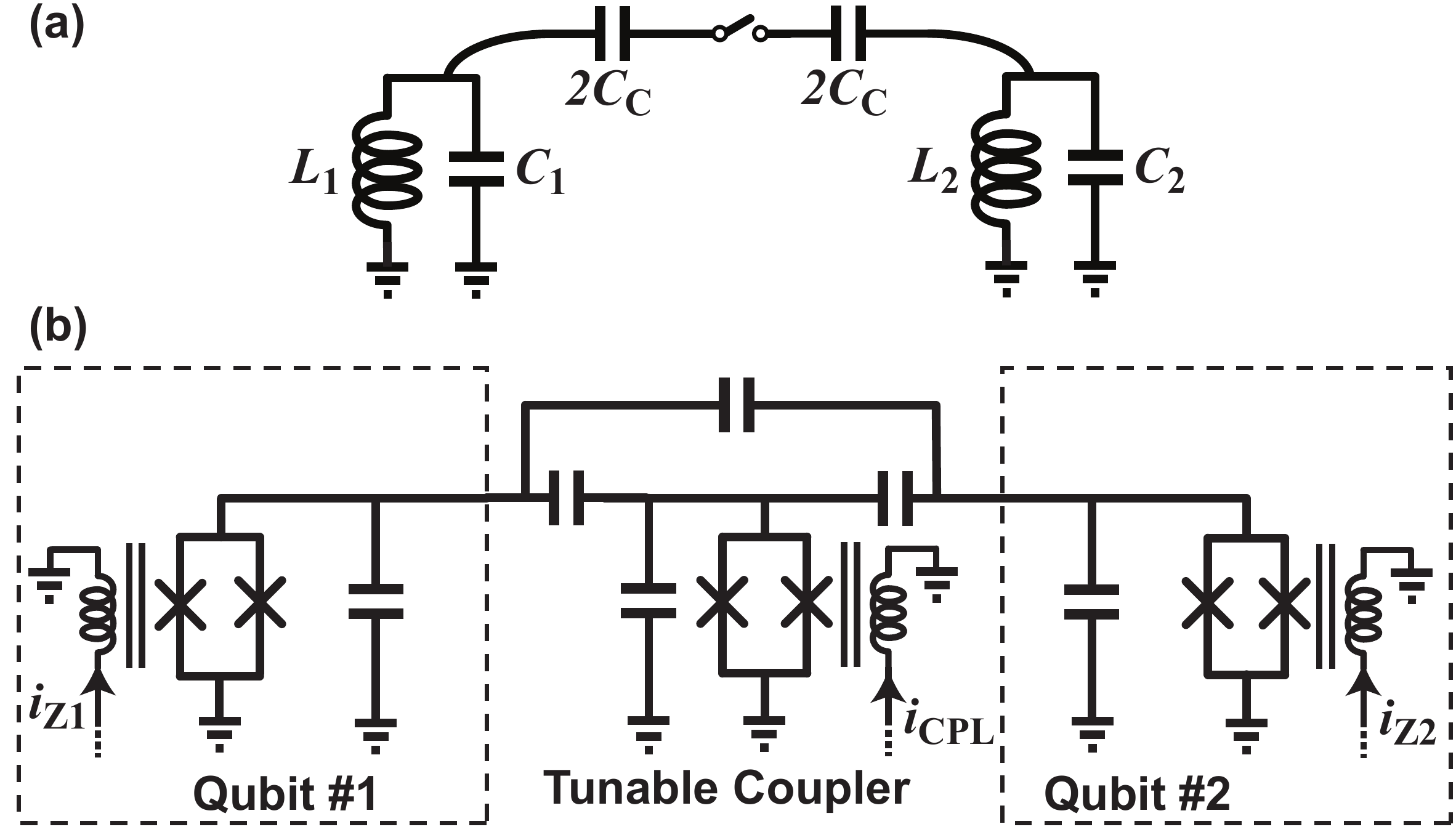}
    \caption{(a) Coupling of linear resonators via a switched capacitance. (b) Coupling of transmon qubits via a tunable coupler. Two qubit gates can be mediated using the qubit and coupler bias currents ($i_\text{Z1}$, $i_\text{Z2}$, and $i_\text{CPL}$).}
    \label{fig:LRcpl}
\end{figure}
While a library of single qubit gates allowing arbitrary control on the Bloch sphere is required for universal quantum computing, this alone is not sufficient. As described earlier, we also need to create entangled states if we are going to exploit the exponentially large computational space. To do this, we must carry out operations involving multiple qubits, which requires entangling the qubits in a deterministic manner. To understand how one might interact qubits, we can begin by considering the circuit of Fig.~\ref{fig:LRcpl}(a), in which two linear resonators are coupled through a switched capacitor network. Assuming an ideal switch in the open position, the resonators are isolated and the natural frequencies of the circuit are just those of the individual resonators: $\omega_{1}=1/\sqrt{L_1C_1}$ and $\omega_{2}=1/\sqrt{L_2C_2}$. However, when the switch is closed, the resonators pull each other and the natural frequencies of the circuit shift. If each resonator was initially oscillating at the respective resonant frequency, the oscillation dynamics would shift from a distinct frequency at each node to the joint modes of the coupled resonator system (which will be shifted from the bare frequencies so long as $\omega_{1}$ and $\omega_{2}$ are distinct). In the case where the coupling is weak, we can think of the coupling as a perturbation applied to each of the isolated resonance frequencies. As such, if we energize one of the resonators and pulse the switch closed for a short period of time, the oscillation in the resonator will acquire a phase shift due to the temporary shift in resonance frequency.

Now, imagine both of these resonators have a nonlinear relationship between their isolated natural frequencies and the amplitude of oscillation (similar to a qubit). In this case, if we perform the same experiment where we energize one of the resonators, pulse the coupling on, and look at the acquired phase, the result will depend upon the state of the second resonator, since the degree to which it pulls the first resonator is state dependent. This behavior describes one way the basic interactions required to perform entanglement generating two-qubit gates can be carried out. 

Two-qubit gates require a mechanism for interacting qubits in a deterministic manner, and many different approaches have been proposed and demonstrated. For instance, interactions between superconducting qubits can be engineered at the circuit level by introducing either static~\cite{barends2014superconducting} or tunable couplings~\cite{arute2019quantum}, which may be implemented inductively or capacitively. One approach is shown in Fig.~\ref{fig:LRcpl}(b), where a tunable coupler allows interactions between a pair of frequency tunable transmons~\cite{yan2018tunable}. The coupler itself consists of a transmon---which serves as a frequency tunable \emph{LC} resonator---that is capacitively coupled to each of the qubits. With this structure, it is possible to realize coupling strengths ranging from completely off to tens of MHz. One can engineer a wide range of two-qubit gates through proper design of the three current bias waveforms for this circuit~\cite{foxen2020demonstrating}. For instance, if the qubits are tuned into resonance and the coupling is enabled, a single excitation will oscillate back and forth between the qubits (that is, oscillation will occur between the probability amplitudes of the $|01\rangle$ and $|10\rangle$ states), and by properly setting the gate duration it is possible to engineer a gate where $\alpha_{01}$ and $\alpha_{10}$ are swapped.

While two-qubit gates for superconducting qubits can be carried out through control of the qubit frequencies as described above, they can also be performed without changing the qubit frequency, using an additional microwave drive tone instead. Such approaches are necessary in architectures that use fixed-frequency qubits~(see for example Ref.~\cite{jurcevic2020demonstration}). One example of an all-microwave two qubit gate is the cross-resonance (CR) gate~\cite{rigetti2010fully}, in which one of a pair of reactively-coupled qubits is driven by a microwave tone at the qubit frequency of the other qubit. With the appropriate drive amplitude and duration along with the addition of a single qubit gate applied to each of the qubits, the CR gate can be used to implement a CNOT operation, which swaps $\alpha_{10}$ and $\alpha_{11}$ while leaving $\alpha_{00}$ and $\alpha_{01}$ unchanged.  The gates described here are just two of a wide assortment of gates that can be applied to superconducting qubits; for further discussion of two-qubit gates in superconducting qubits, we refer the reader to Ref.~\cite{krantz2019quantum}.

Two-qubit entangling gates in semiconductor spin qubits are generally implemented via the Heisenberg exchange interaction between electrons on neighbouring dots of an array\cite{Loss1998}. The charge dipole associated with the two-electron system opens the prospect of coupling S-T qubits capacitively, since the relative spin orientation of one qubit can lead to charge rearrangement that effectively gates another qubit \cite{Shulman2012}. The advantage of exchange coupling is its controllability, modulating the tunnel coupling between two adjacent quantum dots for a controlled amount of time. The evolution of the two-qubit system then depends on the wave-function overlap of the electron states, resulting in the physical exchange of the electron positions to execute a $\sqrt{\mathrm{SWAP}}$ entangling gate. A major technical challenge for exchange coupled qubits is the  sensitivity of the tunnel rate to gate voltage and the requirement that electrons must be brought within   nanometers of each other for coupling. This later aspect leads to crowding of gate electrodes and challenges for crosstalk mitigation. Alternative coupling schemes making use of intermediate electron states \cite{jellybean} are presently an active area of research.  A further fruitful direction is to couple remote S-T qubits via a cavity resonator, following similar approaches to superconducting qubits \cite{hybrid}. To date, two-qubit gates using E-O qubits have not been demonstrated, although qubit coupling schemes are likely to be similar to S-T qubits.

Trapped ion hyperfine and Zeeman qubits have negligible direct interaction with each other due to very weak spin-spin coupling.  However, the motion of multiple ions in a single trapping potential is very strongly coupled.  As a result, almost all entangling gates carried out between trapped ion qubits are realized using the quantum motion of the trapped ions as an intermediary ``bus''.  An effective ion-ion interaction is realized by coupling the trapped ion spin to its motion in an appropriate way, using external control fields.  Typically this is done with laser beams, but it is also possible to do using rf and/or microwave fields.  Crucially, spin-motion coupling requires a spatial gradient of the control field over the spatial extent of the ions' quantum mechanical zero-point motion in the trap, typical $\sim10$ nm.  The magnetic field gradient of free-space microwaves near $\omega_{01}$ (usually a few GHz) is very small over this length scale.  However, microwave magnetic fields with negligible gradient strength can be combined with additional static~\cite{Mintert2001, Johanning2009a} or few-MHz~\cite{Sutherland2019, Srinivas2019} magnetic field gradients to produce the desired spin-motion coupling.  Magnetic field gradients at microwave frequencies near $\omega_{01}$, or near resonance with the ion motional frequency, can also be used for spin-motion coupling and entangling gates~\cite{Wineland1998, Ospelkaus2008}.  These gate protocols are often carried out in surface-electrode traps, where larger near-field magnetic field gradients can be generated~\cite{Ospelkaus2011, Warring2013a, Welzel2018}.  

A number of experimental demonstrations of high-fidelity microwave-based entangling gates between ions have been carried out~\cite{Harty2016, Weidt2016, Hahn2019, Zarantonello2019, Srinivas2020}.  In general, these gates are slower ($\sim$ ms duration) than laser-based entangling gates, which can be performed in tens or hundreds of $\mu$s (and some as fast as a few $\mu$s~\cite{Schafer2018}).  However, the fidelities reported for microwave-based gates, with errors in the few $10^{-3}$ range per gate, are competitive with the fidelities of laser-based gates.  In addition, laser-based gates have fundamental fidelity limitations due to off-resonant scattering from excited electronic states in the ion~\cite{Ozeri2007}, making microwave-based gates, which do not have this fundamental limit, an appealing alternative.

\subsection{Hardware for quantum state control}

\begin{figure}
    \centering
    \includegraphics[width=\columnwidth]{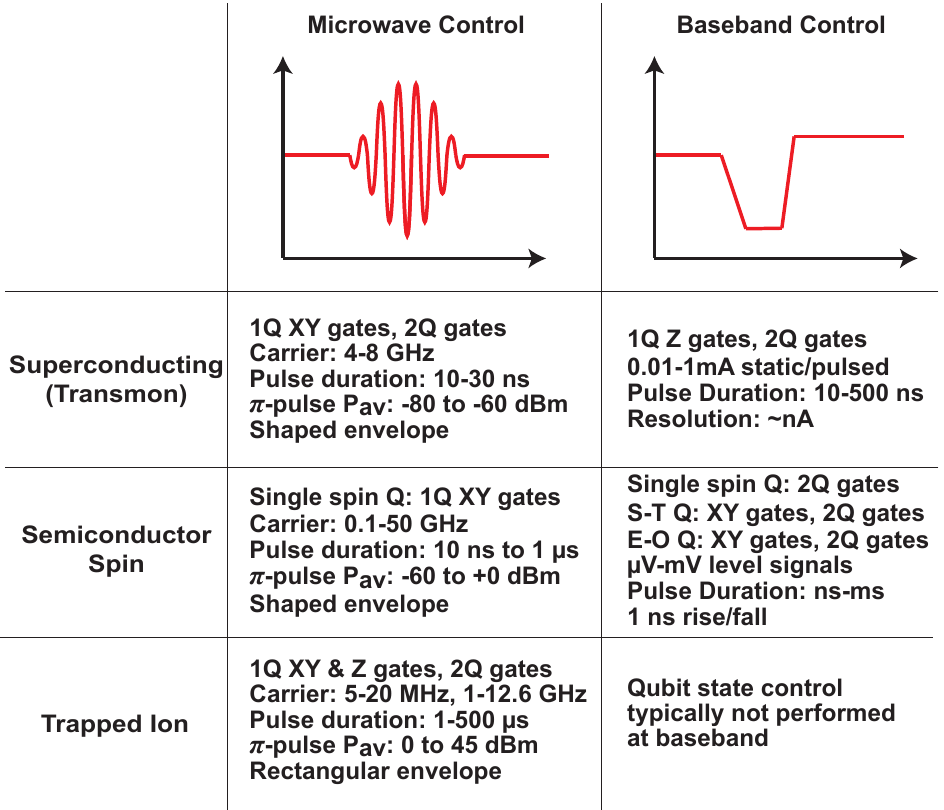}
    \caption{Summary of microwave/baseband control requirements for each of the qubit technologies. Abbreviations: Q--qubit, $P_\text{av}$--available power at qubit drive port.}  
    \label{fig:waves}
\end{figure}

\begin{figure}
    \centering
    \includegraphics[width=\columnwidth]{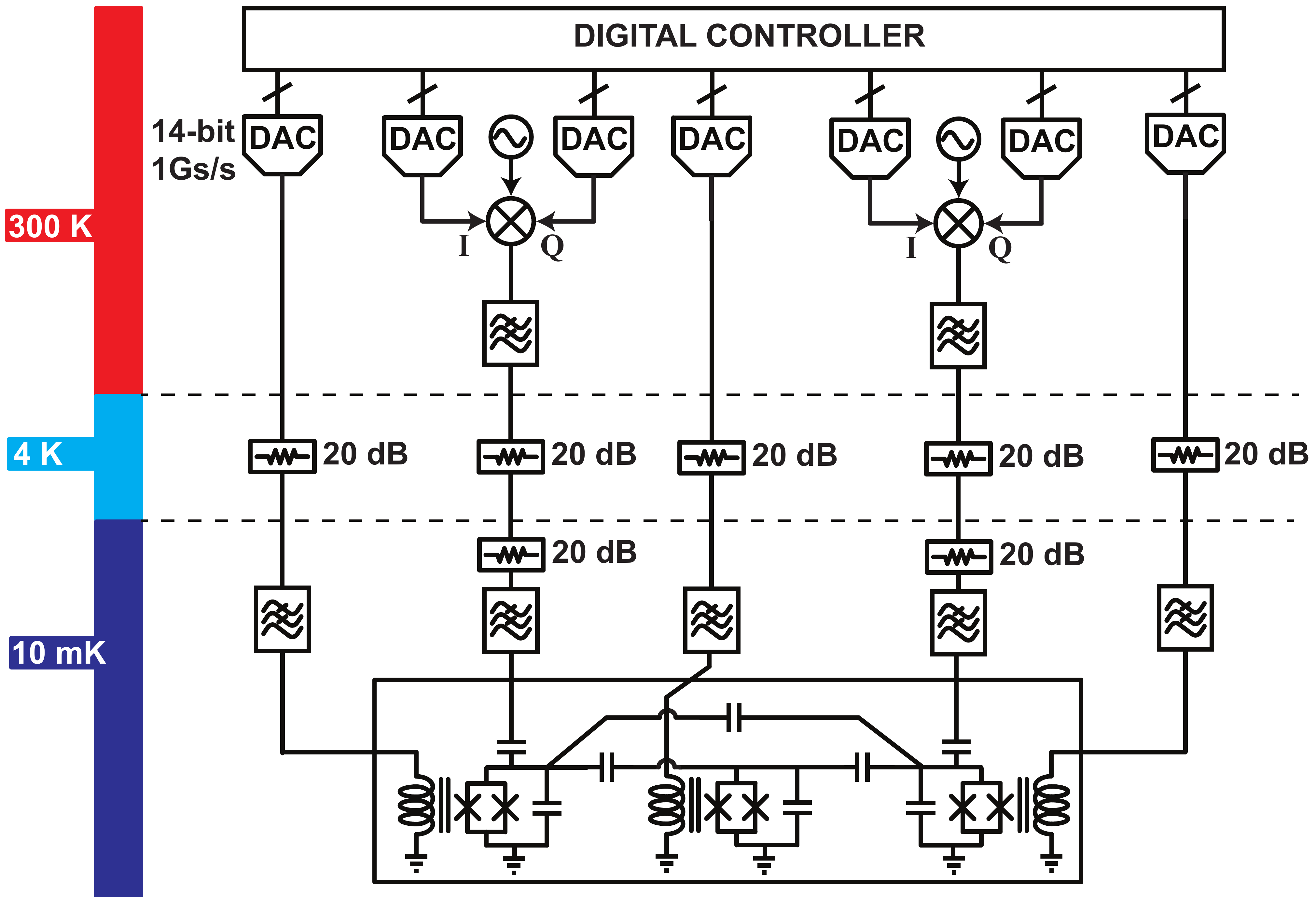}
    \caption{Simplified control hardware for a two-qubit transmon-based quantum processor with tunable coupler.}
    \label{fig:sc_control}
\end{figure}

High-level control requirements for each of the three qubit technologies are summarized in Fig.~\ref{fig:waves}. While the details vary greatly among the technologies, some form of pulsed RF waveforms and/or baseband control signals are required to run a quantum processor. An exemplary control system for a pair of superconducting qubits connected via a tunable coupler appears in Fig.~\ref{fig:sc_control}. Each \emph{XY} control signal is generated using single-sideband mixing, with the complex envelope generated using a pair of high-speed DACs. Alternatively, the RF \emph{XY} signals are sometimes directly generated using high speed DACs~\cite{raftery2017direct,kalfus2020high}, obviating the need for analog mixing. The \emph{XY} signals are heavily attenuated to suppress thermal noise at the qubit drive port. An additional three DACs generate the broadband control signals required to drive the \emph{Z} control lines and coupler bias port. These DACs feature 14 bits of resolution, as required for the frequency control of the qubits and coupler. The digital waveforms are generated using a field-programmable gate array (FPGA), which is configured to orchestrate quantum algorithms. Architectures such as this are extensible to hundreds of qubits and widely in use among the quantum computing community. Part of the reason that the field of quantum computing has experienced rapid growth over the past ten years is that components required to build control systems such as this have become commodity items, thanks to the wireless communications revolution.


\section{Measuring the state of a qubit\label{sec:measurement}}

To behave quantum mechanically, as required for quantum computation, qubits must be well-isolated from sources of noise or dissipation in their environment.  We have described this requirement thus far by stating that $Q_d$, $Q_i$, and $Q_c$ must be large ($>10^6$, and ideally even higher).  However, at some point in a quantum algorithm, the state of the qubit must be measured, a task which necessarily requires coupling the qubit to the outside world strongly enough that the qubit can influence the state of the measurement apparatus and thus allow us to determine the qubit state.  This coupling is characterized by a quality factor $Q_m$ which describes the effective dissipation bath seen by the qubit due to coupling with the measurement apparatus.

Because the energy difference between the two qubit states $\hbar\omega_{01}$ is very small for microwave-frequency qubits, some form of amplification must be used to convert it into a classical-level signal strong enough to be digitized and analyzed to read out the state of the qubit.  Multiple stages of amplification are used; as with standard low-noise microwave receivers, the overall performance is most sensitive to the first amplification stages.

Superconducting qubits, and increasingly semiconducting qubits as well, typically use a technique called dispersive readout to measure the qubit state~\cite{Blais2004, Wallraff2005,mi2017strong,stockklauser2017strong,WestDGS}.  This approach relies on coupling the qubit to a microwave-frequency superconducting resonator.  The resonator frequency $\omega_r$ is detuned from $\omega_{01}$ by an amount $\Delta\gg g$, where $g$ is the qubit-resonator coupling strength.  In this so-called ``dispersive regime'' of the coupling, the qubit and resonator modes do not exchange energy, but the frequency of the resonator is shifted by an amount $2\chi$, called the ``dispersive shift'', depending on the qubit state.  The resonator is overcoupled to a feedline, giving a resonator linewidth $\kappa$ which is ideally {${\sim2\chi}$} (typically $\kappa\sim 2\chi\sim5$ MHz).  The resonator is typically probed in reflection, as shown in Fig.~\ref{fig:XMONctrl}c, with a probe tone near $\omega_r$.  The reflected signal will depend strongly on whether the resonance frequency has been shifted, which in turn indicates the qubit state.  Thus the measurement of the qubit state can be realized by measuring $S_{11}$ of the readout resonator at a single frequency near $\omega_r$.   For strongly overcoupled resonators, $|S_{11}|\approx 1$, but $\arg(S_{11})$ will be qubit-state-dependent near $\omega_r$.  One can think of this as a binary phase-shift-keyed signal, whose symbols (typically separated by $\approx180^\circ$, but potentially separated by a smaller angle) correspond to the two qubit states.  We note that this process also constitutes a kind of amplification: the single-photon qubit energy $\hbar\omega_{01}$ has been turned into a phase shift on many cw probe photons.  

The challenge here lies in the fact that the probe tone must be very weak, so as not to scramble the qubit state after initially collapsing it to $\ket{0}$ or $\ket{1}$~\cite{Boissonneault2009, Slichter2012}.  Otherwise, the resonator frequency would jump back and forth along with the qubit state, and the readout signal would stop providing information about the state to which the qubit initially collapsed.  The limit on the probe tone amplitude depends on the vector difference in the IQ plane between the reflected probe tone corresponding to state $\ket{0}$ and the reflected probe tone corresponding to state $\ket{1}$\textemdash in other words, the size of the phase-shift-keyed signal.  For superconducting qubits this difference signal should generally be kept below $\sim300\,\mathrm{nV}_{\mathrm{peak}}$, which corresponds to a signal power of around $-120$ dBm. For spins, probe powers up to $-80$ dBm have been used. 

This weak probe tone must be amplified, but the amplifier noise must be small enough not to drown out the signal.  While one can average in time (in other words, decrease the resolution bandwidth) to remove amplifier noise and recover the weak signal, 
we must perform our measurement in a time $\ll\mathrm{T}_1$, so that the qubit state does not decay during readout and thus corrupt the readout result.  Practically speaking, this means that readout must be performed in several hundred nanoseconds, or equivalently that the resolution bandwidth must be $\gtrsim10$\,MHz. The duration of measurement also poses a limit to the clock rate of a quantum algorithm or error correcting code. 

The solution is to use an ultra-low-noise amplifier as the first stage of the receiver, such that even the weak readout signal can be amplified with signal-to-noise ratio well above 1.  Superconducting parametric amplifiers based on Josephson junctions as nonlinear elements~\cite{Aumentado2020} can reach the quantum limit for noise, where the amplifier noise temperature $T_N$ at frequency $\omega$ is equal to $\hbar\omega/2$.  For 6 GHz signals, this corresponds to $T_N=144$ mK.  Dispersive qubit readout using superconducting parametric amplifiers was first implemented a decade ago~\cite{Vijay2011}, and readout fidelities of $\sim99$ \% have been reported~\cite{Jeffrey2014}.  Progress in superconducting parametric amplifiers in the past 15 years has been driven by quantum computing applications; we discuss this in more detail in Section~\ref{sec:ql_amps}.  

As discussed above, it is important that $Q_m$ be high, so that the measurement process not couple the qubit to a lossy environment that would cause $\mathrm{T}_1$ decay.  The readout resonator acts as a bandpass filter between the qubit and the lossy (real) impedance of the readout transmission line; since the resonator is far detuned from the qubit, this suppresses coupling between the qubit and this source of loss.  However, for some qubits this coupling is still the dominant source of loss.  A standard way to mitigate this loss is to place a bandpass filter (usually a second resonator) between the readout resonator and the transmission line~\cite{Reed2010, Jeffrey2014}.  This type of filter is known as a \emph{Purcell filter} and can be used to boost $Q_m$ by an additional two orders of magnitude or more, depending on the design parameters.  The bandwidth of the Purcell filter can be much larger than $\kappa$, so that it does not affect the speed of the readout.  

It is also possible to read out a superconducting qubit by turning on a strong coupling to a microwave photon counter.  If the qubit were in state $\ket{1}$, it would decay to $\ket{0}$ and emit a photon of energy $\hbar\omega_{01}$ into the photon counter.  It is challenging in practice to realize a microwave photon counter because the single photon energy is so small in the few-GHz regime, but such a device (which is essentially a modified superconducting phase qubit) has been demonstrated with readout fidelity as high as 98.4 \%~\cite{Opremcak2018, opremcak2020high}.

For trapped ion qubits, the very weak coupling of the qubit state to microwave fields (much higher $Q_d$ that the other technologies) means that it would be very difficult and inefficient to extract information from the ion at microwave frequencies.  However, trapped ions possess optical transitions, allowing easy extraction of optical photons, which have the added benefit of being easily detected with low background noise by room-temperature single-photon counters\footnote{This is primarily because of their high energy relative to the available thermal energy ($\hbar\omega\gg k_BT$ for optical photons at $T=300$ K).}.  

Trapped ions are generally read out using the so-called electron shelving technique~\cite{Dehmelt1982}, where the probability amplitudes in states $\ket{0}$ and $\ket{1}$ are mapped to two suitable states $\ket{b}$ and $\ket{d}$ using coherent control pulses of the same types used for qubit manipulation (microwave pulses and/or laser pulses).  When illuminated with a laser beam of appropriate wavelength and polarization, an ion in state $\ket{b}$ will fluoresce, absorbing photons and re-emitting them in all directions.  The laser beam is chosen to drive a so-called cycling transition, where the ion is excited from the state $\ket{b}$ and then emits a photon, always returning to $\ket{b}$ after the emission.  This enables repeated rounds of excitation and emission.  It is sometimes necessary to use multiple laser beams to ``close'' this cycle.  In contrast, an ion ``shelved'' in the state $\ket{d}$ (chosen such that all transitions out of $\ket{d}$ are far off resonance with the readout laser) will not interact with the laser beam and thus will not give off fluorescence photons.  By collecting a fraction of the fluorescence photons with an imaging objective and counting them with a single photon counter, it is possible to distinguish between a fluorescing ``bright'' ion in state $\ket{b}$ and a ``dark'' ion in state $\ket{d}$, as long as the mean number of photons counted for bright and dark ions is sufficiently different.  In practice, a readout duration of several hundred microseconds typically gives tens of counts for a bright ion, and $\sim1-4$ counts for a dark ion, although the duration and count rates can vary by an order of magnitude or more depending on the specifics of the setup.  Readout fidelities as high as 99.99 \% have been demonstrated using this technique~\cite{Myerson2008}.

Electron shelving readout implements two forms of amplification.  First, mapping into the states $\ket{b}$ and $\ket{d}$ and scattering a single photon can be thought of as turning the qubit energy difference $\hbar\omega_{01}$ into the energy of an emitted UV or visible photon, which is typically between $10^5$ and $10^8$ times larger for hyperfine or Zeeman qubits.  Secondly, the cycling transition allows up to $\sim10^6$ such photons to be scattered during the readout operation, giving further gain.

\section{Other applications of microwave technology\label{sec:otherapps}}
Beyond the direct use of microwave signals and techniques for qubit state control and measurement, there are many other critical applications of microwave and rf technology in quantum computing.  This section details several of the essential supporting roles played by microwave technology, beyond direct control and measurement of qubits.  

\subsection{RF for ion trapping\label{sec:traprf}}

Trapped ion qubits for quantum computing applications are almost always confined using a combination of radio-frequency and static electric fields in a so-called Paul trap~\cite{Wineland1998} (Penning traps, which use a combination of static magnetic and electric fields to confine ions, can also be used for quantum computing but are more typically employed for quantum simulation or precision spectroscopy experiments~\cite{Britton2012, Dilling2018, Jain2020}).  The rf electric fields for trapping are generated by applying rf voltages to specific electrodes of the trap, such than an rf electric field quadrupole is formed in vacuum at a distance of typically 30 $\mu$m to several mm from the electrodes.  The oscillating quadrupolar rf fields provide an effective confining potential for ions.  We refer the reader to Refs.~\cite{Wineland1998, Leibfried2003} for further details on Paul traps.  

The frequency and amplitude of the applied rf necessary for stable trapping of ions depend on the physical dimensions of the trap, the species of ion, and the desired strength of the confinement.  Larger traps, heavier ions, and stronger confinement all require larger rf voltages on the trap electrodes.  Reducing the rf frequency enables trapping with lower rf voltages, but if the frequency is reduced too much then the trap will become unstable and the ions cannot be held reliably.  Typical rf Paul traps for quantum computing applications operate at rf frequencies between 20 MHz and $\sim100$ MHz, and with peak rf voltages on the trap electrodes between 10 V and $\sim500$ V.  

These combinations of frequency and amplitude are generally quite difficult to achieve with direct driving, so the rf electrodes are typically incorporated into an rf resonant structure, which provides rf voltage step-up and filters noise from the rf drive electronics that could excite the ion motion.  Many different classes of resonators have been used, including helical resonators~\cite{Macalpine1959, Siverns2012}, coaxial resonators~\cite{Jefferts1995}, and lumped element resonators~\cite{Gandolfi2012, Brandl2016}.  These resonators typically have loaded quality factors of several hundred, although some cryogenic resonators can approach $10^4$.  Toroidal transformers on ferrite cores, operating as flux-coupled impedance-transforming ununs, can also be used to provide voltage step-up~\cite{Allcock2011}.  

Qubit-qubit coupling for trapped ions relies on the use of the ions' shared motional degree of freedom as a coupling ``bus''.  Multi-qubit gates typically have durations $\tau_g$ from $\sim10$ $\mu$s up to several ms~\cite{Bruzewicz2019}, with high-fidelity microwave-based gates typically in the $\sim$ms range.  To achieve high gate fidelity, the motional frequency must be stable to $\ll\frac{1}{\tau_g}$.  In practice, this means that the ion motional frequencies, typically a few MHz, should be stable at the $\sim10$ Hz level or below, or better than a part in $10^5$.  In a typical linear rf Paul trap, $1/3$ of the motional modes (called ``axial'' modes) have frequencies determined purely by static potentials applied to trap electrodes and $2/3$ of the modes (called ``radial'' modes) have frequencies that depend linearly on the rf voltage applied to the trap electrodes.  Many multi-qubit gate schemes use radial modes, so achieving the required motional frequency stability of $\sim10$ ppm or better means that the rf voltage on the trap electrodes must be stable at the $\sim10$ ppm level as well.  This requires high stability of the resonator frequency and quality factor, high gain stability for the rf amplifier driving the resonator, and low amplitude and phase noise for the rf generator.  If the resonator $Q$ is too high, this can cause increased voltage fluctuations at the trap through FM-to-AM conversion, giving some intuition for why trap rf resonators are not designed simply for maximum $Q$.  Recent work in the field has demonstrated rf limiting amplifiers~\cite{Harty2013} and active feedback methods for rf amplitude stabilization~\cite{Harty2013, Johnson2016} which can meet the desired $\sim10$ ppm amplitude stability of the rf voltage at the trap.  

\subsection{Applications to laser systems for QC\label{sec:lasers}}

Numerous qubit technologies, including trapped ion qubits, neutral atom qubits~\cite{Saffman2016, Henriet2020}, and optically-active defect centers in solids~\cite{Dobrovitski2013, Doherty2013}, rely on laser-based control and readout methods.  These laser systems use microwave technology for a number of tasks.  
Most laser beams for quantum information experiments need to be pulsed on and off with microsecond rise/fall times, too fast for a mechanical shutter.  In addition, sometimes a laser beam with amplitude modulation, rapidly tunable frequency, or multiple frequency components in a single beam, is desired.  For these tasks, acousto-optic modulators (AOMs)~\cite{Korpel1981} and/or electro-optic modulators (EOMs)~\cite{Desmarais1997} are used.  An EOM can be thought of like a microwave mixer, except the LO port uses an optical signal while the IF port accepts an rf or microwave signal; signals at the IF port modulate the optical LO and give rise to sidebands on the optical output.  AOMs are more akin to single-sideband mixers.  They perform frequency-shifting of the laser light passing through them, but can also be used for amplitude modulation (including on/off switching) of laser beams.  The ``carrier leakage'' and ``spurious sideband'' laser beams have different directions of propagation from the desired ``sideband'' at the AOM output, enabling them to be filtered out spatially.  Both AOMs and EOMs require rf or microwave signals at around $+30$ dBm to function; generally AOMs are driven at frequencies from $\sim50$ to $\sim500$ MHz, with 10s of MHz bandwidth, while EOMs can accept drive tones from near dc up to many GHz, and have widely varying bandwidths depending on the choice of resonant rf/microwave enhancement circuits inside.    

In addition to modulation of laser light, microwave technology is also important for stabilization of the lasers themselves.  A famous method for locking a microwave oscillator to a stable reference cavity using modulation sidebands, due to Pound~\cite{Pound1946}, was extended to operation with laser oscillators and stable optical references (cavities or spectral lines) by Drever, Hall, and coworkers~\cite{Drever1983}, and is known in the laser community as Pound-Drever-Hall (PDH) locking.  PDH lock circuits, which operate at rf/microwave frequencies, are ubiquitous in laser systems.  There are also a wide variety of servo loops for laser amplitude stabilization which rely on modulating the amplitude of the rf drive to an AOM as the feedback signal.  Mode-locked lasers, which emit periodic short ($\lesssim$~ps) pulses of light with an extremely stable repetition rate, forming a frequency comb, are used in trapped ion applications to drive stimulated Raman transitions between states with very large detuning, such as the 12.6 GHz hyperfine qubit states in $^{171}$Yb$^+$~\cite{Mizrahi2014}.  For optimum performance, the repetition rate of the mode locked laser (typically $\sim100$ MHz, but potentially as high as a few GHz) must be stabilized, which requires implementing a phase-locked loop between a stable reference oscillator at the desired repetition rate and the signal from the laser pulse train on a fast photodiode.

\section{Microwave innovations from quantum computing\label{sec:innovations}}
Quantum computing relies heavily on microwave technologies already developed for other applications.  However, quantum computing also requires operation in new performance regimes, and has inspired the development of novel microwave technologies and systems to meet those challenges.  We describe two major advances in microwave technology that have arisen from quantum computing research: quantum-limited microwave amplifiers, and cryogenic non-reciprocal microwave devices, including chip-scale non-reciprocal devices.  

\subsection{Quantum-limited amplifiers\label{sec:ql_amps}}
As described in Section~\ref{sec:measurement}, the readout of superconducting qubits and semiconducting qubits involves microwave signals so weak that they are not far above the quantum noise floor of half a photon per unit bandwidth, $S(\omega)=\hbar\omega/2$.  Faithful amplification of these signals requires amplifiers with noise performance at or near the quantum limit, which at 6 GHz corresponds to a noise temperature of 144\,mK.

To achieve this noise performance, the field has turned to superconducting parametric amplifiers.  Parametric amplifiers rely on a nonlinear element or elements whose parameters (inductance or capacitance) are modulated in time by a strong pump.  This modulation transfers power from the pump into other modes (known as the signal and the idler), coherently amplifying the energy in those modes~\cite{Louisell1960, Aumentado2020}.  

Before the advent of high-electron-mobility transistor (HEMT) amplifiers, microwave amplification was sometimes accomplished using parametric amplifiers based on varactor diodes or inductors with saturable cores~\cite{Mumford1960}.  Parametric amplifiers based on the Josephson effect in superconductors have been studied intermittently since the 1960s, but the field was revitalized by work in the late 2000s on Josephson parametric amplifiers with near-quantum-limited noise performance and bandwidths of $\sim10$ MHz~\cite{Castellanos-Beltran2007, Yamamoto2008, Vijay2008, Bergeal2010a}.  The viability of these amplifiers for high-fidelity qubit readout was demonstrated shortly thereafter~\cite{Vijay2011}, and they were rapidly adopted as the state of the art~\cite{Hatridge2013, Jeffrey2014}.  Subsequent work has used impedance engineering to increase bandwidths to the $\sim$ GHz range~\cite{Mutus2014, Roy2015, Naaman2019}, and more complex designs with many Josephson junctions have shown substantial improvement in saturation powers~\cite{Macklin2015, Naaman2019}.  Most designs operate in reflection, typically requiring bulky circulators to separate the output signal from the input signal, and to protect the qubit circuit from the strong pump tone, so transmission-mode parametric amplifiers providing directional, non-reciprocal amplification have been developed~\cite{Abdo2013, Macklin2015, Lecocq2017, Ranzani2019}.  A considerably more extensive review of the literature than is provided here can be found in Ref.~\cite{Aumentado2020}.  

The bandwidth, saturation power, and noise performance of current state-of-the-art parametric amplifiers allows simultaneous readout of multiple qubits by frequency multiplexing their readout resonators, which share a common feedline to the parametric amplifier.  The gain and noise performance of these parametric amplifiers are such that they set the overall receiver gain for the amplification chain (the rest of which consists of ultra-low-noise cryogenic and room-temperature microwave transistor amplifiers) out to the room temperature demodulation and digitization circuitry.

\subsection{Non-reciprocal devices}
Microwave circuit elements exhibiting non-reciprocity are currently used heavily in quantum computing, mostly in the amplification of readout signals from qubits. In this context they serve two main purposes. First, they function as isolation devices that prevent noise originating in the readout amplification chain from impinging on the qubits. Their second use is in the context of parametric amplifiers (see above) that operate in reflective mode. 
Here, circulators are configured to separate input and output signals as well as to isolate the qubits from the pump tones that supply energy to the amplifier.

\begin{figure}[t!]
\centering
\includegraphics[width=\columnwidth]{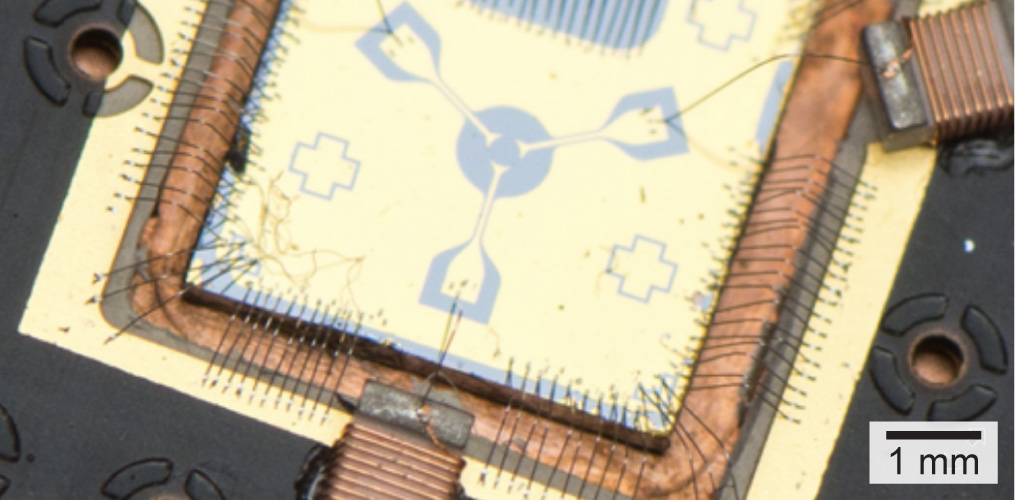}
\caption{On-chip microwave circulator based on the quantum Hall effect. Microwave excitations travel at a velocity 1000 times slower than in vacuum, enabling an equivalent  reduction in the signal wavelength and device footprint~\cite{Mahoney2017}.}
\label{fig:circulator}
\end{figure}
To achieve non-reciprocity, traditional microwave circulators exploit ferromagnetic materials. These devices are necessarily large and bulky components, since they make use of interference effects that occur over a length scale comparable to the microwave wavelength. Further limitations include their typically narrow-band performance, insertion loss, limited non-reciprocity, and variation in specifications or failure when operated at cryogenic temperatures or in large magnetic fields. From the perspective of scaling up to the number of readout channels needed for a large quantum system, the footprint alone of conventional circulators is a significant barrier to the development of tightly integrated systems.

Recently, efforts have focused on realising miniaturized devices that exhibit nonreciprocity.  One avenue uses active devices, specifically Josephson parametric amplifiers designed to achieve directional gain and reverse isolation while maintaining quantum-limited noise performance~\cite{Abdo2013, Macklin2015, Sliwa2015, Lecocq2017, Lecocq2020b}.  Ref.~\cite{Ranzani2019} provides an in-depth review of nonreciprocal active circuits with quantum-limited noise.  Such devices have recently been demonstrated for direct, high-fidelity readout of superconducting qubits without external circulators~\cite{Abdo2020, Rosenthal2020, Lecocq2020a}.

Miniaturized passive circulators also hold promise for quantum computing applications. For example, Fig.~\ref{fig:circulator} shows a 3-port circulator device based on the inherent non-reciprocity of the quantum Hall effect~\cite{Mahoney2017}. Here, the size of the device is 1/1000th of the free-space wavelength of the microwave signals it handles.  This small size is achieved by exploiting the chiral, ``slow-light'' response of a two-dimensional electron gas in the quantum Hall regime. For an integrated GaAs device with 330 $\mu$m diameter and 1 GHz center frequency, a nonreciprocity of 25 dB is observed over a 50 MHz bandwidth. Furthermore, the nonreciprocity can be dynamically tuned by varying the voltage at the port, an aspect that may enable reconfigurable passive routing of microwave signals on chip. The forward transmission of this particular device was limited to -20 dB due to the impedance mismatch between 50 $\Omega$ environment and the device impedance, which is set by the resistance quantum $R_K\approx25\,\mathrm{k}\Omega$. However, improved impedance matching design could be used to increase the forward transmission~\cite{Bosco2017}.  Although the quantum Hall device requires a large external magnetic field, magnetically-doped topological insulator materials offer a further route to realizing non-reciprocal devices without the need for such external magnetic fields~\cite{MahoneyTI17}.  Beyond non-reciprocal behavior, the underlying physics of these systems suggests their use as compact and tunable delay lines, microwave interferometers, and low-loss, high impedance transmission lines~\cite{Bosco}. 

For future quantum computing systems, miniaturized, on-chip non-reciprocal microwave devices of the kinds described above are likely to play a key role.

\section{Microwave challenges in reaching the full potential of quantum computing}

In order to realize the full promise of quantum computing, it will be necessary to build systems capable of executing QEC protocols. In general, these protocols use ensembles of error-prone physical qubits to implement ``logical qubits,'' which have reduced rates of error in comparison to their constituent physical qubits. The degree to which errors are suppressed depends both upon the error rates of the physical qubits and the degree of redundancy. However, for such a protocol to work in the first place, the physical qubit error rates must be on average below a threshold, which depends on the specifics of the QEC protocol. There are many different QEC schemes~\cite{roffe2019quantum}, with physical error thresholds ranging from $\sim10^{-4}$ up to about 1 \%. Once the physical error rates are below the threshold, the degree of redundancy (the number of physical qubits) required to achieve a specific logical error rate scales logarithmically with further reduction in the physical error rates. A practical target of exceeding the threshold by an order of magnitude (0.1 \% physical qubit error rates for some codes) appears feasible, but still requires about 1,000 physical qubits per logical qubit. Thus, it is estimated that realizing a practical system with 1,000 logical qubits (far below the number required to break RSA encryption) will require building a million-qubit-scale system~\cite{Fowler2012}. There are numerous microwave-related challenges associated with building such a computer. Here we provide a discussion of some of these challenges.

Today's quantum processors are controlled and measured using systems that mostly comprise room temperature electronics, with high-performance microwave interconnects used to interface to the quantum processor. Let us begin by considering the feasibility of scaling this approach to the million qubit level. It will be helpful to start by quantifying the number of control and measurement channels required for a brute force scaling approach. For a large-scale quantum processor implemented using superconducting transmon qubits arranged in the architecture of~\cite{arute2019quantum}, processor operation requires one \emph{XY} channel and three \emph{Z} control channels per qubit. In addition, one readout channel is required for every $\sim$10 qubits. In the case of semiconductor spin qubits a handful of dc bias wires per qubit are needed to define tunnel barriers and chemical potentials. Depending on the flavor of spin qubit at least one high bandwidth channel is required per qubit, with additional channels, shifted in phase, often used to cancel crosstalk. Readout of spin qubits based on rf reflectometry requires a single wideband line per $\sim$ 10 qubits using frequency multiplexing.
Trapped ion systems generally rely on their long qubit coherence times, as well as the ability to reconfigure ion and/or laser beam positions in the trap, to perform logic operations on different subsets of qubits in series, rather than performing simultaneous operations on all qubits.  This means that the scaling of microwave control hardware does not have a precise relationship to the scaling of qubit numbers.  However, to avoid excessive slowing of computations, the serialization would ideally be at most a factor of tens, not hundreds or larger.

Given these constraints, is does not appear feasible to scale existing technology to the level demanded by large-scale error-corrected quantum computers. For instance, if one were to build a quantum control and measurement system for a million transmon qubits using present Google technology~\cite{arute2019quantum}, it would occupy over 15,000\,m$^2$ of floor space and dissipate about 40\,MW (not including the dissipation of the amplifiers that would be required to compensate for loss in the 
cable runs). 

\subsection{Scaling of interconnects}
Beyond the sheer volume of the quantum control and measurement electronics, one must also consider the feasibility of connecting the electronics to a quantum processor. Here, the fundamental constraint is going to be the ratio of the required number of interconnects to the chip surface area, which sets the required interconnect density. For large-scale quantum processors, this simplifies to the ratio of the number of I/O signals required per qubit to the qubit pitch squared (assuming a 2D array). For superconducting qubits, which are arranged on a pitch of about 1\,mm and require about 4.2 lines per qubit, an interconnect density of about 4.2 lines per square millimeter is required; while this is believed to be feasible, it will require significant microwave engineering to deliver these signals while maintaining low crosstalk and avoiding dissipation. For instance, a million qubit quantum processor will require about 3 million \emph{Z} control lines, and to keep heating due to the $\approx50\,$\,$\mu$A static \emph{Z} currents to below 10\,$\mu$W (necessary because of the limited available cooling power at mK temperatures), the contact resistance between the qubit package and the cable assembly must be kept to the m$\Omega$ level. Developing qubit packaging techniques that enable high coherence and low crosstalk while achieving this specification will require significant research.  Superconducting interconnects, which offer near-lossless and near-dispersionless electrical propagation~\cite{herr2002high} while providing very low thermal conductivity, can be used to transport signals from 10\,mK to 4\,K. However, achieving the required interconnect density between 4\,K and room temperature (where superconducting interconnects are not an option) may be challenging due to dispersion and losses, both of which distort control waveforms. As such, the control and measurement system may have to reside partially or entirely at 4\,K. Regardless, considerable work is required to develop interconnect systems between 4\,K and 10\,mK which offer reproducible performance and the stringent cross-talk performance that is required for a large-scale quantum computer.   

For semiconductor spin qubits, the nanoscale qubit dimensions make interfacing a large scale quantum processor to an external control system impractical without a large degree of multiplexing~\cite{franke2019rent,Hornibrook2015}. Consider, for instance, that as many as 10 wires are required within the $100\,\mathrm{nm} \times 100\,\mathrm{nm}$ footprint of a spin qubit. This geometric I/O bottleneck motivates the integration of cryogenic electronics with the qubit platform to handle signal generation and multiplexing~\cite{PaukaPRApp} without needing large cable assemblies that carry signals to room temperature electronics~\cite{Reilly_IEDM}. The power dissipation of these classical electronics can be significant, however, leading to heating of the qubits and a degradation in fidelity if tightly integrated in a monolithic configuration. 

Two distinct approaches are currently being pursued to address this trade-off between the I/O bottleneck and qubit temperature. The first involves relocating the spin qubit platform and integrated control circuits to higher temperatures (near 1 K), where substantially more cooling power is available by using pumped helium-4~\cite{petit2020universal,yang2020operation}. However, this approach causes reductions in qubit fidelity, eventually requiring orders of magnitude more noisy physical qubits to encode a logical qubit. The increased qubit count then requires additional I/O and control electronics that may effectively cancel the gains from operating at higher temperature. An alternate approach involves operating the control sub-systems at the same millikelvin temperature as the qubits, but on a separate chip that is thermally decoupled~\cite{Pauka_arxiv}. Taking advantage of the high impedance (open circuit) nature of gate electrodes and leveraging lithographically defined chip-to-chip interconnect strategies, this millikelvin approach alleviates the power dissipation otherwise required to drive low impedance cables between temperature stages.

Large-scale trapped-ion quantum computers will likely operate cryogenically, in the 4 K to 10 K temperature range, because of superior vacuum pressure (enabling ions to remain trapped for much longer times) and reduced electric field noise (improving motional coherence and entangling gate fidelities).  The much larger cooling powers available at these temperatures make the interconnect problem less formidable, although still not easy.  Because many ions can be addressed with a single microwave control line, a relatively small number of control lines will be needed compared to the other qubit technologies.  However, because the control signals have much higher power, on-chip dissipation may present an issue.  Ion traps rely on a very large number of static or slowly-varying control voltages for defining trap potentials and moving ions around the trap, with roughly 10 times as many control voltages as potential wells (each of which may hold single or multiple ions) in the trap.  However, these lines draw zero or minimal current and do not require high bandwidth ($<1$ MHz is generally ample), so they can be made from very thin, low-thermal-conductivity wires.  Research into generating these voltages with DACs fabricated in the trap substrate itself is being pursued by some groups~\cite{Stuart2019}.  

\subsection{Scaling of control systems}

Today's quantum computers are essentially research devices with individual qubit performance just approaching the edge of what is required to implement QEC protocols. Developing control protocols for use on these prototype systems has necessitated the use of flexible, high-speed arbitrary waveform generators so that a researcher can quickly test new control paradigms without developing new hardware. While this is a logical approach for operating today's relatively small-scale quantum processors, the cost, size, and power associated with scaling this approach to run a million-qubit quantum processor motivate the development of a more optimized approach.
 
Several research groups are currently investigating the integration of quantum control circuits, targeting operation at a physical temperature of 4\,K~\cite{patra2017cryo,patra2020scalable,bardin201929,bardin2019design,bashir2019mixed,8702442}. However, the requirements for these systems are stringent and co-optimization of the classical controller and the quantum processor will likely be required. For a device thermalized at 4\,K, the power consumption will be limited to well below 1\,mW per control channel, and this tight power budget must be met without increasing control errors. For \emph{XY} controllers, strategies must be developed to minimize the number of microwave carriers required, mitigate crosstalk, optimize pulse waveforms, and maintain phase coherence among the $\sim$1 million \emph{XY} control channels.  The same considerations, albeit with somewhat relaxed power restrictions, are also relevant for room-temperature control electronics.

Similar challenges are present for the development of the scalable baseband waveform generators required for \emph{Z} control.  In particular, losses and reflections along the interconnects between the \emph{Z} controllers and the qubit/coupler control ports cause distortion in the same way that non-return-to-zero signals experience distortion when traveling over a backplane. In today's systems, the long cables between the room temperature electronics and the quantum processor lead to settling times that can be on the order of microseconds~\cite{Reilly_IEDM,langford2017experimentally}. To compensate for these  long settling times, this response is often characterized and deconvolved  from the transmitted \emph{Z} signal, similar to the use of pre-emphasis in wireline applications. While the use of  bipolar \emph{Z} pulses reduces the impact of long-term settling, the short time-scale system response must still be compensated~\cite{rol2019fast}. As such, research will be required to architect \emph{Z} control systems that are both precise and low power.

\subsection{Scaling of readout}

Scaling of readout is also of critical importance and microwave innovations will certaintly be needed. For superconducting and semiconductor spin qubits using dispersive readout, scalable amplification chains achieving near-quantum-limited noise performance are required. Assuming a frequency domain multiplexing factor of 10$\times$, a million qubit quantum controller will require 100,000 readout channels. In today's systems, each readout channel contains a parametric amplifier and several circulators (up to 5) thermalized to 10\,mK, a HEMT-based low noise amplifier thermalized to 4\,K, and further amplification and digitization at room temperature. In order to scale to the levels required to implement a practical error-corrected quantum computer, each of these technologies will have to be optimized for manufacturability, cost, performance, and size. For instance, the semiconductor low noise amplifiers used at 4\,K (e.g.,~\cite{LNF}) achieve excellent noise temperatures, but are hand assembled, hand tested, and dissipate about an order of magnitude too much power for use in a large-scale system. As with the case of control systems, it is necessary that scalable readout systems be integrated to the maximum extent possible. Amplifiers that can be mass manufactured in a silicon technology platform and achieve similar performance to today's HEMT LNAs while dissipating $\ll1\,$mW should be developed so that electronics currently at room temperature can be directly integrated with the low noise amplifier and thermalized to 4\,K. While SiGe processes appear to be a promising technology for meeting these goals~\cite{montazeri2015ultra,wongIMS2020}, research is still required to determine if it is possible to achieve both the performance and repeatability required using SiGe technologies. Additionally, work is required to minimize the volume of the microwave electronics that must be thermalized to the 10\,mK stage (i.e. the parametric amplifiers and circulators).

\subsection{Other microwave challenges}

Beyond the classical to quantum interface, there are significant microwave challenges in the design of the quantum processor itself. For instance, considering superconducting qubits, as the dimensions of the quantum processor grow, designers will have to rely more heavily on electromagnetic simulation tools to predict and avoid undesired moding in what will eventually become wafer-scale devices. Modeling these effects requires incorporating superconducting physics and the cryogenic properties of dielectrics into electromagnetic tools while developing methods to efficiently solve for high-Q resonances in large structures. Microwave expertise will be required, both in developing efficient simulation tools tailored to this purpose as well as in developing techniques to mitigate these undesired modes.

As the system size grows, it may also become important to develop techniques to predict system performance from within a single design environment (similar to the infrastructure that has been developed for digital design). This may involve developing qubit models which are compatible with commercial circuit solvers (e.g. \cite{8791334}) and tying the classical and quantum systems together in a circuit solver tool. For systems such as superconducting quantum computers, which are spatially distributed, it may also be necessary to develop techniques to accurately model the interconnects between the classical control system and the quantum processor.

\section{Conclusion}
Microwave technology has played a key role in the rise of quantum computing, and the two fields will continue to interact synergistically in the years ahead. The pursuit of an error-corrected quantum computer has been enabled by microwave technologies made available by the explosive growth of the wireless communication industry. However, reaching the performance currently achieved by today's state-of-the-art quantum processors has also necessitated the development of new microwave technologies, such as quantum-limited parametric amplifiers.

Today, there is a growing race to implement a fault-tolerant quantum computer.  As described in this article, numerous microwave-related challenges must be overcome to build such a device, and continued microwave innovation will certainly be required. Critical areas for further research range from the development of new ultra-low-loss interconnect systems to the design of highly efficient quantum control systems. The active engagement of microwave engineers in this exciting and important effort will be essential to its success.

\section*{Acknowledgment}
We thank J. Aumentado, K. Beloy, O. Naaman, H. Neven, and L. J. Stephenson for a careful reading of the manuscript, and M. Foss-Feig, D. Sank, and K. C. Young for helpful discussions.  This work was partially supported by National Science Foundation Grant \#1809114, and by the NIST Quantum Information Program.


\begin{thebibliography}{161}%
\makeatletter
\providecommand \@ifxundefined [1]{%
 \@ifx{#1\undefined}
}%
\providecommand \@ifnum [1]{%
 \ifnum #1\expandafter \@firstoftwo
 \else \expandafter \@secondoftwo
 \fi
}%
\providecommand \@ifx [1]{%
 \ifx #1\expandafter \@firstoftwo
 \else \expandafter \@secondoftwo
 \fi
}%
\providecommand \natexlab [1]{#1}%
\providecommand \enquote  [1]{``#1''}%
\providecommand \bibnamefont  [1]{#1}%
\providecommand \bibfnamefont [1]{#1}%
\providecommand \citenamefont [1]{#1}%
\providecommand \href@noop [0]{\@secondoftwo}%
\providecommand \href [0]{\begingroup \@sanitize@url \@href}%
\providecommand \@href[1]{\@@startlink{#1}\@@href}%
\providecommand \@@href[1]{\endgroup#1\@@endlink}%
\providecommand \@sanitize@url [0]{\catcode `\\12\catcode `\$12\catcode
  `\&12\catcode `\#12\catcode `\^12\catcode `\_12\catcode `\%12\relax}%
\providecommand \@@startlink[1]{}%
\providecommand \@@endlink[0]{}%
\providecommand \url  [0]{\begingroup\@sanitize@url \@url }%
\providecommand \@url [1]{\endgroup\@href {#1}{\urlprefix }}%
\providecommand \urlprefix  [0]{URL }%
\providecommand \Eprint [0]{\href }%
\providecommand \doibase [0]{https://doi.org/}%
\providecommand \selectlanguage [0]{\@gobble}%
\providecommand \bibinfo  [0]{\@secondoftwo}%
\providecommand \bibfield  [0]{\@secondoftwo}%
\providecommand \translation [1]{[#1]}%
\providecommand \BibitemOpen [0]{}%
\providecommand \bibitemStop [0]{}%
\providecommand \bibitemNoStop [0]{.\EOS\space}%
\providecommand \EOS [0]{\spacefactor3000\relax}%
\providecommand \BibitemShut  [1]{\csname bibitem#1\endcsname}%
\let\auto@bib@innerbib\@empty
\bibitem [{\citenamefont {Buderi}(1996)}]{buderi}%
  \BibitemOpen
  \bibfield  {author} {\bibinfo {author} {\bibfnamefont {R.}~\bibnamefont
  {Buderi}},\ }\href@noop {} {\emph {\bibinfo {title} {The Invention that
  Changed the World: How a Small Group of Radar Pioneers Won the Second World
  War and Launched a Technological Revolution}}},\ Sloan technology series\
  (\bibinfo  {publisher} {Simon \& Schuster},\ \bibinfo {year}
  {1996})\BibitemShut {NoStop}%
\bibitem [{\citenamefont {Townes}\ and\ \citenamefont
  {Schawlow}(1975)}]{Townes1975}%
  \BibitemOpen
  \bibfield  {author} {\bibinfo {author} {\bibfnamefont {C.~H.}\ \bibnamefont
  {Townes}}\ and\ \bibinfo {author} {\bibfnamefont {A.~L.}\ \bibnamefont
  {Schawlow}},\ }\href@noop {} {\emph {\bibinfo {title} {{Microwave
  spectroscopy}}}}\ (\bibinfo  {publisher} {Dover},\ \bibinfo {address}
  {Mineola, NY},\ \bibinfo {year} {1975})\BibitemShut {NoStop}%
\bibitem [{\citenamefont {Slichter}(1996)}]{slichter1996principles}%
  \BibitemOpen
  \bibfield  {author} {\bibinfo {author} {\bibfnamefont {C.~P.}\ \bibnamefont
  {Slichter}},\ }\href@noop {} {\emph {\bibinfo {title} {Principles of Magnetic
  Resonance}}},\ Springer Series in Solid-State Sciences\ (\bibinfo
  {publisher} {Springer},\ \bibinfo {year} {1996})\BibitemShut {NoStop}%
\bibitem [{\citenamefont {Purcell}\ \emph {et~al.}(1946)\citenamefont
  {Purcell}, \citenamefont {Torrey},\ and\ \citenamefont
  {Pound}}]{Purcell1946}%
  \BibitemOpen
  \bibfield  {author} {\bibinfo {author} {\bibfnamefont {E.~M.}\ \bibnamefont
  {Purcell}}, \bibinfo {author} {\bibfnamefont {H.~C.}\ \bibnamefont
  {Torrey}},\ and\ \bibinfo {author} {\bibfnamefont {R.~V.}\ \bibnamefont
  {Pound}},\ }\bibfield  {title} {\bibinfo {title} {Resonance absorption by
  nuclear magnetic moments in a solid},\ }\href
  {https://doi.org/10.1103/PhysRev.69.37} {\bibfield  {journal} {\bibinfo
  {journal} {Phys. Rev.}\ }\textbf {\bibinfo {volume} {69}},\ \bibinfo {pages}
  {37} (\bibinfo {year} {1946})}\BibitemShut {NoStop}%
\bibitem [{\citenamefont {Bloch}\ \emph {et~al.}(1946)\citenamefont {Bloch},
  \citenamefont {Hansen},\ and\ \citenamefont {Packard}}]{Bloch1946}%
  \BibitemOpen
  \bibfield  {author} {\bibinfo {author} {\bibfnamefont {F.}~\bibnamefont
  {Bloch}}, \bibinfo {author} {\bibfnamefont {W.~W.}\ \bibnamefont {Hansen}},\
  and\ \bibinfo {author} {\bibfnamefont {M.}~\bibnamefont {Packard}},\
  }\bibfield  {title} {\bibinfo {title} {Nuclear induction},\ }\href
  {https://doi.org/10.1103/PhysRev.69.127} {\bibfield  {journal} {\bibinfo
  {journal} {Phys. Rev.}\ }\textbf {\bibinfo {volume} {69}},\ \bibinfo {pages}
  {127} (\bibinfo {year} {1946})}\BibitemShut {NoStop}%
\bibitem [{\citenamefont {Bardin}\ \emph {et~al.}(2020)\citenamefont {Bardin},
  \citenamefont {Sank}, \citenamefont {Naaman},\ and\ \citenamefont
  {Jeffrey}}]{bardin2020quantum}%
  \BibitemOpen
  \bibfield  {author} {\bibinfo {author} {\bibfnamefont {J.~C.}\ \bibnamefont
  {Bardin}}, \bibinfo {author} {\bibfnamefont {D.}~\bibnamefont {Sank}},
  \bibinfo {author} {\bibfnamefont {O.}~\bibnamefont {Naaman}},\ and\ \bibinfo
  {author} {\bibfnamefont {E.}~\bibnamefont {Jeffrey}},\ }\bibfield  {title}
  {\bibinfo {title} {Quantum computing: An introduction for microwave
  engineers},\ }\href@noop {} {\bibfield  {journal} {\bibinfo  {journal} {IEEE
  Microwave Magazine}\ }\textbf {\bibinfo {volume} {21}},\ \bibinfo {pages}
  {24} (\bibinfo {year} {2020})}\BibitemShut {NoStop}%
\bibitem [{\citenamefont {Arute}\ \emph {et~al.}(2019)\citenamefont {Arute},
  \citenamefont {Arya}, \citenamefont {Babbush}, \citenamefont {Bacon},
  \citenamefont {Bardin}, \citenamefont {Barends}, \citenamefont {Biswas},
  \citenamefont {Boixo}, \citenamefont {Brandao}, \citenamefont {Buell} \emph
  {et~al.}}]{arute2019quantum}%
  \BibitemOpen
  \bibfield  {author} {\bibinfo {author} {\bibfnamefont {F.}~\bibnamefont
  {Arute}}, \bibinfo {author} {\bibfnamefont {K.}~\bibnamefont {Arya}},
  \bibinfo {author} {\bibfnamefont {R.}~\bibnamefont {Babbush}}, \bibinfo
  {author} {\bibfnamefont {D.}~\bibnamefont {Bacon}}, \bibinfo {author}
  {\bibfnamefont {J.~C.}\ \bibnamefont {Bardin}}, \bibinfo {author}
  {\bibfnamefont {R.}~\bibnamefont {Barends}}, \bibinfo {author} {\bibfnamefont
  {R.}~\bibnamefont {Biswas}}, \bibinfo {author} {\bibfnamefont
  {S.}~\bibnamefont {Boixo}}, \bibinfo {author} {\bibfnamefont {F.~G.}\
  \bibnamefont {Brandao}}, \bibinfo {author} {\bibfnamefont {D.~A.}\
  \bibnamefont {Buell}}, \emph {et~al.},\ }\bibfield  {title} {\bibinfo {title}
  {Quantum supremacy using a programmable superconducting processor},\
  }\href@noop {} {\bibfield  {journal} {\bibinfo  {journal} {Nature}\ }\textbf
  {\bibinfo {volume} {574}},\ \bibinfo {pages} {505} (\bibinfo {year}
  {2019})}\BibitemShut {NoStop}%
\bibitem [{\citenamefont {Nielsen}\ and\ \citenamefont
  {Chuang}(2000)}]{Nielsen2000}%
  \BibitemOpen
  \bibfield  {author} {\bibinfo {author} {\bibfnamefont {M.~A.}\ \bibnamefont
  {Nielsen}}\ and\ \bibinfo {author} {\bibfnamefont {I.~L.}\ \bibnamefont
  {Chuang}},\ }\href@noop {} {\emph {\bibinfo {title} {{Quantum Computation and
  Quantum Information}}}}\ (\bibinfo  {publisher} {Cambridge University
  Press},\ \bibinfo {address} {Cambridge},\ \bibinfo {year} {2000})\BibitemShut
  {NoStop}%
\bibitem [{\citenamefont {Shor}(1999)}]{Shor1999}%
  \BibitemOpen
  \bibfield  {author} {\bibinfo {author} {\bibfnamefont {P.~W.}\ \bibnamefont
  {Shor}},\ }\bibfield  {title} {\bibinfo {title} {{Polynomial-Time Algorithms
  for Prime Factorization and Discrete Logarithms on a Quantum Computer}},\
  }\href {https://doi.org/10.1137/S0036144598347011} {\bibfield  {journal}
  {\bibinfo  {journal} {SIAM Rev.}\ }\textbf {\bibinfo {volume} {41}},\
  \bibinfo {pages} {303} (\bibinfo {year} {1999})}\BibitemShut {NoStop}%
\bibitem [{\citenamefont {Wineland}\ \emph {et~al.}(1998)\citenamefont
  {Wineland}, \citenamefont {Monroe}, \citenamefont {Itano}, \citenamefont
  {Leibfried}, \citenamefont {King},\ and\ \citenamefont
  {Meekhof}}]{Wineland1998}%
  \BibitemOpen
  \bibfield  {author} {\bibinfo {author} {\bibfnamefont {D.~J.}\ \bibnamefont
  {Wineland}}, \bibinfo {author} {\bibfnamefont {C.}~\bibnamefont {Monroe}},
  \bibinfo {author} {\bibfnamefont {W.~M.}\ \bibnamefont {Itano}}, \bibinfo
  {author} {\bibfnamefont {D.}~\bibnamefont {Leibfried}}, \bibinfo {author}
  {\bibfnamefont {B.~E.}\ \bibnamefont {King}},\ and\ \bibinfo {author}
  {\bibfnamefont {D.~M.}\ \bibnamefont {Meekhof}},\ }\bibfield  {title}
  {\bibinfo {title} {{Experimental issues in coherent quantum-state
  manipulation of trapped atomic ions}},\ }\href@noop {} {\bibfield  {journal}
  {\bibinfo  {journal} {J. Res. Natl. Inst. Stand. Technol.}\ }\textbf
  {\bibinfo {volume} {103}},\ \bibinfo {pages} {259} (\bibinfo {year}
  {1998})}\BibitemShut {NoStop}%
\bibitem [{\citenamefont {Leibfried}\ \emph {et~al.}(2003)\citenamefont
  {Leibfried}, \citenamefont {Blatt}, \citenamefont {Monroe},\ and\
  \citenamefont {Wineland}}]{Leibfried2003}%
  \BibitemOpen
  \bibfield  {author} {\bibinfo {author} {\bibfnamefont {D.}~\bibnamefont
  {Leibfried}}, \bibinfo {author} {\bibfnamefont {R.}~\bibnamefont {Blatt}},
  \bibinfo {author} {\bibfnamefont {C.}~\bibnamefont {Monroe}},\ and\ \bibinfo
  {author} {\bibfnamefont {D.}~\bibnamefont {Wineland}},\ }\bibfield  {title}
  {\bibinfo {title} {{Quantum dynamics of single trapped ions}},\ }\href
  {https://doi.org/10.1103/RevModPhys.75.281} {\bibfield  {journal} {\bibinfo
  {journal} {Rev. Mod. Phys.}\ }\textbf {\bibinfo {volume} {75}},\ \bibinfo
  {pages} {281} (\bibinfo {year} {2003})}\BibitemShut {NoStop}%
\bibitem [{\citenamefont {Bruzewicz}\ \emph {et~al.}(2019)\citenamefont
  {Bruzewicz}, \citenamefont {Chiaverini}, \citenamefont {McConnell},\ and\
  \citenamefont {Sage}}]{Bruzewicz2019}%
  \BibitemOpen
  \bibfield  {author} {\bibinfo {author} {\bibfnamefont {C.~D.}\ \bibnamefont
  {Bruzewicz}}, \bibinfo {author} {\bibfnamefont {J.}~\bibnamefont
  {Chiaverini}}, \bibinfo {author} {\bibfnamefont {R.}~\bibnamefont
  {McConnell}},\ and\ \bibinfo {author} {\bibfnamefont {J.~M.}\ \bibnamefont
  {Sage}},\ }\bibfield  {title} {\bibinfo {title} {{Trapped-ion quantum
  computing: Progress and challenges}},\ }\href
  {https://doi.org/10.1063/1.5088164} {\bibfield  {journal} {\bibinfo
  {journal} {Appl. Phys. Rev.}\ }\textbf {\bibinfo {volume} {6}},\ \bibinfo
  {pages} {021314} (\bibinfo {year} {2019})}\BibitemShut {NoStop}%
\bibitem [{\citenamefont {Seidelin}\ \emph {et~al.}(2006)\citenamefont
  {Seidelin}, \citenamefont {Chiaverini}, \citenamefont {Reichle},
  \citenamefont {Bollinger}, \citenamefont {Leibfried}, \citenamefont
  {Britton}, \citenamefont {Wesenberg}, \citenamefont {Blakestad},
  \citenamefont {Epstein}, \citenamefont {Hume}, \citenamefont {Itano},
  \citenamefont {Jost}, \citenamefont {Langer}, \citenamefont {Ozeri},
  \citenamefont {Shiga},\ and\ \citenamefont {Wineland}}]{Seidelin2006a}%
  \BibitemOpen
  \bibfield  {author} {\bibinfo {author} {\bibfnamefont {S.}~\bibnamefont
  {Seidelin}}, \bibinfo {author} {\bibfnamefont {J.}~\bibnamefont
  {Chiaverini}}, \bibinfo {author} {\bibfnamefont {R.}~\bibnamefont {Reichle}},
  \bibinfo {author} {\bibfnamefont {J.~J.}\ \bibnamefont {Bollinger}}, \bibinfo
  {author} {\bibfnamefont {D.}~\bibnamefont {Leibfried}}, \bibinfo {author}
  {\bibfnamefont {J.}~\bibnamefont {Britton}}, \bibinfo {author} {\bibfnamefont
  {J.~H.}\ \bibnamefont {Wesenberg}}, \bibinfo {author} {\bibfnamefont {R.~B.}\
  \bibnamefont {Blakestad}}, \bibinfo {author} {\bibfnamefont {R.~J.}\
  \bibnamefont {Epstein}}, \bibinfo {author} {\bibfnamefont {D.~B.}\
  \bibnamefont {Hume}}, \bibinfo {author} {\bibfnamefont {W.~M.}\ \bibnamefont
  {Itano}}, \bibinfo {author} {\bibfnamefont {J.~D.}\ \bibnamefont {Jost}},
  \bibinfo {author} {\bibfnamefont {C.}~\bibnamefont {Langer}}, \bibinfo
  {author} {\bibfnamefont {R.}~\bibnamefont {Ozeri}}, \bibinfo {author}
  {\bibfnamefont {N.}~\bibnamefont {Shiga}},\ and\ \bibinfo {author}
  {\bibfnamefont {D.~J.}\ \bibnamefont {Wineland}},\ }\bibfield  {title}
  {\bibinfo {title} {{Microfabricated Surface-Electrode Ion Trap for Scalable
  Quantum Information Processing}},\ }\href
  {https://doi.org/10.1103/PhysRevLett.96.253003} {\bibfield  {journal}
  {\bibinfo  {journal} {Phys. Rev. Lett.}\ }\textbf {\bibinfo {volume} {96}},\
  \bibinfo {pages} {253003} (\bibinfo {year} {2006})}\BibitemShut {NoStop}%
\bibitem [{\citenamefont {Wright}\ \emph {et~al.}(2019)\citenamefont {Wright},
  \citenamefont {Beck}, \citenamefont {Debnath}, \citenamefont {Amini},
  \citenamefont {Nam}, \citenamefont {Grzesiak}, \citenamefont {Chen},
  \citenamefont {Pisenti}, \citenamefont {Chmielewski}, \citenamefont
  {Collins}, \citenamefont {Hudek}, \citenamefont {Mizrahi}, \citenamefont
  {Wong-Campos}, \citenamefont {Allen}, \citenamefont {Apisdorf}, \citenamefont
  {Solomon}, \citenamefont {Williams}, \citenamefont {Ducore}, \citenamefont
  {Blinov}, \citenamefont {Kreikemeier}, \citenamefont {Chaplin}, \citenamefont
  {Keesan}, \citenamefont {Monroe},\ and\ \citenamefont {Kim}}]{Wright2019}%
  \BibitemOpen
  \bibfield  {author} {\bibinfo {author} {\bibfnamefont {K.}~\bibnamefont
  {Wright}}, \bibinfo {author} {\bibfnamefont {K.~M.}\ \bibnamefont {Beck}},
  \bibinfo {author} {\bibfnamefont {S.}~\bibnamefont {Debnath}}, \bibinfo
  {author} {\bibfnamefont {J.~M.}\ \bibnamefont {Amini}}, \bibinfo {author}
  {\bibfnamefont {Y.}~\bibnamefont {Nam}}, \bibinfo {author} {\bibfnamefont
  {N.}~\bibnamefont {Grzesiak}}, \bibinfo {author} {\bibfnamefont {J.-S.}\
  \bibnamefont {Chen}}, \bibinfo {author} {\bibfnamefont {N.~C.}\ \bibnamefont
  {Pisenti}}, \bibinfo {author} {\bibfnamefont {M.}~\bibnamefont
  {Chmielewski}}, \bibinfo {author} {\bibfnamefont {C.}~\bibnamefont
  {Collins}}, \bibinfo {author} {\bibfnamefont {K.~M.}\ \bibnamefont {Hudek}},
  \bibinfo {author} {\bibfnamefont {J.}~\bibnamefont {Mizrahi}}, \bibinfo
  {author} {\bibfnamefont {J.~D.}\ \bibnamefont {Wong-Campos}}, \bibinfo
  {author} {\bibfnamefont {S.}~\bibnamefont {Allen}}, \bibinfo {author}
  {\bibfnamefont {J.}~\bibnamefont {Apisdorf}}, \bibinfo {author}
  {\bibfnamefont {P.}~\bibnamefont {Solomon}}, \bibinfo {author} {\bibfnamefont
  {M.}~\bibnamefont {Williams}}, \bibinfo {author} {\bibfnamefont {A.~M.}\
  \bibnamefont {Ducore}}, \bibinfo {author} {\bibfnamefont {A.}~\bibnamefont
  {Blinov}}, \bibinfo {author} {\bibfnamefont {S.~M.}\ \bibnamefont
  {Kreikemeier}}, \bibinfo {author} {\bibfnamefont {V.}~\bibnamefont
  {Chaplin}}, \bibinfo {author} {\bibfnamefont {M.}~\bibnamefont {Keesan}},
  \bibinfo {author} {\bibfnamefont {C.}~\bibnamefont {Monroe}},\ and\ \bibinfo
  {author} {\bibfnamefont {J.}~\bibnamefont {Kim}},\ }\bibfield  {title}
  {\bibinfo {title} {{Benchmarking an 11-qubit quantum computer}},\ }\href
  {https://doi.org/10.1038/s41467-019-13534-2} {\bibfield  {journal} {\bibinfo
  {journal} {Nat. Commun.}\ }\textbf {\bibinfo {volume} {10}},\ \bibinfo
  {pages} {5464} (\bibinfo {year} {2019})}\BibitemShut {NoStop}%
\bibitem [{\citenamefont {Pino}\ \emph {et~al.}(2020)\citenamefont {Pino},
  \citenamefont {Dreiling}, \citenamefont {Figgatt}, \citenamefont {Gaebler},
  \citenamefont {Moses}, \citenamefont {Allman}, \citenamefont {Baldwin},
  \citenamefont {Foss-Feig}, \citenamefont {Hayes}, \citenamefont {Mayer},
  \citenamefont {Ryan-Anderson},\ and\ \citenamefont {Neyenhuis}}]{Pino2020}%
  \BibitemOpen
  \bibfield  {author} {\bibinfo {author} {\bibfnamefont {J.~M.}\ \bibnamefont
  {Pino}}, \bibinfo {author} {\bibfnamefont {J.~M.}\ \bibnamefont {Dreiling}},
  \bibinfo {author} {\bibfnamefont {C.}~\bibnamefont {Figgatt}}, \bibinfo
  {author} {\bibfnamefont {J.~P.}\ \bibnamefont {Gaebler}}, \bibinfo {author}
  {\bibfnamefont {S.~A.}\ \bibnamefont {Moses}}, \bibinfo {author}
  {\bibfnamefont {M.~S.}\ \bibnamefont {Allman}}, \bibinfo {author}
  {\bibfnamefont {C.~H.}\ \bibnamefont {Baldwin}}, \bibinfo {author}
  {\bibfnamefont {M.}~\bibnamefont {Foss-Feig}}, \bibinfo {author}
  {\bibfnamefont {D.}~\bibnamefont {Hayes}}, \bibinfo {author} {\bibfnamefont
  {K.}~\bibnamefont {Mayer}}, \bibinfo {author} {\bibfnamefont
  {C.}~\bibnamefont {Ryan-Anderson}},\ and\ \bibinfo {author} {\bibfnamefont
  {B.}~\bibnamefont {Neyenhuis}},\ }\bibfield  {title} {\bibinfo {title}
  {{Demonstration of the QCCD trapped-ion quantum computer architecture}},\
  }\href@noop {} {\bibfield  {journal} {\bibinfo  {journal} {arXiv:2003.01293}\
  } (\bibinfo {year} {2020})}\BibitemShut {NoStop}%
\bibitem [{\citenamefont {Eschner}\ \emph {et~al.}(2003)\citenamefont
  {Eschner}, \citenamefont {Morigi}, \citenamefont {Schmidt-Kaler},\ and\
  \citenamefont {Blatt}}]{Eschner2003}%
  \BibitemOpen
  \bibfield  {author} {\bibinfo {author} {\bibfnamefont {J.}~\bibnamefont
  {Eschner}}, \bibinfo {author} {\bibfnamefont {G.}~\bibnamefont {Morigi}},
  \bibinfo {author} {\bibfnamefont {F.}~\bibnamefont {Schmidt-Kaler}},\ and\
  \bibinfo {author} {\bibfnamefont {R.}~\bibnamefont {Blatt}},\ }\bibfield
  {title} {\bibinfo {title} {{Laser cooling of trapped ions}},\ }\href
  {https://doi.org/10.1364/JOSAB.20.001003} {\bibfield  {journal} {\bibinfo
  {journal} {J. Opt. Soc. Am. B}\ }\textbf {\bibinfo {volume} {20}},\ \bibinfo
  {pages} {1003} (\bibinfo {year} {2003})}\BibitemShut {NoStop}%
\bibitem [{\citenamefont {Ruster}\ \emph {et~al.}(2016)\citenamefont {Ruster},
  \citenamefont {Schmiegelow}, \citenamefont {Kaufmann}, \citenamefont
  {Warschburger}, \citenamefont {Schmidt-Kaler},\ and\ \citenamefont
  {Poschinger}}]{Ruster2016}%
  \BibitemOpen
  \bibfield  {author} {\bibinfo {author} {\bibfnamefont {T.}~\bibnamefont
  {Ruster}}, \bibinfo {author} {\bibfnamefont {C.~T.}\ \bibnamefont
  {Schmiegelow}}, \bibinfo {author} {\bibfnamefont {H.}~\bibnamefont
  {Kaufmann}}, \bibinfo {author} {\bibfnamefont {C.}~\bibnamefont
  {Warschburger}}, \bibinfo {author} {\bibfnamefont {F.}~\bibnamefont
  {Schmidt-Kaler}},\ and\ \bibinfo {author} {\bibfnamefont {U.~G.}\
  \bibnamefont {Poschinger}},\ }\bibfield  {title} {\bibinfo {title} {{A
  long-lived Zeeman trapped-ion qubit}},\ }\href
  {https://doi.org/10.1007/s00340-016-6527-4} {\bibfield  {journal} {\bibinfo
  {journal} {Appl. Phys. B}\ }\textbf {\bibinfo {volume} {122}},\ \bibinfo
  {pages} {254} (\bibinfo {year} {2016})}\BibitemShut {NoStop}%
\bibitem [{\citenamefont {Harty}\ \emph {et~al.}(2014)\citenamefont {Harty},
  \citenamefont {Allcock}, \citenamefont {Ballance}, \citenamefont {Guidoni},
  \citenamefont {Janacek}, \citenamefont {Linke}, \citenamefont {Stacey},\ and\
  \citenamefont {Lucas}}]{Harty2014a}%
  \BibitemOpen
  \bibfield  {author} {\bibinfo {author} {\bibfnamefont {T.~P.}\ \bibnamefont
  {Harty}}, \bibinfo {author} {\bibfnamefont {D.~T.~C.}\ \bibnamefont
  {Allcock}}, \bibinfo {author} {\bibfnamefont {C.~J.}\ \bibnamefont
  {Ballance}}, \bibinfo {author} {\bibfnamefont {L.}~\bibnamefont {Guidoni}},
  \bibinfo {author} {\bibfnamefont {H.~A.}\ \bibnamefont {Janacek}}, \bibinfo
  {author} {\bibfnamefont {N.~M.}\ \bibnamefont {Linke}}, \bibinfo {author}
  {\bibfnamefont {D.~N.}\ \bibnamefont {Stacey}},\ and\ \bibinfo {author}
  {\bibfnamefont {D.~M.}\ \bibnamefont {Lucas}},\ }\bibfield  {title} {\bibinfo
  {title} {{High-Fidelity Preparation, Gates, Memory, and Readout of a
  Trapped-Ion Quantum Bit}},\ }\href
  {https://doi.org/10.1103/PhysRevLett.113.220501} {\bibfield  {journal}
  {\bibinfo  {journal} {Phys. Rev. Lett.}\ }\textbf {\bibinfo {volume} {113}},\
  \bibinfo {pages} {220501} (\bibinfo {year} {2014})}\BibitemShut {NoStop}%
\bibitem [{\citenamefont {Wang}\ \emph {et~al.}(2020)\citenamefont {Wang},
  \citenamefont {Luan}, \citenamefont {Qiao}, \citenamefont {Um}, \citenamefont
  {Zhang}, \citenamefont {Wang}, \citenamefont {Yuan}, \citenamefont {Gu},
  \citenamefont {Zhang},\ and\ \citenamefont {Kim}}]{Wang2020}%
  \BibitemOpen
  \bibfield  {author} {\bibinfo {author} {\bibfnamefont {P.}~\bibnamefont
  {Wang}}, \bibinfo {author} {\bibfnamefont {C.-Y.}\ \bibnamefont {Luan}},
  \bibinfo {author} {\bibfnamefont {M.}~\bibnamefont {Qiao}}, \bibinfo {author}
  {\bibfnamefont {M.}~\bibnamefont {Um}}, \bibinfo {author} {\bibfnamefont
  {J.}~\bibnamefont {Zhang}}, \bibinfo {author} {\bibfnamefont
  {Y.}~\bibnamefont {Wang}}, \bibinfo {author} {\bibfnamefont {X.}~\bibnamefont
  {Yuan}}, \bibinfo {author} {\bibfnamefont {M.}~\bibnamefont {Gu}}, \bibinfo
  {author} {\bibfnamefont {J.}~\bibnamefont {Zhang}},\ and\ \bibinfo {author}
  {\bibfnamefont {K.}~\bibnamefont {Kim}},\ }\bibfield  {title} {\bibinfo
  {title} {{Single ion-qubit exceeding one hour coherence time}},\ }\href@noop
  {} {\bibfield  {journal} {\bibinfo  {journal} {arXiv:2008.00251}\ } (\bibinfo
  {year} {2020})}\BibitemShut {NoStop}%
\bibitem [{\citenamefont {Ball}\ \emph {et~al.}(2016)\citenamefont {Ball},
  \citenamefont {Oliver},\ and\ \citenamefont {Biercuk}}]{Ball2016}%
  \BibitemOpen
  \bibfield  {author} {\bibinfo {author} {\bibfnamefont {H.}~\bibnamefont
  {Ball}}, \bibinfo {author} {\bibfnamefont {W.~D.}\ \bibnamefont {Oliver}},\
  and\ \bibinfo {author} {\bibfnamefont {M.~J.}\ \bibnamefont {Biercuk}},\
  }\bibfield  {title} {\bibinfo {title} {{The role of master clock stability in
  quantum information processing}},\ }\href
  {https://doi.org/10.1038/npjqi.2016.33} {\bibfield  {journal} {\bibinfo
  {journal} {npj Quantum Inf.}\ }\textbf {\bibinfo {volume} {2}},\ \bibinfo
  {pages} {16033} (\bibinfo {year} {2016})}\BibitemShut {NoStop}%
\bibitem [{\citenamefont {Sepiol}\ \emph {et~al.}(2019)\citenamefont {Sepiol},
  \citenamefont {Hughes}, \citenamefont {Tarlton}, \citenamefont {Nadlinger},
  \citenamefont {Ballance}, \citenamefont {Ballance}, \citenamefont {Harty},
  \citenamefont {Steane}, \citenamefont {Goodwin},\ and\ \citenamefont
  {Lucas}}]{Sepiol2019}%
  \BibitemOpen
  \bibfield  {author} {\bibinfo {author} {\bibfnamefont {M.~A.}\ \bibnamefont
  {Sepiol}}, \bibinfo {author} {\bibfnamefont {A.~C.}\ \bibnamefont {Hughes}},
  \bibinfo {author} {\bibfnamefont {J.~E.}\ \bibnamefont {Tarlton}}, \bibinfo
  {author} {\bibfnamefont {D.~P.}\ \bibnamefont {Nadlinger}}, \bibinfo {author}
  {\bibfnamefont {T.~G.}\ \bibnamefont {Ballance}}, \bibinfo {author}
  {\bibfnamefont {C.~J.}\ \bibnamefont {Ballance}}, \bibinfo {author}
  {\bibfnamefont {T.~P.}\ \bibnamefont {Harty}}, \bibinfo {author}
  {\bibfnamefont {A.~M.}\ \bibnamefont {Steane}}, \bibinfo {author}
  {\bibfnamefont {J.~F.}\ \bibnamefont {Goodwin}},\ and\ \bibinfo {author}
  {\bibfnamefont {D.~M.}\ \bibnamefont {Lucas}},\ }\bibfield  {title} {\bibinfo
  {title} {{Probing Qubit Memory Errors at the Part-per-Million Level}},\
  }\href {https://doi.org/10.1103/PhysRevLett.123.110503} {\bibfield  {journal}
  {\bibinfo  {journal} {Phys. Rev. Lett.}\ }\textbf {\bibinfo {volume} {123}},\
  \bibinfo {pages} {110503} (\bibinfo {year} {2019})}\BibitemShut {NoStop}%
\bibitem [{\citenamefont {Kane}(1998)}]{Kane1998}%
  \BibitemOpen
  \bibfield  {author} {\bibinfo {author} {\bibfnamefont {B.~E.}\ \bibnamefont
  {Kane}},\ }\bibfield  {title} {\bibinfo {title} {A silicon-based nuclear spin
  quantum computer},\ }\href {https://doi.org/10.1038/30156} {\bibfield
  {journal} {\bibinfo  {journal} {Nature}\ }\textbf {\bibinfo {volume} {393}},\
  \bibinfo {pages} {133} (\bibinfo {year} {1998})}\BibitemShut {NoStop}%
\bibitem [{\citenamefont {Zwanenburg}\ \emph {et~al.}(2013)\citenamefont
  {Zwanenburg}, \citenamefont {Dzurak}, \citenamefont {Morello}, \citenamefont
  {Simmons}, \citenamefont {Hollenberg}, \citenamefont {Klimeck}, \citenamefont
  {Rogge}, \citenamefont {Coppersmith},\ and\ \citenamefont
  {Eriksson}}]{RMPSilicon}%
  \BibitemOpen
  \bibfield  {author} {\bibinfo {author} {\bibfnamefont {F.~A.}\ \bibnamefont
  {Zwanenburg}}, \bibinfo {author} {\bibfnamefont {A.~S.}\ \bibnamefont
  {Dzurak}}, \bibinfo {author} {\bibfnamefont {A.}~\bibnamefont {Morello}},
  \bibinfo {author} {\bibfnamefont {M.~Y.}\ \bibnamefont {Simmons}}, \bibinfo
  {author} {\bibfnamefont {L.~C.~L.}\ \bibnamefont {Hollenberg}}, \bibinfo
  {author} {\bibfnamefont {G.}~\bibnamefont {Klimeck}}, \bibinfo {author}
  {\bibfnamefont {S.}~\bibnamefont {Rogge}}, \bibinfo {author} {\bibfnamefont
  {S.~N.}\ \bibnamefont {Coppersmith}},\ and\ \bibinfo {author} {\bibfnamefont
  {M.~A.}\ \bibnamefont {Eriksson}},\ }\bibfield  {title} {\bibinfo {title}
  {Silicon quantum electronics},\ }\href
  {https://doi.org/10.1103/RevModPhys.85.961} {\bibfield  {journal} {\bibinfo
  {journal} {Rev. Mod. Phys.}\ }\textbf {\bibinfo {volume} {85}},\ \bibinfo
  {pages} {961} (\bibinfo {year} {2013})}\BibitemShut {NoStop}%
\bibitem [{\citenamefont {Sigillito}\ \emph {et~al.}(2019)\citenamefont
  {Sigillito}, \citenamefont {Loy}, \citenamefont {Zajac}, \citenamefont
  {Gullans}, \citenamefont {Edge},\ and\ \citenamefont {Petta}}]{SiGe}%
  \BibitemOpen
  \bibfield  {author} {\bibinfo {author} {\bibfnamefont {A.~J.}\ \bibnamefont
  {Sigillito}}, \bibinfo {author} {\bibfnamefont {J.~C.}\ \bibnamefont {Loy}},
  \bibinfo {author} {\bibfnamefont {D.~M.}\ \bibnamefont {Zajac}}, \bibinfo
  {author} {\bibfnamefont {M.~J.}\ \bibnamefont {Gullans}}, \bibinfo {author}
  {\bibfnamefont {L.~F.}\ \bibnamefont {Edge}},\ and\ \bibinfo {author}
  {\bibfnamefont {J.~R.}\ \bibnamefont {Petta}},\ }\bibfield  {title} {\bibinfo
  {title} {Site-selective quantum control in an isotopically enriched
  $^{28}\mathrm{Si}/\mathrm{Si}_{0.7}\mathrm{Ge}_{0.3}$ quadruple quantum
  dot},\ }\href {https://doi.org/10.1103/PhysRevApplied.11.061006} {\bibfield
  {journal} {\bibinfo  {journal} {Phys. Rev. Applied}\ }\textbf {\bibinfo
  {volume} {11}},\ \bibinfo {pages} {061006} (\bibinfo {year}
  {2019})}\BibitemShut {NoStop}%
\bibitem [{\citenamefont {Wrachtrup}\ and\ \citenamefont
  {Jelezko}(2006)}]{diamond}%
  \BibitemOpen
  \bibfield  {author} {\bibinfo {author} {\bibfnamefont {J.}~\bibnamefont
  {Wrachtrup}}\ and\ \bibinfo {author} {\bibfnamefont {F.}~\bibnamefont
  {Jelezko}},\ }\bibfield  {title} {\bibinfo {title} {Processing quantum
  information in diamond},\ }\href
  {https://doi.org/10.1088/0953-8984/18/21/s08} {\bibfield  {journal} {\bibinfo
   {journal} {Journal of Physics: Condensed Matter}\ }\textbf {\bibinfo
  {volume} {18}},\ \bibinfo {pages} {S807} (\bibinfo {year}
  {2006})}\BibitemShut {NoStop}%
\bibitem [{\citenamefont {Kouwenhoven}\ \emph {et~al.}(1997)\citenamefont
  {Kouwenhoven}, \citenamefont {Marcus}, \citenamefont {McEuen}, \citenamefont
  {Tarucha}, \citenamefont {Westervelt},\ and\ \citenamefont
  {Wingreen}}]{Kouwenhoven1997}%
  \BibitemOpen
  \bibfield  {author} {\bibinfo {author} {\bibfnamefont {L.~P.}\ \bibnamefont
  {Kouwenhoven}}, \bibinfo {author} {\bibfnamefont {C.~M.}\ \bibnamefont
  {Marcus}}, \bibinfo {author} {\bibfnamefont {P.~L.}\ \bibnamefont {McEuen}},
  \bibinfo {author} {\bibfnamefont {S.}~\bibnamefont {Tarucha}}, \bibinfo
  {author} {\bibfnamefont {R.~M.}\ \bibnamefont {Westervelt}},\ and\ \bibinfo
  {author} {\bibfnamefont {N.~S.}\ \bibnamefont {Wingreen}},\ }\bibinfo {title}
  {Electron transport in quantum dots},\ in\ \href
  {https://doi.org/10.1007/978-94-015-8839-3\_4} {\emph {\bibinfo {booktitle}
  {Mesoscopic Electron Transport}}},\ \bibinfo {editor} {edited by\ \bibinfo
  {editor} {\bibfnamefont {L.~L.}\ \bibnamefont {Sohn}}, \bibinfo {editor}
  {\bibfnamefont {L.~P.}\ \bibnamefont {Kouwenhoven}},\ and\ \bibinfo {editor}
  {\bibfnamefont {G.}~\bibnamefont {Sch{\"o}n}}}\ (\bibinfo  {publisher}
  {Springer Netherlands},\ \bibinfo {address} {Dordrecht},\ \bibinfo {year}
  {1997})\ pp.\ \bibinfo {pages} {105--214}\BibitemShut {NoStop}%
\bibitem [{\citenamefont {Yang}\ \emph {et~al.}(2020)\citenamefont {Yang},
  \citenamefont {Leon}, \citenamefont {Hwang}, \citenamefont {Saraiva},
  \citenamefont {Tanttu}, \citenamefont {Huang}, \citenamefont {Lemyre},
  \citenamefont {Chan}, \citenamefont {Tan}, \citenamefont {Hudson} \emph
  {et~al.}}]{yang2020operation}%
  \BibitemOpen
  \bibfield  {author} {\bibinfo {author} {\bibfnamefont {C.~H.}\ \bibnamefont
  {Yang}}, \bibinfo {author} {\bibfnamefont {R.}~\bibnamefont {Leon}}, \bibinfo
  {author} {\bibfnamefont {J.}~\bibnamefont {Hwang}}, \bibinfo {author}
  {\bibfnamefont {A.}~\bibnamefont {Saraiva}}, \bibinfo {author} {\bibfnamefont
  {T.}~\bibnamefont {Tanttu}}, \bibinfo {author} {\bibfnamefont
  {W.}~\bibnamefont {Huang}}, \bibinfo {author} {\bibfnamefont {J.~C.}\
  \bibnamefont {Lemyre}}, \bibinfo {author} {\bibfnamefont {K.~W.}\
  \bibnamefont {Chan}}, \bibinfo {author} {\bibfnamefont {K.}~\bibnamefont
  {Tan}}, \bibinfo {author} {\bibfnamefont {F.~E.}\ \bibnamefont {Hudson}},
  \emph {et~al.},\ }\bibfield  {title} {\bibinfo {title} {Operation of a
  silicon quantum processor unit cell above one kelvin},\ }\href@noop {}
  {\bibfield  {journal} {\bibinfo  {journal} {Nature}\ }\textbf {\bibinfo
  {volume} {580}},\ \bibinfo {pages} {350} (\bibinfo {year}
  {2020})}\BibitemShut {NoStop}%
\bibitem [{\citenamefont {Petit}\ \emph {et~al.}(2020)\citenamefont {Petit},
  \citenamefont {Eenink}, \citenamefont {Russ}, \citenamefont {Lawrie},
  \citenamefont {Hendrickx}, \citenamefont {Philips}, \citenamefont {Clarke},
  \citenamefont {Vandersypen},\ and\ \citenamefont
  {Veldhorst}}]{petit2020universal}%
  \BibitemOpen
  \bibfield  {author} {\bibinfo {author} {\bibfnamefont {L.}~\bibnamefont
  {Petit}}, \bibinfo {author} {\bibfnamefont {H.~G.~J.}\ \bibnamefont
  {Eenink}}, \bibinfo {author} {\bibfnamefont {M.}~\bibnamefont {Russ}},
  \bibinfo {author} {\bibfnamefont {W.~I.~L.}\ \bibnamefont {Lawrie}}, \bibinfo
  {author} {\bibfnamefont {N.~W.}\ \bibnamefont {Hendrickx}}, \bibinfo {author}
  {\bibfnamefont {S.~G.~J.}\ \bibnamefont {Philips}}, \bibinfo {author}
  {\bibfnamefont {J.~S.}\ \bibnamefont {Clarke}}, \bibinfo {author}
  {\bibfnamefont {L.~M.~K.}\ \bibnamefont {Vandersypen}},\ and\ \bibinfo
  {author} {\bibfnamefont {M.}~\bibnamefont {Veldhorst}},\ }\bibfield  {title}
  {\bibinfo {title} {Universal quantum logic in hot silicon qubits},\
  }\href@noop {} {\bibfield  {journal} {\bibinfo  {journal} {Nature}\ }\textbf
  {\bibinfo {volume} {580}},\ \bibinfo {pages} {355} (\bibinfo {year}
  {2020})}\BibitemShut {NoStop}%
\bibitem [{\citenamefont {Loss}\ and\ \citenamefont
  {DiVincenzo}(1998)}]{Loss1998}%
  \BibitemOpen
  \bibfield  {author} {\bibinfo {author} {\bibfnamefont {D.}~\bibnamefont
  {Loss}}\ and\ \bibinfo {author} {\bibfnamefont {D.~P.}\ \bibnamefont
  {DiVincenzo}},\ }\bibfield  {title} {\bibinfo {title} {Quantum computation
  with quantum dots},\ }\href {https://doi.org/10.1103/PhysRevA.57.120}
  {\bibfield  {journal} {\bibinfo  {journal} {Phys. Rev. A}\ }\textbf {\bibinfo
  {volume} {57}},\ \bibinfo {pages} {120} (\bibinfo {year} {1998})}\BibitemShut
  {NoStop}%
\bibitem [{\citenamefont {Morello}\ \emph {et~al.}(2010)\citenamefont
  {Morello}, \citenamefont {Pla}, \citenamefont {Zwanenburg}, \citenamefont
  {Chan}, \citenamefont {Tan}, \citenamefont {Huebl}, \citenamefont
  {M{\"{o}}tt{\"{o}}nen}, \citenamefont {Nugroho}, \citenamefont {Yang},
  \citenamefont {van Donkelaar}, \citenamefont {Alves}, \citenamefont
  {Jamieson}, \citenamefont {Escott}, \citenamefont {Hollenberg}, \citenamefont
  {Clark},\ and\ \citenamefont {Dzurak}}]{Morello2010}%
  \BibitemOpen
  \bibfield  {author} {\bibinfo {author} {\bibfnamefont {A.}~\bibnamefont
  {Morello}}, \bibinfo {author} {\bibfnamefont {J.~J.}\ \bibnamefont {Pla}},
  \bibinfo {author} {\bibfnamefont {F.~A.}\ \bibnamefont {Zwanenburg}},
  \bibinfo {author} {\bibfnamefont {K.~W.}\ \bibnamefont {Chan}}, \bibinfo
  {author} {\bibfnamefont {K.~Y.}\ \bibnamefont {Tan}}, \bibinfo {author}
  {\bibfnamefont {H.}~\bibnamefont {Huebl}}, \bibinfo {author} {\bibfnamefont
  {M.}~\bibnamefont {M{\"{o}}tt{\"{o}}nen}}, \bibinfo {author} {\bibfnamefont
  {C.~D.}\ \bibnamefont {Nugroho}}, \bibinfo {author} {\bibfnamefont
  {C.}~\bibnamefont {Yang}}, \bibinfo {author} {\bibfnamefont {J.~A.}\
  \bibnamefont {van Donkelaar}}, \bibinfo {author} {\bibfnamefont {A.~D.~C.}\
  \bibnamefont {Alves}}, \bibinfo {author} {\bibfnamefont {D.~N.}\ \bibnamefont
  {Jamieson}}, \bibinfo {author} {\bibfnamefont {C.~C.}\ \bibnamefont
  {Escott}}, \bibinfo {author} {\bibfnamefont {L.~C.~L.}\ \bibnamefont
  {Hollenberg}}, \bibinfo {author} {\bibfnamefont {R.~G.}\ \bibnamefont
  {Clark}},\ and\ \bibinfo {author} {\bibfnamefont {A.~S.}\ \bibnamefont
  {Dzurak}},\ }\bibfield  {title} {\bibinfo {title} {{Single-shot readout of an
  electron spin in silicon.}},\ }\href {https://doi.org/10.1038/nature09392}
  {\bibfield  {journal} {\bibinfo  {journal} {Nature}\ }\textbf {\bibinfo
  {volume} {467}},\ \bibinfo {pages} {687} (\bibinfo {year}
  {2010})}\BibitemShut {NoStop}%
\bibitem [{\citenamefont {Levy}(2002)}]{Levy2002}%
  \BibitemOpen
  \bibfield  {author} {\bibinfo {author} {\bibfnamefont {J.}~\bibnamefont
  {Levy}},\ }\bibfield  {title} {\bibinfo {title} {Universal quantum
  computation with spin-$1/2$ pairs and {H}eisenberg exchange},\ }\href
  {https://doi.org/10.1103/PhysRevLett.89.147902} {\bibfield  {journal}
  {\bibinfo  {journal} {Phys. Rev. Lett.}\ }\textbf {\bibinfo {volume} {89}},\
  \bibinfo {pages} {147902} (\bibinfo {year} {2002})}\BibitemShut {NoStop}%
\bibitem [{\citenamefont {Croot}\ \emph {et~al.}(2018)\citenamefont {Croot},
  \citenamefont {Pauka}, \citenamefont {Watson}, \citenamefont {Gardner},
  \citenamefont {Fallahi}, \citenamefont {Manfra},\ and\ \citenamefont
  {Reilly}}]{jellybean}%
  \BibitemOpen
  \bibfield  {author} {\bibinfo {author} {\bibfnamefont {X.~G.}\ \bibnamefont
  {Croot}}, \bibinfo {author} {\bibfnamefont {S.~J.}\ \bibnamefont {Pauka}},
  \bibinfo {author} {\bibfnamefont {J.~D.}\ \bibnamefont {Watson}}, \bibinfo
  {author} {\bibfnamefont {G.~C.}\ \bibnamefont {Gardner}}, \bibinfo {author}
  {\bibfnamefont {S.}~\bibnamefont {Fallahi}}, \bibinfo {author} {\bibfnamefont
  {M.~J.}\ \bibnamefont {Manfra}},\ and\ \bibinfo {author} {\bibfnamefont
  {D.~J.}\ \bibnamefont {Reilly}},\ }\bibfield  {title} {\bibinfo {title}
  {Device architecture for coupling spin qubits via an intermediate quantum
  state},\ }\href {https://doi.org/10.1103/PhysRevApplied.10.044058} {\bibfield
   {journal} {\bibinfo  {journal} {Phys. Rev. Applied}\ }\textbf {\bibinfo
  {volume} {10}},\ \bibinfo {pages} {044058} (\bibinfo {year}
  {2018})}\BibitemShut {NoStop}%
\bibitem [{\citenamefont {DiVincenzo}\ \emph {et~al.}(2000)\citenamefont
  {DiVincenzo}, \citenamefont {Bacon}, \citenamefont {Kempe}, \citenamefont
  {Burkard},\ and\ \citenamefont {Whaley}}]{DiVincenzo2000a}%
  \BibitemOpen
  \bibfield  {author} {\bibinfo {author} {\bibfnamefont {D.~P.}\ \bibnamefont
  {DiVincenzo}}, \bibinfo {author} {\bibfnamefont {D.}~\bibnamefont {Bacon}},
  \bibinfo {author} {\bibfnamefont {J.}~\bibnamefont {Kempe}}, \bibinfo
  {author} {\bibfnamefont {G.}~\bibnamefont {Burkard}},\ and\ \bibinfo {author}
  {\bibfnamefont {K.~B.}\ \bibnamefont {Whaley}},\ }\bibfield  {title}
  {\bibinfo {title} {{Universal quantum computation with the exchange
  interaction}},\ }\href {https://doi.org/10.1038/35042541} {\bibfield
  {journal} {\bibinfo  {journal} {Nature}\ }\textbf {\bibinfo {volume} {408}},\
  \bibinfo {pages} {339} (\bibinfo {year} {2000})}\BibitemShut {NoStop}%
\bibitem [{\citenamefont {Medford}\ \emph {et~al.}(2013)\citenamefont
  {Medford}, \citenamefont {Beil}, \citenamefont {Taylor}, \citenamefont
  {Rashba}, \citenamefont {Lu}, \citenamefont {Gossard},\ and\ \citenamefont
  {Marcus}}]{Medford2013}%
  \BibitemOpen
  \bibfield  {author} {\bibinfo {author} {\bibfnamefont {J.}~\bibnamefont
  {Medford}}, \bibinfo {author} {\bibfnamefont {J.}~\bibnamefont {Beil}},
  \bibinfo {author} {\bibfnamefont {J.~M.}\ \bibnamefont {Taylor}}, \bibinfo
  {author} {\bibfnamefont {E.~I.}\ \bibnamefont {Rashba}}, \bibinfo {author}
  {\bibfnamefont {H.}~\bibnamefont {Lu}}, \bibinfo {author} {\bibfnamefont
  {A.~C.}\ \bibnamefont {Gossard}},\ and\ \bibinfo {author} {\bibfnamefont
  {C.~M.}\ \bibnamefont {Marcus}},\ }\bibfield  {title} {\bibinfo {title}
  {Quantum-dot-based resonant exchange qubit},\ }\href
  {https://doi.org/10.1103/PhysRevLett.111.050501} {\bibfield  {journal}
  {\bibinfo  {journal} {Phys. Rev. Lett.}\ }\textbf {\bibinfo {volume} {111}},\
  \bibinfo {pages} {050501} (\bibinfo {year} {2013})}\BibitemShut {NoStop}%
\bibitem [{\citenamefont {Hanson}\ \emph {et~al.}(2007)\citenamefont {Hanson},
  \citenamefont {Kouwenhoven}, \citenamefont {Petta}, \citenamefont {Tarucha},\
  and\ \citenamefont {Vandersypen}}]{HansonRMP}%
  \BibitemOpen
  \bibfield  {author} {\bibinfo {author} {\bibfnamefont {R.}~\bibnamefont
  {Hanson}}, \bibinfo {author} {\bibfnamefont {L.~P.}\ \bibnamefont
  {Kouwenhoven}}, \bibinfo {author} {\bibfnamefont {J.~R.}\ \bibnamefont
  {Petta}}, \bibinfo {author} {\bibfnamefont {S.}~\bibnamefont {Tarucha}},\
  and\ \bibinfo {author} {\bibfnamefont {L.~M.~K.}\ \bibnamefont
  {Vandersypen}},\ }\bibfield  {title} {\bibinfo {title} {Spins in few-electron
  quantum dots},\ }\href {https://doi.org/10.1103/RevModPhys.79.1217}
  {\bibfield  {journal} {\bibinfo  {journal} {Rev. Mod. Phys.}\ }\textbf
  {\bibinfo {volume} {79}},\ \bibinfo {pages} {1217} (\bibinfo {year}
  {2007})}\BibitemShut {NoStop}%
\bibitem [{\citenamefont {Martinis}\ \emph {et~al.}(1985)\citenamefont
  {Martinis}, \citenamefont {Devoret},\ and\ \citenamefont
  {Clarke}}]{martinis1985energy}%
  \BibitemOpen
  \bibfield  {author} {\bibinfo {author} {\bibfnamefont {J.~M.}\ \bibnamefont
  {Martinis}}, \bibinfo {author} {\bibfnamefont {M.~H.}\ \bibnamefont
  {Devoret}},\ and\ \bibinfo {author} {\bibfnamefont {J.}~\bibnamefont
  {Clarke}},\ }\bibfield  {title} {\bibinfo {title} {Energy-level quantization
  in the zero-voltage state of a current-biased josephson junction},\
  }\href@noop {} {\bibfield  {journal} {\bibinfo  {journal} {Phys. Rev. Lett.}\
  }\textbf {\bibinfo {volume} {55}},\ \bibinfo {pages} {1543} (\bibinfo {year}
  {1985})}\BibitemShut {NoStop}%
\bibitem [{\citenamefont {Nakamura}\ \emph {et~al.}(1999)\citenamefont
  {Nakamura}, \citenamefont {Pashkin},\ and\ \citenamefont
  {Tsai}}]{nakamura1999coherent}%
  \BibitemOpen
  \bibfield  {author} {\bibinfo {author} {\bibfnamefont {Y.}~\bibnamefont
  {Nakamura}}, \bibinfo {author} {\bibfnamefont {Y.~A.}\ \bibnamefont
  {Pashkin}},\ and\ \bibinfo {author} {\bibfnamefont {J.~S.}\ \bibnamefont
  {Tsai}},\ }\bibfield  {title} {\bibinfo {title} {Coherent control of
  macroscopic quantum states in a single-{C}ooper-pair box},\ }\href@noop {}
  {\bibfield  {journal} {\bibinfo  {journal} {nature}\ }\textbf {\bibinfo
  {volume} {398}},\ \bibinfo {pages} {786} (\bibinfo {year}
  {1999})}\BibitemShut {NoStop}%
\bibitem [{\citenamefont {Koch}\ \emph {et~al.}(2007)\citenamefont {Koch},
  \citenamefont {Terri}, \citenamefont {Gambetta}, \citenamefont {Houck},
  \citenamefont {Schuster}, \citenamefont {Majer}, \citenamefont {Blais},
  \citenamefont {Devoret}, \citenamefont {Girvin},\ and\ \citenamefont
  {Schoelkopf}}]{koch2007charge}%
  \BibitemOpen
  \bibfield  {author} {\bibinfo {author} {\bibfnamefont {J.}~\bibnamefont
  {Koch}}, \bibinfo {author} {\bibfnamefont {M.~Y.}\ \bibnamefont {Terri}},
  \bibinfo {author} {\bibfnamefont {J.}~\bibnamefont {Gambetta}}, \bibinfo
  {author} {\bibfnamefont {A.~A.}\ \bibnamefont {Houck}}, \bibinfo {author}
  {\bibfnamefont {D.}~\bibnamefont {Schuster}}, \bibinfo {author}
  {\bibfnamefont {J.}~\bibnamefont {Majer}}, \bibinfo {author} {\bibfnamefont
  {A.}~\bibnamefont {Blais}}, \bibinfo {author} {\bibfnamefont {M.~H.}\
  \bibnamefont {Devoret}}, \bibinfo {author} {\bibfnamefont {S.~M.}\
  \bibnamefont {Girvin}},\ and\ \bibinfo {author} {\bibfnamefont {R.~J.}\
  \bibnamefont {Schoelkopf}},\ }\bibfield  {title} {\bibinfo {title}
  {Charge-insensitive qubit design derived from the cooper pair box},\
  }\href@noop {} {\bibfield  {journal} {\bibinfo  {journal} {Phys. Rev. A}\
  }\textbf {\bibinfo {volume} {76}},\ \bibinfo {pages} {042319} (\bibinfo
  {year} {2007})}\BibitemShut {NoStop}%
\bibitem [{\citenamefont {Krantz}\ \emph {et~al.}(2019)\citenamefont {Krantz},
  \citenamefont {Kjaergaard}, \citenamefont {Yan}, \citenamefont {Orlando},
  \citenamefont {Gustavsson},\ and\ \citenamefont
  {Oliver}}]{krantz2019quantum}%
  \BibitemOpen
  \bibfield  {author} {\bibinfo {author} {\bibfnamefont {P.}~\bibnamefont
  {Krantz}}, \bibinfo {author} {\bibfnamefont {M.}~\bibnamefont {Kjaergaard}},
  \bibinfo {author} {\bibfnamefont {F.}~\bibnamefont {Yan}}, \bibinfo {author}
  {\bibfnamefont {T.~P.}\ \bibnamefont {Orlando}}, \bibinfo {author}
  {\bibfnamefont {S.}~\bibnamefont {Gustavsson}},\ and\ \bibinfo {author}
  {\bibfnamefont {W.~D.}\ \bibnamefont {Oliver}},\ }\bibfield  {title}
  {\bibinfo {title} {A quantum engineer's guide to superconducting qubits},\
  }\href@noop {} {\bibfield  {journal} {\bibinfo  {journal} {Applied Physics
  Reviews}\ }\textbf {\bibinfo {volume} {6}},\ \bibinfo {pages} {021318}
  (\bibinfo {year} {2019})}\BibitemShut {NoStop}%
\bibitem [{\citenamefont {Chen}(2018)}]{Jimmythesis}%
  \BibitemOpen
  \bibfield  {author} {\bibinfo {author} {\bibfnamefont {Z.}~\bibnamefont
  {Chen}},\ }\emph {\bibinfo {title} {Metrology of quantum control and
  measurement in superconducting qubits}},\ \href@noop {} {Ph.D. thesis},\
  \bibinfo  {school} {UC Santa Barbara} (\bibinfo {year} {2018})\BibitemShut
  {NoStop}%
\bibitem [{\citenamefont {Place}\ \emph {et~al.}(2020)\citenamefont {Place},
  \citenamefont {Rodgers}, \citenamefont {Mundada}, \citenamefont {Smitham},
  \citenamefont {Fitzpatrick}, \citenamefont {Leng}, \citenamefont {Premkumar},
  \citenamefont {Bryon}, \citenamefont {Sussman}, \citenamefont {Cheng} \emph
  {et~al.}}]{place2020new}%
  \BibitemOpen
  \bibfield  {author} {\bibinfo {author} {\bibfnamefont {A.~P.}\ \bibnamefont
  {Place}}, \bibinfo {author} {\bibfnamefont {L.~V.}\ \bibnamefont {Rodgers}},
  \bibinfo {author} {\bibfnamefont {P.}~\bibnamefont {Mundada}}, \bibinfo
  {author} {\bibfnamefont {B.~M.}\ \bibnamefont {Smitham}}, \bibinfo {author}
  {\bibfnamefont {M.}~\bibnamefont {Fitzpatrick}}, \bibinfo {author}
  {\bibfnamefont {Z.}~\bibnamefont {Leng}}, \bibinfo {author} {\bibfnamefont
  {A.}~\bibnamefont {Premkumar}}, \bibinfo {author} {\bibfnamefont
  {J.}~\bibnamefont {Bryon}}, \bibinfo {author} {\bibfnamefont
  {S.}~\bibnamefont {Sussman}}, \bibinfo {author} {\bibfnamefont
  {G.}~\bibnamefont {Cheng}}, \emph {et~al.},\ }\bibfield  {title} {\bibinfo
  {title} {New material platform for superconducting transmon qubits with
  coherence times exceeding 0.3 milliseconds},\ }\href@noop {} {\bibfield
  {journal} {\bibinfo  {journal} {arXiv:2003.00024}\ } (\bibinfo {year}
  {2020})}\BibitemShut {NoStop}%
\bibitem [{\citenamefont {Jurcevic}\ \emph {et~al.}(2020)\citenamefont
  {Jurcevic}, \citenamefont {Javadi-Abhari}, \citenamefont {Bishop},
  \citenamefont {Lauer}, \citenamefont {Bogorin}, \citenamefont {Brink},
  \citenamefont {Capelluto}, \citenamefont {G{\"u}nl{\"u}k}, \citenamefont
  {Itoko}, \citenamefont {Kanazawa} \emph
  {et~al.}}]{jurcevic2020demonstration}%
  \BibitemOpen
  \bibfield  {author} {\bibinfo {author} {\bibfnamefont {P.}~\bibnamefont
  {Jurcevic}}, \bibinfo {author} {\bibfnamefont {A.}~\bibnamefont
  {Javadi-Abhari}}, \bibinfo {author} {\bibfnamefont {L.~S.}\ \bibnamefont
  {Bishop}}, \bibinfo {author} {\bibfnamefont {I.}~\bibnamefont {Lauer}},
  \bibinfo {author} {\bibfnamefont {D.~F.}\ \bibnamefont {Bogorin}}, \bibinfo
  {author} {\bibfnamefont {M.}~\bibnamefont {Brink}}, \bibinfo {author}
  {\bibfnamefont {L.}~\bibnamefont {Capelluto}}, \bibinfo {author}
  {\bibfnamefont {O.}~\bibnamefont {G{\"u}nl{\"u}k}}, \bibinfo {author}
  {\bibfnamefont {T.}~\bibnamefont {Itoko}}, \bibinfo {author} {\bibfnamefont
  {N.}~\bibnamefont {Kanazawa}}, \emph {et~al.},\ }\bibfield  {title} {\bibinfo
  {title} {Demonstration of quantum volume 64 on a superconducting quantum
  computing system},\ }\href@noop {} {\bibfield  {journal} {\bibinfo  {journal}
  {arXiv:2008.08571}\ } (\bibinfo {year} {2020})}\BibitemShut {NoStop}%
\bibitem [{\citenamefont {Klimov}\ \emph {et~al.}(2018)\citenamefont {Klimov},
  \citenamefont {Kelly}, \citenamefont {Chen}, \citenamefont {Neeley},
  \citenamefont {Megrant}, \citenamefont {Burkett}, \citenamefont {Barends},
  \citenamefont {Arya}, \citenamefont {Chiaro}, \citenamefont {Chen} \emph
  {et~al.}}]{klimov2018fluctuations}%
  \BibitemOpen
  \bibfield  {author} {\bibinfo {author} {\bibfnamefont {P.}~\bibnamefont
  {Klimov}}, \bibinfo {author} {\bibfnamefont {J.}~\bibnamefont {Kelly}},
  \bibinfo {author} {\bibfnamefont {Z.}~\bibnamefont {Chen}}, \bibinfo {author}
  {\bibfnamefont {M.}~\bibnamefont {Neeley}}, \bibinfo {author} {\bibfnamefont
  {A.}~\bibnamefont {Megrant}}, \bibinfo {author} {\bibfnamefont
  {B.}~\bibnamefont {Burkett}}, \bibinfo {author} {\bibfnamefont
  {R.}~\bibnamefont {Barends}}, \bibinfo {author} {\bibfnamefont
  {K.}~\bibnamefont {Arya}}, \bibinfo {author} {\bibfnamefont {B.}~\bibnamefont
  {Chiaro}}, \bibinfo {author} {\bibfnamefont {Y.}~\bibnamefont {Chen}}, \emph
  {et~al.},\ }\bibfield  {title} {\bibinfo {title} {Fluctuations of
  energy-relaxation times in superconducting qubits},\ }\href@noop {}
  {\bibfield  {journal} {\bibinfo  {journal} {Phys. Rev. Lett.}\ }\textbf
  {\bibinfo {volume} {121}},\ \bibinfo {pages} {090502} (\bibinfo {year}
  {2018})}\BibitemShut {NoStop}%
\bibitem [{\citenamefont {Koppens}\ \emph {et~al.}(2006)\citenamefont
  {Koppens}, \citenamefont {Buizert}, \citenamefont {Tielrooij}, \citenamefont
  {Vink}, \citenamefont {Nowack}, \citenamefont {Meunier}, \citenamefont
  {Kouwenhoven},\ and\ \citenamefont {Vandersypen}}]{Koppens2006}%
  \BibitemOpen
  \bibfield  {author} {\bibinfo {author} {\bibfnamefont {F.~H.~L.}\
  \bibnamefont {Koppens}}, \bibinfo {author} {\bibfnamefont {C.}~\bibnamefont
  {Buizert}}, \bibinfo {author} {\bibfnamefont {K.~J.}\ \bibnamefont
  {Tielrooij}}, \bibinfo {author} {\bibfnamefont {I.~T.}\ \bibnamefont {Vink}},
  \bibinfo {author} {\bibfnamefont {K.~C.}\ \bibnamefont {Nowack}}, \bibinfo
  {author} {\bibfnamefont {T.}~\bibnamefont {Meunier}}, \bibinfo {author}
  {\bibfnamefont {L.~P.}\ \bibnamefont {Kouwenhoven}},\ and\ \bibinfo {author}
  {\bibfnamefont {L.~M.~K.}\ \bibnamefont {Vandersypen}},\ }\bibfield  {title}
  {\bibinfo {title} {{Driven coherent oscillations of a single electron spin in
  a quantum dot}},\ }\href {https://doi.org/10.1038/nature05065} {\bibfield
  {journal} {\bibinfo  {journal} {Nature}\ }\textbf {\bibinfo {volume} {442}},\
  \bibinfo {pages} {766} (\bibinfo {year} {2006})}\BibitemShut {NoStop}%
\bibitem [{\citenamefont {van Dijk}\ \emph {et~al.}(2019)\citenamefont {van
  Dijk}, \citenamefont {Kawakami}, \citenamefont {Schouten}, \citenamefont
  {Veldhorst}, \citenamefont {Vandersypen}, \citenamefont {Babaie},
  \citenamefont {Charbon},\ and\ \citenamefont {Sebastiano}}]{VanDijk2019}%
  \BibitemOpen
  \bibfield  {author} {\bibinfo {author} {\bibfnamefont {J.}~\bibnamefont {van
  Dijk}}, \bibinfo {author} {\bibfnamefont {E.}~\bibnamefont {Kawakami}},
  \bibinfo {author} {\bibfnamefont {R.}~\bibnamefont {Schouten}}, \bibinfo
  {author} {\bibfnamefont {M.}~\bibnamefont {Veldhorst}}, \bibinfo {author}
  {\bibfnamefont {L.}~\bibnamefont {Vandersypen}}, \bibinfo {author}
  {\bibfnamefont {M.}~\bibnamefont {Babaie}}, \bibinfo {author} {\bibfnamefont
  {E.}~\bibnamefont {Charbon}},\ and\ \bibinfo {author} {\bibfnamefont
  {F.}~\bibnamefont {Sebastiano}},\ }\bibfield  {title} {\bibinfo {title}
  {{Impact of Classical Control Electronics on Qubit Fidelity}},\ }\href
  {https://doi.org/10.1103/PhysRevApplied.12.044054} {\bibfield  {journal}
  {\bibinfo  {journal} {Phys. Rev. Appl.}\ }\textbf {\bibinfo {volume} {12}},\
  \bibinfo {pages} {044054} (\bibinfo {year} {2019})}\BibitemShut {NoStop}%
\bibitem [{\citenamefont {Ospelkaus}\ \emph {et~al.}(2011)\citenamefont
  {Ospelkaus}, \citenamefont {Warring}, \citenamefont {Colombe}, \citenamefont
  {Brown}, \citenamefont {Amini}, \citenamefont {Leibfried},\ and\
  \citenamefont {Wineland}}]{Ospelkaus2011}%
  \BibitemOpen
  \bibfield  {author} {\bibinfo {author} {\bibfnamefont {C.}~\bibnamefont
  {Ospelkaus}}, \bibinfo {author} {\bibfnamefont {U.}~\bibnamefont {Warring}},
  \bibinfo {author} {\bibfnamefont {Y.}~\bibnamefont {Colombe}}, \bibinfo
  {author} {\bibfnamefont {K.~R.}\ \bibnamefont {Brown}}, \bibinfo {author}
  {\bibfnamefont {J.~M.}\ \bibnamefont {Amini}}, \bibinfo {author}
  {\bibfnamefont {D.}~\bibnamefont {Leibfried}},\ and\ \bibinfo {author}
  {\bibfnamefont {D.~J.}\ \bibnamefont {Wineland}},\ }\bibfield  {title}
  {\bibinfo {title} {{Microwave quantum logic gates for trapped ions}},\ }\href
  {https://doi.org/10.1038/nature10290} {\bibfield  {journal} {\bibinfo
  {journal} {Nature}\ }\textbf {\bibinfo {volume} {476}},\ \bibinfo {pages}
  {181} (\bibinfo {year} {2011})}\BibitemShut {NoStop}%
\bibitem [{\citenamefont {Clerk}\ \emph {et~al.}(2010)\citenamefont {Clerk},
  \citenamefont {Devoret}, \citenamefont {Girvin}, \citenamefont {Marquardt},\
  and\ \citenamefont {Schoelkopf}}]{clerk2010introduction}%
  \BibitemOpen
  \bibfield  {author} {\bibinfo {author} {\bibfnamefont {A.~A.}\ \bibnamefont
  {Clerk}}, \bibinfo {author} {\bibfnamefont {M.~H.}\ \bibnamefont {Devoret}},
  \bibinfo {author} {\bibfnamefont {S.~M.}\ \bibnamefont {Girvin}}, \bibinfo
  {author} {\bibfnamefont {F.}~\bibnamefont {Marquardt}},\ and\ \bibinfo
  {author} {\bibfnamefont {R.~J.}\ \bibnamefont {Schoelkopf}},\ }\bibfield
  {title} {\bibinfo {title} {Introduction to quantum noise, measurement, and
  amplification},\ }\href@noop {} {\bibfield  {journal} {\bibinfo  {journal}
  {Rev. Mod. Phys.}\ }\textbf {\bibinfo {volume} {82}},\ \bibinfo {pages}
  {1155} (\bibinfo {year} {2010})}\BibitemShut {NoStop}%
\bibitem [{\citenamefont {Brun}(2020)}]{Brun2019}%
  \BibitemOpen
  \bibfield  {author} {\bibinfo {author} {\bibfnamefont {T.~A.}\ \bibnamefont
  {Brun}},\ }\bibfield  {title} {\bibinfo {title} {Quantum error correction},\
  }\href@noop {} {\bibfield  {journal} {\bibinfo  {journal} {arXiv:1910.03672}\
  } (\bibinfo {year} {2020})}\BibitemShut {NoStop}%
\bibitem [{\citenamefont {Roffe}(2019)}]{roffe2019quantum}%
  \BibitemOpen
  \bibfield  {author} {\bibinfo {author} {\bibfnamefont {J.}~\bibnamefont
  {Roffe}},\ }\bibfield  {title} {\bibinfo {title} {Quantum error correction:
  an introductory guide},\ }\href@noop {} {\bibfield  {journal} {\bibinfo
  {journal} {Contemporary Physics}\ }\textbf {\bibinfo {volume} {60}},\
  \bibinfo {pages} {226} (\bibinfo {year} {2019})}\BibitemShut {NoStop}%
\bibitem [{\citenamefont {Knill}(2005)}]{Knill2005}%
  \BibitemOpen
  \bibfield  {author} {\bibinfo {author} {\bibfnamefont {E.}~\bibnamefont
  {Knill}},\ }\bibfield  {title} {\bibinfo {title} {{Quantum computing with
  realistically noisy devices}},\ }\href {https://doi.org/10.1038/nature03350}
  {\bibfield  {journal} {\bibinfo  {journal} {Nature}\ }\textbf {\bibinfo
  {volume} {434}},\ \bibinfo {pages} {39} (\bibinfo {year} {2005})}\BibitemShut
  {NoStop}%
\bibitem [{\citenamefont {Fowler}\ \emph {et~al.}(2012)\citenamefont {Fowler},
  \citenamefont {Mariantoni}, \citenamefont {Martinis},\ and\ \citenamefont
  {Cleland}}]{Fowler2012}%
  \BibitemOpen
  \bibfield  {author} {\bibinfo {author} {\bibfnamefont {A.~G.}\ \bibnamefont
  {Fowler}}, \bibinfo {author} {\bibfnamefont {M.}~\bibnamefont {Mariantoni}},
  \bibinfo {author} {\bibfnamefont {J.~M.}\ \bibnamefont {Martinis}},\ and\
  \bibinfo {author} {\bibfnamefont {A.~N.}\ \bibnamefont {Cleland}},\
  }\bibfield  {title} {\bibinfo {title} {{Surface codes: Towards practical
  large-scale quantum computation}},\ }\href
  {https://doi.org/10.1103/PhysRevA.86.032324} {\bibfield  {journal} {\bibinfo
  {journal} {Phys. Rev. A}\ }\textbf {\bibinfo {volume} {86}},\ \bibinfo
  {pages} {032324} (\bibinfo {year} {2012})}\BibitemShut {NoStop}%
\bibitem [{\citenamefont {Steffen}\ \emph {et~al.}(2003)\citenamefont
  {Steffen}, \citenamefont {Martinis},\ and\ \citenamefont
  {Chuang}}]{steffen2003accurate}%
  \BibitemOpen
  \bibfield  {author} {\bibinfo {author} {\bibfnamefont {M.}~\bibnamefont
  {Steffen}}, \bibinfo {author} {\bibfnamefont {J.~M.}\ \bibnamefont
  {Martinis}},\ and\ \bibinfo {author} {\bibfnamefont {I.~L.}\ \bibnamefont
  {Chuang}},\ }\bibfield  {title} {\bibinfo {title} {Accurate control of
  {J}osephson phase qubits},\ }\href@noop {} {\bibfield  {journal} {\bibinfo
  {journal} {Phys. Rev. B}\ }\textbf {\bibinfo {volume} {68}},\ \bibinfo
  {pages} {224518} (\bibinfo {year} {2003})}\BibitemShut {NoStop}%
\bibitem [{\citenamefont {Motzoi}\ \emph {et~al.}(2009)\citenamefont {Motzoi},
  \citenamefont {Gambetta}, \citenamefont {Rebentrost},\ and\ \citenamefont
  {Wilhelm}}]{motzoi2009simple}%
  \BibitemOpen
  \bibfield  {author} {\bibinfo {author} {\bibfnamefont {F.}~\bibnamefont
  {Motzoi}}, \bibinfo {author} {\bibfnamefont {J.~M.}\ \bibnamefont
  {Gambetta}}, \bibinfo {author} {\bibfnamefont {P.}~\bibnamefont
  {Rebentrost}},\ and\ \bibinfo {author} {\bibfnamefont {F.~K.}\ \bibnamefont
  {Wilhelm}},\ }\bibfield  {title} {\bibinfo {title} {Simple pulses for
  elimination of leakage in weakly nonlinear qubits},\ }\href@noop {}
  {\bibfield  {journal} {\bibinfo  {journal} {Phys. Rev. Lett.}\ }\textbf
  {\bibinfo {volume} {103}},\ \bibinfo {pages} {110501} (\bibinfo {year}
  {2009})}\BibitemShut {NoStop}%
\bibitem [{\citenamefont {Chow}\ \emph {et~al.}(2010)\citenamefont {Chow},
  \citenamefont {DiCarlo}, \citenamefont {Gambetta}, \citenamefont {Motzoi},
  \citenamefont {Frunzio}, \citenamefont {Girvin},\ and\ \citenamefont
  {Schoelkopf}}]{chow2010optimized}%
  \BibitemOpen
  \bibfield  {author} {\bibinfo {author} {\bibfnamefont {J.~M.}\ \bibnamefont
  {Chow}}, \bibinfo {author} {\bibfnamefont {L.}~\bibnamefont {DiCarlo}},
  \bibinfo {author} {\bibfnamefont {J.~M.}\ \bibnamefont {Gambetta}}, \bibinfo
  {author} {\bibfnamefont {F.}~\bibnamefont {Motzoi}}, \bibinfo {author}
  {\bibfnamefont {L.}~\bibnamefont {Frunzio}}, \bibinfo {author} {\bibfnamefont
  {S.~M.}\ \bibnamefont {Girvin}},\ and\ \bibinfo {author} {\bibfnamefont
  {R.~J.}\ \bibnamefont {Schoelkopf}},\ }\bibfield  {title} {\bibinfo {title}
  {Optimized driving of superconducting artificial atoms for improved
  single-qubit gates},\ }\href@noop {} {\bibfield  {journal} {\bibinfo
  {journal} {Phys. Rev. A}\ }\textbf {\bibinfo {volume} {82}},\ \bibinfo
  {pages} {040305} (\bibinfo {year} {2010})}\BibitemShut {NoStop}%
\bibitem [{\citenamefont {Chen}\ \emph {et~al.}(2016)\citenamefont {Chen},
  \citenamefont {Kelly}, \citenamefont {Quintana}, \citenamefont {Barends},
  \citenamefont {Campbell}, \citenamefont {Chen}, \citenamefont {Chiaro},
  \citenamefont {Dunsworth}, \citenamefont {Fowler}, \citenamefont {Lucero}
  \emph {et~al.}}]{chen2016measuring}%
  \BibitemOpen
  \bibfield  {author} {\bibinfo {author} {\bibfnamefont {Z.}~\bibnamefont
  {Chen}}, \bibinfo {author} {\bibfnamefont {J.}~\bibnamefont {Kelly}},
  \bibinfo {author} {\bibfnamefont {C.}~\bibnamefont {Quintana}}, \bibinfo
  {author} {\bibfnamefont {R.}~\bibnamefont {Barends}}, \bibinfo {author}
  {\bibfnamefont {B.}~\bibnamefont {Campbell}}, \bibinfo {author}
  {\bibfnamefont {Y.}~\bibnamefont {Chen}}, \bibinfo {author} {\bibfnamefont
  {B.}~\bibnamefont {Chiaro}}, \bibinfo {author} {\bibfnamefont
  {A.}~\bibnamefont {Dunsworth}}, \bibinfo {author} {\bibfnamefont
  {A.}~\bibnamefont {Fowler}}, \bibinfo {author} {\bibfnamefont
  {E.}~\bibnamefont {Lucero}}, \emph {et~al.},\ }\bibfield  {title} {\bibinfo
  {title} {Measuring and suppressing quantum state leakage in a superconducting
  qubit},\ }\href@noop {} {\bibfield  {journal} {\bibinfo  {journal} {Phys.
  Rev. Lett.}\ }\textbf {\bibinfo {volume} {116}},\ \bibinfo {pages} {020501}
  (\bibinfo {year} {2016})}\BibitemShut {NoStop}%
\bibitem [{\citenamefont {{Aude Craik}}\ \emph {et~al.}(2017)\citenamefont
  {{Aude Craik}}, \citenamefont {Linke}, \citenamefont {Sepiol}, \citenamefont
  {Harty}, \citenamefont {Goodwin}, \citenamefont {Ballance}, \citenamefont
  {Stacey}, \citenamefont {Steane}, \citenamefont {Lucas},\ and\ \citenamefont
  {Allcock}}]{AudeCraik2017}%
  \BibitemOpen
  \bibfield  {author} {\bibinfo {author} {\bibfnamefont {D.~P.~L.}\
  \bibnamefont {{Aude Craik}}}, \bibinfo {author} {\bibfnamefont {N.~M.}\
  \bibnamefont {Linke}}, \bibinfo {author} {\bibfnamefont {M.~A.}\ \bibnamefont
  {Sepiol}}, \bibinfo {author} {\bibfnamefont {T.~P.}\ \bibnamefont {Harty}},
  \bibinfo {author} {\bibfnamefont {J.~F.}\ \bibnamefont {Goodwin}}, \bibinfo
  {author} {\bibfnamefont {C.~J.}\ \bibnamefont {Ballance}}, \bibinfo {author}
  {\bibfnamefont {D.~N.}\ \bibnamefont {Stacey}}, \bibinfo {author}
  {\bibfnamefont {A.~M.}\ \bibnamefont {Steane}}, \bibinfo {author}
  {\bibfnamefont {D.~M.}\ \bibnamefont {Lucas}},\ and\ \bibinfo {author}
  {\bibfnamefont {D.~T.~C.}\ \bibnamefont {Allcock}},\ }\bibfield  {title}
  {\bibinfo {title} {{High-fidelity spatial and polarization addressing of
  $^{43}$Ca$^+$qubits using near-field microwave control}},\ }\href
  {https://doi.org/10.1103/PhysRevA.95.022337} {\bibfield  {journal} {\bibinfo
  {journal} {Phys. Rev. A}\ }\textbf {\bibinfo {volume} {95}},\ \bibinfo
  {pages} {022337} (\bibinfo {year} {2017})}\BibitemShut {NoStop}%
\bibitem [{\citenamefont {Piltz}\ \emph {et~al.}(2014)\citenamefont {Piltz},
  \citenamefont {Sriarunothai}, \citenamefont {Var{\'{o}}n},\ and\
  \citenamefont {Wunderlich}}]{Piltz2014}%
  \BibitemOpen
  \bibfield  {author} {\bibinfo {author} {\bibfnamefont {C.}~\bibnamefont
  {Piltz}}, \bibinfo {author} {\bibfnamefont {T.}~\bibnamefont {Sriarunothai}},
  \bibinfo {author} {\bibfnamefont {A.~F.}\ \bibnamefont {Var{\'{o}}n}},\ and\
  \bibinfo {author} {\bibfnamefont {C.}~\bibnamefont {Wunderlich}},\ }\bibfield
   {title} {\bibinfo {title} {{A trapped-ion-based quantum byte with 10$^{-5}$
  next-neighbour cross-talk}},\ }\href {https://doi.org/10.1038/ncomms5679}
  {\bibfield  {journal} {\bibinfo  {journal} {Nat. Commun.}\ }\textbf {\bibinfo
  {volume} {5}},\ \bibinfo {pages} {1} (\bibinfo {year} {2014})}\BibitemShut
  {NoStop}%
\bibitem [{\citenamefont {Leibfried}(1999)}]{Leibfried1999}%
  \BibitemOpen
  \bibfield  {author} {\bibinfo {author} {\bibfnamefont {D.}~\bibnamefont
  {Leibfried}},\ }\bibfield  {title} {\bibinfo {title} {{Individual addressing
  and state readout of trapped ions utilizing rf micromotion}},\ }\href
  {https://doi.org/10.1103/PhysRevA.60.R3335} {\bibfield  {journal} {\bibinfo
  {journal} {Phys. Rev. A}\ }\textbf {\bibinfo {volume} {60}},\ \bibinfo
  {pages} {R3335} (\bibinfo {year} {1999})}\BibitemShut {NoStop}%
\bibitem [{\citenamefont {Staanum}\ and\ \citenamefont
  {Drewsen}(2002)}]{Staanum2002}%
  \BibitemOpen
  \bibfield  {author} {\bibinfo {author} {\bibfnamefont {P.}~\bibnamefont
  {Staanum}}\ and\ \bibinfo {author} {\bibfnamefont {M.}~\bibnamefont
  {Drewsen}},\ }\bibfield  {title} {\bibinfo {title} {{Trapped-ion quantum
  logic utilizing position-dependent ac Stark shifts}},\ }\href
  {https://doi.org/10.1103/PhysRevA.66.040302} {\bibfield  {journal} {\bibinfo
  {journal} {Phys. Rev. A}\ }\textbf {\bibinfo {volume} {66}},\ \bibinfo
  {pages} {040302} (\bibinfo {year} {2002})}\BibitemShut {NoStop}%
\bibitem [{\citenamefont {Warring}\ \emph
  {et~al.}(2013{\natexlab{a}})\citenamefont {Warring}, \citenamefont
  {Ospelkaus}, \citenamefont {Colombe}, \citenamefont {J{\"{o}}rdens},
  \citenamefont {Leibfried},\ and\ \citenamefont {Wineland}}]{Warring2013}%
  \BibitemOpen
  \bibfield  {author} {\bibinfo {author} {\bibfnamefont {U.}~\bibnamefont
  {Warring}}, \bibinfo {author} {\bibfnamefont {C.}~\bibnamefont {Ospelkaus}},
  \bibinfo {author} {\bibfnamefont {Y.}~\bibnamefont {Colombe}}, \bibinfo
  {author} {\bibfnamefont {R.}~\bibnamefont {J{\"{o}}rdens}}, \bibinfo {author}
  {\bibfnamefont {D.}~\bibnamefont {Leibfried}},\ and\ \bibinfo {author}
  {\bibfnamefont {D.~J.}\ \bibnamefont {Wineland}},\ }\bibfield  {title}
  {\bibinfo {title} {{Individual-Ion Addressing with Microwave Field
  Gradients}},\ }\href {https://doi.org/10.1103/PhysRevLett.110.173002}
  {\bibfield  {journal} {\bibinfo  {journal} {Phys. Rev. Lett.}\ }\textbf
  {\bibinfo {volume} {110}},\ \bibinfo {pages} {173002} (\bibinfo {year}
  {2013}{\natexlab{a}})}\BibitemShut {NoStop}%
\bibitem [{\citenamefont {Srinivas}(2020)}]{Srinivas2020}%
  \BibitemOpen
  \bibfield  {author} {\bibinfo {author} {\bibfnamefont {R.}~\bibnamefont
  {Srinivas}},\ }\emph {\bibinfo {title} {{Laser-free trapped-ion quantum logic
  with a radiofrequency magnetic field gradient}}},\ \href@noop {} {Ph.D.
  thesis},\ \bibinfo  {school} {University of Colorado, Boulder} (\bibinfo
  {year} {2020})\BibitemShut {NoStop}%
\bibitem [{\citenamefont {Takeda}\ \emph {et~al.}(2020)\citenamefont {Takeda},
  \citenamefont {Noiri}, \citenamefont {Yoneda}, \citenamefont {Nakajima},\
  and\ \citenamefont {Tarucha}}]{Takeda2020}%
  \BibitemOpen
  \bibfield  {author} {\bibinfo {author} {\bibfnamefont {K.}~\bibnamefont
  {Takeda}}, \bibinfo {author} {\bibfnamefont {A.}~\bibnamefont {Noiri}},
  \bibinfo {author} {\bibfnamefont {J.}~\bibnamefont {Yoneda}}, \bibinfo
  {author} {\bibfnamefont {T.}~\bibnamefont {Nakajima}},\ and\ \bibinfo
  {author} {\bibfnamefont {S.}~\bibnamefont {Tarucha}},\ }\bibfield  {title}
  {\bibinfo {title} {Resonantly driven singlet-triplet spin qubit in silicon},\
  }\href {https://doi.org/10.1103/PhysRevLett.124.117701} {\bibfield  {journal}
  {\bibinfo  {journal} {Phys. Rev. Lett.}\ }\textbf {\bibinfo {volume} {124}},\
  \bibinfo {pages} {117701} (\bibinfo {year} {2020})}\BibitemShut {NoStop}%
\bibitem [{\citenamefont {McKay}\ \emph {et~al.}(2017)\citenamefont {McKay},
  \citenamefont {Wood}, \citenamefont {Sheldon}, \citenamefont {Chow},\ and\
  \citenamefont {Gambetta}}]{mckay2017efficient}%
  \BibitemOpen
  \bibfield  {author} {\bibinfo {author} {\bibfnamefont {D.~C.}\ \bibnamefont
  {McKay}}, \bibinfo {author} {\bibfnamefont {C.~J.}\ \bibnamefont {Wood}},
  \bibinfo {author} {\bibfnamefont {S.}~\bibnamefont {Sheldon}}, \bibinfo
  {author} {\bibfnamefont {J.~M.}\ \bibnamefont {Chow}},\ and\ \bibinfo
  {author} {\bibfnamefont {J.~M.}\ \bibnamefont {Gambetta}},\ }\bibfield
  {title} {\bibinfo {title} {Efficient {Z} gates for quantum computing},\
  }\href@noop {} {\bibfield  {journal} {\bibinfo  {journal} {Phys. Rev. A}\
  }\textbf {\bibinfo {volume} {96}},\ \bibinfo {pages} {022330} (\bibinfo
  {year} {2017})}\BibitemShut {NoStop}%
\bibitem [{\citenamefont {Srinivas}\ \emph {et~al.}(2019)\citenamefont
  {Srinivas}, \citenamefont {Burd}, \citenamefont {Sutherland}, \citenamefont
  {Wilson}, \citenamefont {Wineland}, \citenamefont {Leibfried}, \citenamefont
  {Allcock},\ and\ \citenamefont {Slichter}}]{Srinivas2019}%
  \BibitemOpen
  \bibfield  {author} {\bibinfo {author} {\bibfnamefont {R.}~\bibnamefont
  {Srinivas}}, \bibinfo {author} {\bibfnamefont {S.~C.}\ \bibnamefont {Burd}},
  \bibinfo {author} {\bibfnamefont {R.~T.}\ \bibnamefont {Sutherland}},
  \bibinfo {author} {\bibfnamefont {A.~C.}\ \bibnamefont {Wilson}}, \bibinfo
  {author} {\bibfnamefont {D.~J.}\ \bibnamefont {Wineland}}, \bibinfo {author}
  {\bibfnamefont {D.}~\bibnamefont {Leibfried}}, \bibinfo {author}
  {\bibfnamefont {D.~T.~C.}\ \bibnamefont {Allcock}},\ and\ \bibinfo {author}
  {\bibfnamefont {D.~H.}\ \bibnamefont {Slichter}},\ }\bibfield  {title}
  {\bibinfo {title} {{Trapped-Ion Spin-Motion Coupling with Microwaves and a
  Near-Motional Oscillating Magnetic Field Gradient}},\ }\href
  {https://doi.org/10.1103/PhysRevLett.122.163201} {\bibfield  {journal}
  {\bibinfo  {journal} {Phys. Rev. Lett.}\ }\textbf {\bibinfo {volume} {122}},\
  \bibinfo {pages} {163201} (\bibinfo {year} {2019})}\BibitemShut {NoStop}%
\bibitem [{\citenamefont {Barends}\ \emph {et~al.}(2014)\citenamefont
  {Barends}, \citenamefont {Kelly}, \citenamefont {Megrant}, \citenamefont
  {Veitia}, \citenamefont {Sank}, \citenamefont {Jeffrey}, \citenamefont
  {White}, \citenamefont {Mutus}, \citenamefont {Fowler}, \citenamefont
  {Campbell} \emph {et~al.}}]{barends2014superconducting}%
  \BibitemOpen
  \bibfield  {author} {\bibinfo {author} {\bibfnamefont {R.}~\bibnamefont
  {Barends}}, \bibinfo {author} {\bibfnamefont {J.}~\bibnamefont {Kelly}},
  \bibinfo {author} {\bibfnamefont {A.}~\bibnamefont {Megrant}}, \bibinfo
  {author} {\bibfnamefont {A.}~\bibnamefont {Veitia}}, \bibinfo {author}
  {\bibfnamefont {D.}~\bibnamefont {Sank}}, \bibinfo {author} {\bibfnamefont
  {E.}~\bibnamefont {Jeffrey}}, \bibinfo {author} {\bibfnamefont {T.~C.}\
  \bibnamefont {White}}, \bibinfo {author} {\bibfnamefont {J.}~\bibnamefont
  {Mutus}}, \bibinfo {author} {\bibfnamefont {A.~G.}\ \bibnamefont {Fowler}},
  \bibinfo {author} {\bibfnamefont {B.}~\bibnamefont {Campbell}}, \emph
  {et~al.},\ }\bibfield  {title} {\bibinfo {title} {Superconducting quantum
  circuits at the surface code threshold for fault tolerance},\ }\href@noop {}
  {\bibfield  {journal} {\bibinfo  {journal} {Nature}\ }\textbf {\bibinfo
  {volume} {508}},\ \bibinfo {pages} {500} (\bibinfo {year}
  {2014})}\BibitemShut {NoStop}%
\bibitem [{\citenamefont {Yan}\ \emph {et~al.}(2018)\citenamefont {Yan},
  \citenamefont {Krantz}, \citenamefont {Sung}, \citenamefont {Kjaergaard},
  \citenamefont {Campbell}, \citenamefont {Orlando}, \citenamefont
  {Gustavsson},\ and\ \citenamefont {Oliver}}]{yan2018tunable}%
  \BibitemOpen
  \bibfield  {author} {\bibinfo {author} {\bibfnamefont {F.}~\bibnamefont
  {Yan}}, \bibinfo {author} {\bibfnamefont {P.}~\bibnamefont {Krantz}},
  \bibinfo {author} {\bibfnamefont {Y.}~\bibnamefont {Sung}}, \bibinfo {author}
  {\bibfnamefont {M.}~\bibnamefont {Kjaergaard}}, \bibinfo {author}
  {\bibfnamefont {D.~L.}\ \bibnamefont {Campbell}}, \bibinfo {author}
  {\bibfnamefont {T.~P.}\ \bibnamefont {Orlando}}, \bibinfo {author}
  {\bibfnamefont {S.}~\bibnamefont {Gustavsson}},\ and\ \bibinfo {author}
  {\bibfnamefont {W.~D.}\ \bibnamefont {Oliver}},\ }\bibfield  {title}
  {\bibinfo {title} {Tunable coupling scheme for implementing high-fidelity
  two-qubit gates},\ }\href@noop {} {\bibfield  {journal} {\bibinfo  {journal}
  {Phys. Rev. Appl.}\ }\textbf {\bibinfo {volume} {10}},\ \bibinfo {pages}
  {054062} (\bibinfo {year} {2018})}\BibitemShut {NoStop}%
\bibitem [{\citenamefont {Foxen}\ \emph {et~al.}(2020)\citenamefont {Foxen},
  \citenamefont {Neill}, \citenamefont {Dunsworth}, \citenamefont {Roushan},
  \citenamefont {Chiaro}, \citenamefont {Megrant}, \citenamefont {Kelly},
  \citenamefont {Chen}, \citenamefont {Satzinger}, \citenamefont {Barends}
  \emph {et~al.}}]{foxen2020demonstrating}%
  \BibitemOpen
  \bibfield  {author} {\bibinfo {author} {\bibfnamefont {B.}~\bibnamefont
  {Foxen}}, \bibinfo {author} {\bibfnamefont {C.}~\bibnamefont {Neill}},
  \bibinfo {author} {\bibfnamefont {A.}~\bibnamefont {Dunsworth}}, \bibinfo
  {author} {\bibfnamefont {P.}~\bibnamefont {Roushan}}, \bibinfo {author}
  {\bibfnamefont {B.}~\bibnamefont {Chiaro}}, \bibinfo {author} {\bibfnamefont
  {A.}~\bibnamefont {Megrant}}, \bibinfo {author} {\bibfnamefont
  {J.}~\bibnamefont {Kelly}}, \bibinfo {author} {\bibfnamefont
  {Z.}~\bibnamefont {Chen}}, \bibinfo {author} {\bibfnamefont {K.}~\bibnamefont
  {Satzinger}}, \bibinfo {author} {\bibfnamefont {R.}~\bibnamefont {Barends}},
  \emph {et~al.},\ }\bibfield  {title} {\bibinfo {title} {Demonstrating a
  continuous set of two-qubit gates for near-term quantum algorithms},\
  }\href@noop {} {\bibfield  {journal} {\bibinfo  {journal} {arXiv:2001.08343}\
  } (\bibinfo {year} {2020})}\BibitemShut {NoStop}%
\bibitem [{\citenamefont {Rigetti}\ and\ \citenamefont
  {Devoret}(2010)}]{rigetti2010fully}%
  \BibitemOpen
  \bibfield  {author} {\bibinfo {author} {\bibfnamefont {C.}~\bibnamefont
  {Rigetti}}\ and\ \bibinfo {author} {\bibfnamefont {M.}~\bibnamefont
  {Devoret}},\ }\bibfield  {title} {\bibinfo {title} {Fully microwave-tunable
  universal gates in superconducting qubits with linear couplings and fixed
  transition frequencies},\ }\href@noop {} {\bibfield  {journal} {\bibinfo
  {journal} {Phys. Rev. B}\ }\textbf {\bibinfo {volume} {81}},\ \bibinfo
  {pages} {134507} (\bibinfo {year} {2010})}\BibitemShut {NoStop}%
\bibitem [{\citenamefont {Shulman}\ \emph {et~al.}(2012)\citenamefont
  {Shulman}, \citenamefont {Dial}, \citenamefont {Harvey}, \citenamefont
  {Bluhm}, \citenamefont {Umansky},\ and\ \citenamefont
  {Yacoby}}]{Shulman2012}%
  \BibitemOpen
  \bibfield  {author} {\bibinfo {author} {\bibfnamefont {M.~D.}\ \bibnamefont
  {Shulman}}, \bibinfo {author} {\bibfnamefont {O.~E.}\ \bibnamefont {Dial}},
  \bibinfo {author} {\bibfnamefont {S.~P.}\ \bibnamefont {Harvey}}, \bibinfo
  {author} {\bibfnamefont {H.}~\bibnamefont {Bluhm}}, \bibinfo {author}
  {\bibfnamefont {V.}~\bibnamefont {Umansky}},\ and\ \bibinfo {author}
  {\bibfnamefont {A.}~\bibnamefont {Yacoby}},\ }\bibfield  {title} {\bibinfo
  {title} {{Demonstration of Entanglement of Electrostatically Coupled
  Singlet-Triplet Qubits}},\ }\href {https://doi.org/10.1126/science.1217692}
  {\bibfield  {journal} {\bibinfo  {journal} {Science}\ }\textbf {\bibinfo
  {volume} {336}},\ \bibinfo {pages} {202} (\bibinfo {year}
  {2012})}\BibitemShut {NoStop}%
\bibitem [{\citenamefont {Clerk}\ \emph {et~al.}(2020)\citenamefont {Clerk},
  \citenamefont {Lehnert}, \citenamefont {Bertet}, \citenamefont {Petta},\ and\
  \citenamefont {Nakamura}}]{hybrid}%
  \BibitemOpen
  \bibfield  {author} {\bibinfo {author} {\bibfnamefont {A.~A.}\ \bibnamefont
  {Clerk}}, \bibinfo {author} {\bibfnamefont {K.~W.}\ \bibnamefont {Lehnert}},
  \bibinfo {author} {\bibfnamefont {P.}~\bibnamefont {Bertet}}, \bibinfo
  {author} {\bibfnamefont {J.~R.}\ \bibnamefont {Petta}},\ and\ \bibinfo
  {author} {\bibfnamefont {Y.}~\bibnamefont {Nakamura}},\ }\bibfield  {title}
  {\bibinfo {title} {Hybrid quantum systems with circuit quantum
  electrodynamics},\ }\href {https://doi.org/10.1038/s41567-020-0797-9}
  {\bibfield  {journal} {\bibinfo  {journal} {Nature Physics}\ }\textbf
  {\bibinfo {volume} {16}},\ \bibinfo {pages} {257} (\bibinfo {year}
  {2020})}\BibitemShut {NoStop}%
\bibitem [{\citenamefont {Mintert}\ and\ \citenamefont
  {Wunderlich}(2001)}]{Mintert2001}%
  \BibitemOpen
  \bibfield  {author} {\bibinfo {author} {\bibfnamefont {F.}~\bibnamefont
  {Mintert}}\ and\ \bibinfo {author} {\bibfnamefont {C.}~\bibnamefont
  {Wunderlich}},\ }\bibfield  {title} {\bibinfo {title} {{Ion-Trap Quantum
  Logic Using Long-Wavelength Radiation}},\ }\href
  {https://doi.org/10.1103/PhysRevLett.87.257904} {\bibfield  {journal}
  {\bibinfo  {journal} {Phys. Rev. Lett.}\ }\textbf {\bibinfo {volume} {87}},\
  \bibinfo {pages} {257904} (\bibinfo {year} {2001})}\BibitemShut {NoStop}%
\bibitem [{\citenamefont {Johanning}\ \emph {et~al.}(2009)\citenamefont
  {Johanning}, \citenamefont {Braun}, \citenamefont {Timoney}, \citenamefont
  {Elman}, \citenamefont {Neuhauser},\ and\ \citenamefont
  {Wunderlich}}]{Johanning2009a}%
  \BibitemOpen
  \bibfield  {author} {\bibinfo {author} {\bibfnamefont {M.}~\bibnamefont
  {Johanning}}, \bibinfo {author} {\bibfnamefont {A.}~\bibnamefont {Braun}},
  \bibinfo {author} {\bibfnamefont {N.}~\bibnamefont {Timoney}}, \bibinfo
  {author} {\bibfnamefont {V.}~\bibnamefont {Elman}}, \bibinfo {author}
  {\bibfnamefont {W.}~\bibnamefont {Neuhauser}},\ and\ \bibinfo {author}
  {\bibfnamefont {C.}~\bibnamefont {Wunderlich}},\ }\bibfield  {title}
  {\bibinfo {title} {{Individual Addressing of Trapped Ions and Coupling of
  Motional and Spin States Using rf Radiation}},\ }\href
  {https://doi.org/10.1103/PhysRevLett.102.073004} {\bibfield  {journal}
  {\bibinfo  {journal} {Phys. Rev. Lett.}\ }\textbf {\bibinfo {volume} {102}},\
  \bibinfo {pages} {073004} (\bibinfo {year} {2009})}\BibitemShut {NoStop}%
\bibitem [{\citenamefont {Sutherland}\ \emph {et~al.}(2019)\citenamefont
  {Sutherland}, \citenamefont {Srinivas}, \citenamefont {Burd}, \citenamefont
  {Leibfried}, \citenamefont {Wilson}, \citenamefont {Wineland}, \citenamefont
  {Allcock}, \citenamefont {Slichter},\ and\ \citenamefont
  {Libby}}]{Sutherland2019}%
  \BibitemOpen
  \bibfield  {author} {\bibinfo {author} {\bibfnamefont {R.~T.}\ \bibnamefont
  {Sutherland}}, \bibinfo {author} {\bibfnamefont {R.}~\bibnamefont
  {Srinivas}}, \bibinfo {author} {\bibfnamefont {S.~C.}\ \bibnamefont {Burd}},
  \bibinfo {author} {\bibfnamefont {D.}~\bibnamefont {Leibfried}}, \bibinfo
  {author} {\bibfnamefont {A.~C.}\ \bibnamefont {Wilson}}, \bibinfo {author}
  {\bibfnamefont {D.~J.}\ \bibnamefont {Wineland}}, \bibinfo {author}
  {\bibfnamefont {D.~T.~C.}\ \bibnamefont {Allcock}}, \bibinfo {author}
  {\bibfnamefont {D.~H.}\ \bibnamefont {Slichter}},\ and\ \bibinfo {author}
  {\bibfnamefont {S.~B.}\ \bibnamefont {Libby}},\ }\bibfield  {title} {\bibinfo
  {title} {{Versatile laser-free trapped-ion entangling gates}},\ }\href
  {https://doi.org/10.1088/1367-2630/ab0be5} {\bibfield  {journal} {\bibinfo
  {journal} {New J. Phys.}\ }\textbf {\bibinfo {volume} {21}},\ \bibinfo
  {pages} {033033} (\bibinfo {year} {2019})}\BibitemShut {NoStop}%
\bibitem [{\citenamefont {Ospelkaus}\ \emph {et~al.}(2008)\citenamefont
  {Ospelkaus}, \citenamefont {Langer}, \citenamefont {Amini}, \citenamefont
  {Brown}, \citenamefont {Leibfried},\ and\ \citenamefont
  {Wineland}}]{Ospelkaus2008}%
  \BibitemOpen
  \bibfield  {author} {\bibinfo {author} {\bibfnamefont {C.}~\bibnamefont
  {Ospelkaus}}, \bibinfo {author} {\bibfnamefont {C.}~\bibnamefont {Langer}},
  \bibinfo {author} {\bibfnamefont {J.}~\bibnamefont {Amini}}, \bibinfo
  {author} {\bibfnamefont {K.}~\bibnamefont {Brown}}, \bibinfo {author}
  {\bibfnamefont {D.}~\bibnamefont {Leibfried}},\ and\ \bibinfo {author}
  {\bibfnamefont {D.}~\bibnamefont {Wineland}},\ }\bibfield  {title} {\bibinfo
  {title} {{Trapped-Ion Quantum Logic Gates Based on Oscillating Magnetic
  Fields}},\ }\href {https://doi.org/10.1103/PhysRevLett.101.090502} {\bibfield
   {journal} {\bibinfo  {journal} {Phys. Rev. Lett.}\ }\textbf {\bibinfo
  {volume} {101}},\ \bibinfo {pages} {090502} (\bibinfo {year}
  {2008})}\BibitemShut {NoStop}%
\bibitem [{\citenamefont {Warring}\ \emph
  {et~al.}(2013{\natexlab{b}})\citenamefont {Warring}, \citenamefont
  {Ospelkaus}, \citenamefont {Colombe}, \citenamefont {Brown}, \citenamefont
  {Amini}, \citenamefont {Carsjens}, \citenamefont {Leibfried},\ and\
  \citenamefont {Wineland}}]{Warring2013a}%
  \BibitemOpen
  \bibfield  {author} {\bibinfo {author} {\bibfnamefont {U.}~\bibnamefont
  {Warring}}, \bibinfo {author} {\bibfnamefont {C.}~\bibnamefont {Ospelkaus}},
  \bibinfo {author} {\bibfnamefont {Y.}~\bibnamefont {Colombe}}, \bibinfo
  {author} {\bibfnamefont {K.~R.}\ \bibnamefont {Brown}}, \bibinfo {author}
  {\bibfnamefont {J.~M.}\ \bibnamefont {Amini}}, \bibinfo {author}
  {\bibfnamefont {M.}~\bibnamefont {Carsjens}}, \bibinfo {author}
  {\bibfnamefont {D.}~\bibnamefont {Leibfried}},\ and\ \bibinfo {author}
  {\bibfnamefont {D.~J.}\ \bibnamefont {Wineland}},\ }\bibfield  {title}
  {\bibinfo {title} {{Techniques for microwave near-field quantum control of
  trapped ions}},\ }\href {https://doi.org/10.1103/PhysRevA.87.013437}
  {\bibfield  {journal} {\bibinfo  {journal} {Phys. Rev. A}\ }\textbf {\bibinfo
  {volume} {87}},\ \bibinfo {pages} {013437} (\bibinfo {year}
  {2013}{\natexlab{b}})}\BibitemShut {NoStop}%
\bibitem [{\citenamefont {Welzel}\ \emph {et~al.}(2019)\citenamefont {Welzel},
  \citenamefont {Stopp},\ and\ \citenamefont {Schmidt-Kaler}}]{Welzel2018}%
  \BibitemOpen
  \bibfield  {author} {\bibinfo {author} {\bibfnamefont {J.}~\bibnamefont
  {Welzel}}, \bibinfo {author} {\bibfnamefont {F.}~\bibnamefont {Stopp}},\ and\
  \bibinfo {author} {\bibfnamefont {F.}~\bibnamefont {Schmidt-Kaler}},\
  }\bibfield  {title} {\bibinfo {title} {{Spin and motion dynamics with zigzag
  ion crystals in transverse magnetic gradients}},\ }\href
  {https://doi.org/10.1088/1361-6455/aaf347} {\bibfield  {journal} {\bibinfo
  {journal} {J. Phys. B At. Mol. Opt. Phys.}\ }\textbf {\bibinfo {volume}
  {52}},\ \bibinfo {pages} {025301} (\bibinfo {year} {2019})}\BibitemShut
  {NoStop}%
\bibitem [{\citenamefont {Harty}\ \emph {et~al.}(2016)\citenamefont {Harty},
  \citenamefont {Sepiol}, \citenamefont {Allcock}, \citenamefont {Ballance},
  \citenamefont {Tarlton},\ and\ \citenamefont {Lucas}}]{Harty2016}%
  \BibitemOpen
  \bibfield  {author} {\bibinfo {author} {\bibfnamefont {T.~P.}\ \bibnamefont
  {Harty}}, \bibinfo {author} {\bibfnamefont {M.~A.}\ \bibnamefont {Sepiol}},
  \bibinfo {author} {\bibfnamefont {D.~T.~C.}\ \bibnamefont {Allcock}},
  \bibinfo {author} {\bibfnamefont {C.~J.}\ \bibnamefont {Ballance}}, \bibinfo
  {author} {\bibfnamefont {J.~E.}\ \bibnamefont {Tarlton}},\ and\ \bibinfo
  {author} {\bibfnamefont {D.~M.}\ \bibnamefont {Lucas}},\ }\bibfield  {title}
  {\bibinfo {title} {{High-Fidelity Trapped-Ion Quantum Logic Using Near-Field
  Microwaves}},\ }\href {https://doi.org/10.1103/PhysRevLett.117.140501}
  {\bibfield  {journal} {\bibinfo  {journal} {Phys. Rev. Lett.}\ }\textbf
  {\bibinfo {volume} {117}},\ \bibinfo {pages} {140501} (\bibinfo {year}
  {2016})}\BibitemShut {NoStop}%
\bibitem [{\citenamefont {Weidt}\ \emph {et~al.}(2016)\citenamefont {Weidt},
  \citenamefont {Randall}, \citenamefont {Webster}, \citenamefont {Lake},
  \citenamefont {Webb}, \citenamefont {Cohen}, \citenamefont {Navickas},
  \citenamefont {Lekitsch}, \citenamefont {Retzker},\ and\ \citenamefont
  {Hensinger}}]{Weidt2016}%
  \BibitemOpen
  \bibfield  {author} {\bibinfo {author} {\bibfnamefont {S.}~\bibnamefont
  {Weidt}}, \bibinfo {author} {\bibfnamefont {J.}~\bibnamefont {Randall}},
  \bibinfo {author} {\bibfnamefont {S.~C.}\ \bibnamefont {Webster}}, \bibinfo
  {author} {\bibfnamefont {K.}~\bibnamefont {Lake}}, \bibinfo {author}
  {\bibfnamefont {A.~E.}\ \bibnamefont {Webb}}, \bibinfo {author}
  {\bibfnamefont {I.}~\bibnamefont {Cohen}}, \bibinfo {author} {\bibfnamefont
  {T.}~\bibnamefont {Navickas}}, \bibinfo {author} {\bibfnamefont
  {B.}~\bibnamefont {Lekitsch}}, \bibinfo {author} {\bibfnamefont
  {A.}~\bibnamefont {Retzker}},\ and\ \bibinfo {author} {\bibfnamefont {W.~K.}\
  \bibnamefont {Hensinger}},\ }\bibfield  {title} {\bibinfo {title}
  {{Trapped-Ion Quantum Logic with Global Radiation Fields}},\ }\href
  {https://doi.org/10.1103/PhysRevLett.117.220501} {\bibfield  {journal}
  {\bibinfo  {journal} {Phys. Rev. Lett.}\ }\textbf {\bibinfo {volume} {117}},\
  \bibinfo {pages} {220501} (\bibinfo {year} {2016})}\BibitemShut {NoStop}%
\bibitem [{\citenamefont {Hahn}\ \emph {et~al.}(2019)\citenamefont {Hahn},
  \citenamefont {Zarantonello}, \citenamefont {Schulte}, \citenamefont
  {Bautista-Salvador}, \citenamefont {Hammerer},\ and\ \citenamefont
  {Ospelkaus}}]{Hahn2019}%
  \BibitemOpen
  \bibfield  {author} {\bibinfo {author} {\bibfnamefont {H.}~\bibnamefont
  {Hahn}}, \bibinfo {author} {\bibfnamefont {G.}~\bibnamefont {Zarantonello}},
  \bibinfo {author} {\bibfnamefont {M.}~\bibnamefont {Schulte}}, \bibinfo
  {author} {\bibfnamefont {A.}~\bibnamefont {Bautista-Salvador}}, \bibinfo
  {author} {\bibfnamefont {K.}~\bibnamefont {Hammerer}},\ and\ \bibinfo
  {author} {\bibfnamefont {C.}~\bibnamefont {Ospelkaus}},\ }\bibfield  {title}
  {\bibinfo {title} {{Integrated $^9$Be$^+$ multi-qubit gate device for the
  ion-trap quantum computer}},\ }\href
  {https://doi.org/10.1038/s41534-019-0184-5} {\bibfield  {journal} {\bibinfo
  {journal} {npj Quantum Inf.}\ }\textbf {\bibinfo {volume} {5}},\ \bibinfo
  {pages} {70} (\bibinfo {year} {2019})}\BibitemShut {NoStop}%
\bibitem [{\citenamefont {Zarantonello}\ \emph {et~al.}(2019)\citenamefont
  {Zarantonello}, \citenamefont {Hahn}, \citenamefont {Morgner}, \citenamefont
  {Schulte}, \citenamefont {Bautista-Salvador}, \citenamefont {Werner},
  \citenamefont {Hammerer},\ and\ \citenamefont
  {Ospelkaus}}]{Zarantonello2019}%
  \BibitemOpen
  \bibfield  {author} {\bibinfo {author} {\bibfnamefont {G.}~\bibnamefont
  {Zarantonello}}, \bibinfo {author} {\bibfnamefont {H.}~\bibnamefont {Hahn}},
  \bibinfo {author} {\bibfnamefont {J.}~\bibnamefont {Morgner}}, \bibinfo
  {author} {\bibfnamefont {M.}~\bibnamefont {Schulte}}, \bibinfo {author}
  {\bibfnamefont {A.}~\bibnamefont {Bautista-Salvador}}, \bibinfo {author}
  {\bibfnamefont {R.~F.}\ \bibnamefont {Werner}}, \bibinfo {author}
  {\bibfnamefont {K.}~\bibnamefont {Hammerer}},\ and\ \bibinfo {author}
  {\bibfnamefont {C.}~\bibnamefont {Ospelkaus}},\ }\bibfield  {title} {\bibinfo
  {title} {{Robust and Resource-Efficient Microwave Near-Field Entangling
  $^9$Be$^+$ Gate}},\ }\href {https://doi.org/10.1103/PhysRevLett.123.260503}
  {\bibfield  {journal} {\bibinfo  {journal} {Phys. Rev. Lett.}\ }\textbf
  {\bibinfo {volume} {123}},\ \bibinfo {pages} {260503} (\bibinfo {year}
  {2019})}\BibitemShut {NoStop}%
\bibitem [{\citenamefont {Sch{\"{a}}fer}\ \emph {et~al.}(2018)\citenamefont
  {Sch{\"{a}}fer}, \citenamefont {Ballance}, \citenamefont {Thirumalai},
  \citenamefont {Stephenson}, \citenamefont {Ballance}, \citenamefont
  {Steane},\ and\ \citenamefont {Lucas}}]{Schafer2018}%
  \BibitemOpen
  \bibfield  {author} {\bibinfo {author} {\bibfnamefont {V.~M.}\ \bibnamefont
  {Sch{\"{a}}fer}}, \bibinfo {author} {\bibfnamefont {C.~J.}\ \bibnamefont
  {Ballance}}, \bibinfo {author} {\bibfnamefont {K.}~\bibnamefont
  {Thirumalai}}, \bibinfo {author} {\bibfnamefont {L.~J.}\ \bibnamefont
  {Stephenson}}, \bibinfo {author} {\bibfnamefont {T.~G.}\ \bibnamefont
  {Ballance}}, \bibinfo {author} {\bibfnamefont {A.~M.}\ \bibnamefont
  {Steane}},\ and\ \bibinfo {author} {\bibfnamefont {D.~M.}\ \bibnamefont
  {Lucas}},\ }\bibfield  {title} {\bibinfo {title} {{Fast quantum logic gates
  with trapped-ion qubits}},\ }\href {https://doi.org/10.1038/nature25737}
  {\bibfield  {journal} {\bibinfo  {journal} {Nature}\ }\textbf {\bibinfo
  {volume} {555}},\ \bibinfo {pages} {75} (\bibinfo {year} {2018})}\BibitemShut
  {NoStop}%
\bibitem [{\citenamefont {Ozeri}\ \emph {et~al.}(2007)\citenamefont {Ozeri},
  \citenamefont {Itano}, \citenamefont {Blakestad}, \citenamefont {Britton},
  \citenamefont {Chiaverini}, \citenamefont {Jost}, \citenamefont {Langer},
  \citenamefont {Leibfried}, \citenamefont {Reichle}, \citenamefont {Seidelin},
  \citenamefont {Wesenberg},\ and\ \citenamefont {Wineland}}]{Ozeri2007}%
  \BibitemOpen
  \bibfield  {author} {\bibinfo {author} {\bibfnamefont {R.}~\bibnamefont
  {Ozeri}}, \bibinfo {author} {\bibfnamefont {W.~M.}\ \bibnamefont {Itano}},
  \bibinfo {author} {\bibfnamefont {R.~B.}\ \bibnamefont {Blakestad}}, \bibinfo
  {author} {\bibfnamefont {J.}~\bibnamefont {Britton}}, \bibinfo {author}
  {\bibfnamefont {J.}~\bibnamefont {Chiaverini}}, \bibinfo {author}
  {\bibfnamefont {J.~D.}\ \bibnamefont {Jost}}, \bibinfo {author}
  {\bibfnamefont {C.}~\bibnamefont {Langer}}, \bibinfo {author} {\bibfnamefont
  {D.}~\bibnamefont {Leibfried}}, \bibinfo {author} {\bibfnamefont
  {R.}~\bibnamefont {Reichle}}, \bibinfo {author} {\bibfnamefont
  {S.}~\bibnamefont {Seidelin}}, \bibinfo {author} {\bibfnamefont {J.~H.}\
  \bibnamefont {Wesenberg}},\ and\ \bibinfo {author} {\bibfnamefont {D.~J.}\
  \bibnamefont {Wineland}},\ }\bibfield  {title} {\bibinfo {title} {{Errors in
  trapped-ion quantum gates due to spontaneous photon scattering}},\ }\href
  {https://doi.org/10.1103/PhysRevA.75.042329} {\bibfield  {journal} {\bibinfo
  {journal} {Phys. Rev. A}\ }\textbf {\bibinfo {volume} {75}},\ \bibinfo
  {pages} {042329} (\bibinfo {year} {2007})}\BibitemShut {NoStop}%
\bibitem [{\citenamefont {Raftery}\ \emph {et~al.}(2017)\citenamefont
  {Raftery}, \citenamefont {Vrajitoarea}, \citenamefont {Zhang}, \citenamefont
  {Leng}, \citenamefont {Srinivasan},\ and\ \citenamefont
  {Houck}}]{raftery2017direct}%
  \BibitemOpen
  \bibfield  {author} {\bibinfo {author} {\bibfnamefont {J.}~\bibnamefont
  {Raftery}}, \bibinfo {author} {\bibfnamefont {A.}~\bibnamefont
  {Vrajitoarea}}, \bibinfo {author} {\bibfnamefont {G.}~\bibnamefont {Zhang}},
  \bibinfo {author} {\bibfnamefont {Z.}~\bibnamefont {Leng}}, \bibinfo {author}
  {\bibfnamefont {S.}~\bibnamefont {Srinivasan}},\ and\ \bibinfo {author}
  {\bibfnamefont {A.}~\bibnamefont {Houck}},\ }\bibfield  {title} {\bibinfo
  {title} {Direct digital synthesis of microwave waveforms for quantum
  computing},\ }\href@noop {} {\bibfield  {journal} {\bibinfo  {journal}
  {arXiv:1703.00942}\ } (\bibinfo {year} {2017})}\BibitemShut {NoStop}%
\bibitem [{\citenamefont {Kalfus}\ \emph {et~al.}(2020)\citenamefont {Kalfus},
  \citenamefont {Lee}, \citenamefont {Ribeill}, \citenamefont {Fallek},
  \citenamefont {Wagner}, \citenamefont {Donovan}, \citenamefont {Rist{\`e}},\
  and\ \citenamefont {Ohki}}]{kalfus2020high}%
  \BibitemOpen
  \bibfield  {author} {\bibinfo {author} {\bibfnamefont {W.~D.}\ \bibnamefont
  {Kalfus}}, \bibinfo {author} {\bibfnamefont {D.~F.}\ \bibnamefont {Lee}},
  \bibinfo {author} {\bibfnamefont {G.~J.}\ \bibnamefont {Ribeill}}, \bibinfo
  {author} {\bibfnamefont {S.~D.}\ \bibnamefont {Fallek}}, \bibinfo {author}
  {\bibfnamefont {A.}~\bibnamefont {Wagner}}, \bibinfo {author} {\bibfnamefont
  {B.}~\bibnamefont {Donovan}}, \bibinfo {author} {\bibfnamefont
  {D.}~\bibnamefont {Rist{\`e}}},\ and\ \bibinfo {author} {\bibfnamefont
  {T.~A.}\ \bibnamefont {Ohki}},\ }\bibfield  {title} {\bibinfo {title}
  {High-fidelity control of superconducting qubits using direct microwave
  synthesis in higher nyquist zones},\ }\href@noop {} {\bibfield  {journal}
  {\bibinfo  {journal} {arXiv:2008.02873}\ } (\bibinfo {year}
  {2020})}\BibitemShut {NoStop}%
\bibitem [{\citenamefont {Blais}\ \emph {et~al.}(2004)\citenamefont {Blais},
  \citenamefont {Huang}, \citenamefont {Wallraff}, \citenamefont {Girvin},\
  and\ \citenamefont {Schoelkopf}}]{Blais2004}%
  \BibitemOpen
  \bibfield  {author} {\bibinfo {author} {\bibfnamefont {A.}~\bibnamefont
  {Blais}}, \bibinfo {author} {\bibfnamefont {R.-S.}\ \bibnamefont {Huang}},
  \bibinfo {author} {\bibfnamefont {A.}~\bibnamefont {Wallraff}}, \bibinfo
  {author} {\bibfnamefont {S.~M.}\ \bibnamefont {Girvin}},\ and\ \bibinfo
  {author} {\bibfnamefont {R.~J.}\ \bibnamefont {Schoelkopf}},\ }\bibfield
  {title} {\bibinfo {title} {{Cavity quantum electrodynamics for
  superconducting electrical circuits: An architecture for quantum
  computation}},\ }\href {https://doi.org/10.1103/PhysRevA.69.062320}
  {\bibfield  {journal} {\bibinfo  {journal} {Phys. Rev. A}\ }\textbf {\bibinfo
  {volume} {69}},\ \bibinfo {pages} {062320} (\bibinfo {year}
  {2004})}\BibitemShut {NoStop}%
\bibitem [{\citenamefont {Wallraff}\ \emph {et~al.}(2005)\citenamefont
  {Wallraff}, \citenamefont {Schuster}, \citenamefont {Blais}, \citenamefont
  {Frunzio}, \citenamefont {Majer}, \citenamefont {Devoret}, \citenamefont
  {Girvin},\ and\ \citenamefont {Schoelkopf}}]{Wallraff2005}%
  \BibitemOpen
  \bibfield  {author} {\bibinfo {author} {\bibfnamefont {A.}~\bibnamefont
  {Wallraff}}, \bibinfo {author} {\bibfnamefont {D.~I.}\ \bibnamefont
  {Schuster}}, \bibinfo {author} {\bibfnamefont {A.}~\bibnamefont {Blais}},
  \bibinfo {author} {\bibfnamefont {L.}~\bibnamefont {Frunzio}}, \bibinfo
  {author} {\bibfnamefont {J.}~\bibnamefont {Majer}}, \bibinfo {author}
  {\bibfnamefont {M.~H.}\ \bibnamefont {Devoret}}, \bibinfo {author}
  {\bibfnamefont {S.~M.}\ \bibnamefont {Girvin}},\ and\ \bibinfo {author}
  {\bibfnamefont {R.~J.}\ \bibnamefont {Schoelkopf}},\ }\bibfield  {title}
  {\bibinfo {title} {{Approaching Unit Visibility for Control of a
  Superconducting Qubit with Dispersive Readout}},\ }\href
  {https://doi.org/10.1103/PhysRevLett.95.060501} {\bibfield  {journal}
  {\bibinfo  {journal} {Phys. Rev. Lett.}\ }\textbf {\bibinfo {volume} {95}},\
  \bibinfo {pages} {060501} (\bibinfo {year} {2005})}\BibitemShut {NoStop}%
\bibitem [{\citenamefont {Mi}\ \emph {et~al.}(2017)\citenamefont {Mi},
  \citenamefont {Cady}, \citenamefont {Zajac}, \citenamefont {Deelman},\ and\
  \citenamefont {Petta}}]{mi2017strong}%
  \BibitemOpen
  \bibfield  {author} {\bibinfo {author} {\bibfnamefont {X.}~\bibnamefont
  {Mi}}, \bibinfo {author} {\bibfnamefont {J.}~\bibnamefont {Cady}}, \bibinfo
  {author} {\bibfnamefont {D.}~\bibnamefont {Zajac}}, \bibinfo {author}
  {\bibfnamefont {P.}~\bibnamefont {Deelman}},\ and\ \bibinfo {author}
  {\bibfnamefont {J.}~\bibnamefont {Petta}},\ }\bibfield  {title} {\bibinfo
  {title} {Strong coupling of a single electron in silicon to a microwave
  photon},\ }\href@noop {} {\bibfield  {journal} {\bibinfo  {journal}
  {Science}\ }\textbf {\bibinfo {volume} {355}},\ \bibinfo {pages} {156}
  (\bibinfo {year} {2017})}\BibitemShut {NoStop}%
\bibitem [{\citenamefont {Stockklauser}\ \emph {et~al.}(2017)\citenamefont
  {Stockklauser}, \citenamefont {Scarlino}, \citenamefont {Koski},
  \citenamefont {Gasparinetti}, \citenamefont {Andersen}, \citenamefont
  {Reichl}, \citenamefont {Wegscheider}, \citenamefont {Ihn}, \citenamefont
  {Ensslin},\ and\ \citenamefont {Wallraff}}]{stockklauser2017strong}%
  \BibitemOpen
  \bibfield  {author} {\bibinfo {author} {\bibfnamefont {A.}~\bibnamefont
  {Stockklauser}}, \bibinfo {author} {\bibfnamefont {P.}~\bibnamefont
  {Scarlino}}, \bibinfo {author} {\bibfnamefont {J.~V.}\ \bibnamefont {Koski}},
  \bibinfo {author} {\bibfnamefont {S.}~\bibnamefont {Gasparinetti}}, \bibinfo
  {author} {\bibfnamefont {C.~K.}\ \bibnamefont {Andersen}}, \bibinfo {author}
  {\bibfnamefont {C.}~\bibnamefont {Reichl}}, \bibinfo {author} {\bibfnamefont
  {W.}~\bibnamefont {Wegscheider}}, \bibinfo {author} {\bibfnamefont
  {T.}~\bibnamefont {Ihn}}, \bibinfo {author} {\bibfnamefont {K.}~\bibnamefont
  {Ensslin}},\ and\ \bibinfo {author} {\bibfnamefont {A.}~\bibnamefont
  {Wallraff}},\ }\bibfield  {title} {\bibinfo {title} {Strong coupling cavity
  qed with gate-defined double quantum dots enabled by a high impedance
  resonator},\ }\href@noop {} {\bibfield  {journal} {\bibinfo  {journal} {Phys.
  Rev. X}\ }\textbf {\bibinfo {volume} {7}},\ \bibinfo {pages} {011030}
  (\bibinfo {year} {2017})}\BibitemShut {NoStop}%
\bibitem [{\citenamefont {West}\ \emph {et~al.}(2019)\citenamefont {West},
  \citenamefont {Hensen}, \citenamefont {Jouan}, \citenamefont {Tanttu},
  \citenamefont {Yang}, \citenamefont {Rossi}, \citenamefont {Gonzalez-Zalba},
  \citenamefont {Hudson}, \citenamefont {Morello}, \citenamefont {Reilly},\
  and\ \citenamefont {Dzurak}}]{WestDGS}%
  \BibitemOpen
  \bibfield  {author} {\bibinfo {author} {\bibfnamefont {A.}~\bibnamefont
  {West}}, \bibinfo {author} {\bibfnamefont {B.}~\bibnamefont {Hensen}},
  \bibinfo {author} {\bibfnamefont {A.}~\bibnamefont {Jouan}}, \bibinfo
  {author} {\bibfnamefont {T.}~\bibnamefont {Tanttu}}, \bibinfo {author}
  {\bibfnamefont {C.-H.}\ \bibnamefont {Yang}}, \bibinfo {author}
  {\bibfnamefont {A.}~\bibnamefont {Rossi}}, \bibinfo {author} {\bibfnamefont
  {M.~F.}\ \bibnamefont {Gonzalez-Zalba}}, \bibinfo {author} {\bibfnamefont
  {F.}~\bibnamefont {Hudson}}, \bibinfo {author} {\bibfnamefont
  {A.}~\bibnamefont {Morello}}, \bibinfo {author} {\bibfnamefont {D.~J.}\
  \bibnamefont {Reilly}},\ and\ \bibinfo {author} {\bibfnamefont {A.~S.}\
  \bibnamefont {Dzurak}},\ }\bibfield  {title} {\bibinfo {title} {Gate-based
  single-shot readout of spins in silicon},\ }\href
  {https://doi.org/10.1038/s41565-019-0400-7} {\bibfield  {journal} {\bibinfo
  {journal} {Nature Nanotechnology}\ }\textbf {\bibinfo {volume} {14}},\
  \bibinfo {pages} {437} (\bibinfo {year} {2019})}\BibitemShut {NoStop}%
\bibitem [{\citenamefont {Boissonneault}\ \emph {et~al.}(2009)\citenamefont
  {Boissonneault}, \citenamefont {Gambetta},\ and\ \citenamefont
  {Blais}}]{Boissonneault2009}%
  \BibitemOpen
  \bibfield  {author} {\bibinfo {author} {\bibfnamefont {M.}~\bibnamefont
  {Boissonneault}}, \bibinfo {author} {\bibfnamefont {J.~M.}\ \bibnamefont
  {Gambetta}},\ and\ \bibinfo {author} {\bibfnamefont {A.}~\bibnamefont
  {Blais}},\ }\bibfield  {title} {\bibinfo {title} {{Dispersive regime of
  circuit QED: Photon-dependent qubit dephasing and relaxation rates}},\ }\href
  {https://doi.org/10.1103/PhysRevA.79.013819} {\bibfield  {journal} {\bibinfo
  {journal} {Phys. Rev. A}\ }\textbf {\bibinfo {volume} {79}},\ \bibinfo
  {pages} {013819} (\bibinfo {year} {2009})}\BibitemShut {NoStop}%
\bibitem [{\citenamefont {Slichter}\ \emph {et~al.}(2012)\citenamefont
  {Slichter}, \citenamefont {Vijay}, \citenamefont {Weber}, \citenamefont
  {Boutin}, \citenamefont {Boissonneault}, \citenamefont {Gambetta},
  \citenamefont {Blais},\ and\ \citenamefont {Siddiqi}}]{Slichter2012}%
  \BibitemOpen
  \bibfield  {author} {\bibinfo {author} {\bibfnamefont {D.~H.}\ \bibnamefont
  {Slichter}}, \bibinfo {author} {\bibfnamefont {R.}~\bibnamefont {Vijay}},
  \bibinfo {author} {\bibfnamefont {S.~J.}\ \bibnamefont {Weber}}, \bibinfo
  {author} {\bibfnamefont {S.}~\bibnamefont {Boutin}}, \bibinfo {author}
  {\bibfnamefont {M.}~\bibnamefont {Boissonneault}}, \bibinfo {author}
  {\bibfnamefont {J.~M.}\ \bibnamefont {Gambetta}}, \bibinfo {author}
  {\bibfnamefont {A.}~\bibnamefont {Blais}},\ and\ \bibinfo {author}
  {\bibfnamefont {I.}~\bibnamefont {Siddiqi}},\ }\bibfield  {title} {\bibinfo
  {title} {{Measurement-Induced Qubit State Mixing in Circuit QED from
  Up-Converted Dephasing Noise}},\ }\href@noop {} {\bibfield  {journal}
  {\bibinfo  {journal} {Phys. Rev. Lett.}\ }\textbf {\bibinfo {volume} {109}},\
  \bibinfo {pages} {153601} (\bibinfo {year} {2012})}\BibitemShut {NoStop}%
\bibitem [{\citenamefont {Aumentado}(2020)}]{Aumentado2020}%
  \BibitemOpen
  \bibfield  {author} {\bibinfo {author} {\bibfnamefont {J.}~\bibnamefont
  {Aumentado}},\ }\bibfield  {title} {\bibinfo {title} {{Superconducting
  Parametric Amplifiers: The State of the Art in Josephson Parametric
  Amplifiers}},\ }\href {https://doi.org/10.1109/MMM.2020.2993476} {\bibfield
  {journal} {\bibinfo  {journal} {IEEE Microw. Mag.}\ }\textbf {\bibinfo
  {volume} {21}},\ \bibinfo {pages} {45} (\bibinfo {year} {2020})}\BibitemShut
  {NoStop}%
\bibitem [{\citenamefont {Vijay}\ \emph {et~al.}(2011)\citenamefont {Vijay},
  \citenamefont {Slichter},\ and\ \citenamefont {Siddiqi}}]{Vijay2011}%
  \BibitemOpen
  \bibfield  {author} {\bibinfo {author} {\bibfnamefont {R.}~\bibnamefont
  {Vijay}}, \bibinfo {author} {\bibfnamefont {D.~H.}\ \bibnamefont
  {Slichter}},\ and\ \bibinfo {author} {\bibfnamefont {I.}~\bibnamefont
  {Siddiqi}},\ }\bibfield  {title} {\bibinfo {title} {{Observation of Quantum
  Jumps in a Superconducting Artificial Atom}},\ }\href
  {https://doi.org/10.1103/PhysRevLett.106.110502} {\bibfield  {journal}
  {\bibinfo  {journal} {Phys. Rev. Lett.}\ }\textbf {\bibinfo {volume} {106}},\
  \bibinfo {pages} {110502} (\bibinfo {year} {2011})}\BibitemShut {NoStop}%
\bibitem [{\citenamefont {Jeffrey}\ \emph {et~al.}(2014)\citenamefont
  {Jeffrey}, \citenamefont {Sank}, \citenamefont {Mutus}, \citenamefont
  {White}, \citenamefont {Kelly}, \citenamefont {Barends}, \citenamefont
  {Chen}, \citenamefont {Chen}, \citenamefont {Chiaro}, \citenamefont
  {Dunsworth}, \citenamefont {Megrant}, \citenamefont {O'Malley}, \citenamefont
  {Neill}, \citenamefont {Roushan}, \citenamefont {Vainsencher}, \citenamefont
  {Wenner}, \citenamefont {Cleland},\ and\ \citenamefont
  {Martinis}}]{Jeffrey2014}%
  \BibitemOpen
  \bibfield  {author} {\bibinfo {author} {\bibfnamefont {E.}~\bibnamefont
  {Jeffrey}}, \bibinfo {author} {\bibfnamefont {D.}~\bibnamefont {Sank}},
  \bibinfo {author} {\bibfnamefont {J.~Y.}\ \bibnamefont {Mutus}}, \bibinfo
  {author} {\bibfnamefont {T.~C.}\ \bibnamefont {White}}, \bibinfo {author}
  {\bibfnamefont {J.}~\bibnamefont {Kelly}}, \bibinfo {author} {\bibfnamefont
  {R.}~\bibnamefont {Barends}}, \bibinfo {author} {\bibfnamefont
  {Y.}~\bibnamefont {Chen}}, \bibinfo {author} {\bibfnamefont {Z.}~\bibnamefont
  {Chen}}, \bibinfo {author} {\bibfnamefont {B.}~\bibnamefont {Chiaro}},
  \bibinfo {author} {\bibfnamefont {A.}~\bibnamefont {Dunsworth}}, \bibinfo
  {author} {\bibfnamefont {A.}~\bibnamefont {Megrant}}, \bibinfo {author}
  {\bibfnamefont {P.~J.~J.}\ \bibnamefont {O'Malley}}, \bibinfo {author}
  {\bibfnamefont {C.}~\bibnamefont {Neill}}, \bibinfo {author} {\bibfnamefont
  {P.}~\bibnamefont {Roushan}}, \bibinfo {author} {\bibfnamefont
  {A.}~\bibnamefont {Vainsencher}}, \bibinfo {author} {\bibfnamefont
  {J.}~\bibnamefont {Wenner}}, \bibinfo {author} {\bibfnamefont {A.~N.}\
  \bibnamefont {Cleland}},\ and\ \bibinfo {author} {\bibfnamefont {J.~M.}\
  \bibnamefont {Martinis}},\ }\bibfield  {title} {\bibinfo {title} {{Fast
  Accurate State Measurement with Superconducting Qubits}},\ }\href
  {https://doi.org/10.1103/PhysRevLett.112.190504} {\bibfield  {journal}
  {\bibinfo  {journal} {Phys. Rev. Lett.}\ }\textbf {\bibinfo {volume} {112}},\
  \bibinfo {pages} {190504} (\bibinfo {year} {2014})}\BibitemShut {NoStop}%
\bibitem [{\citenamefont {Reed}\ \emph {et~al.}(2010)\citenamefont {Reed},
  \citenamefont {Johnson}, \citenamefont {Houck}, \citenamefont {DiCarlo},
  \citenamefont {Chow}, \citenamefont {Schuster}, \citenamefont {Frunzio},\
  and\ \citenamefont {Schoelkopf}}]{Reed2010}%
  \BibitemOpen
  \bibfield  {author} {\bibinfo {author} {\bibfnamefont {M.~D.}\ \bibnamefont
  {Reed}}, \bibinfo {author} {\bibfnamefont {B.~R.}\ \bibnamefont {Johnson}},
  \bibinfo {author} {\bibfnamefont {A.~A.}\ \bibnamefont {Houck}}, \bibinfo
  {author} {\bibfnamefont {L.}~\bibnamefont {DiCarlo}}, \bibinfo {author}
  {\bibfnamefont {J.~M.}\ \bibnamefont {Chow}}, \bibinfo {author}
  {\bibfnamefont {D.~I.}\ \bibnamefont {Schuster}}, \bibinfo {author}
  {\bibfnamefont {L.}~\bibnamefont {Frunzio}},\ and\ \bibinfo {author}
  {\bibfnamefont {R.~J.}\ \bibnamefont {Schoelkopf}},\ }\bibfield  {title}
  {\bibinfo {title} {{Fast reset and suppressings spontaneous emission of a
  superconducting qubit}},\ }\href {https://doi.org/10.1063/1.3435463}
  {\bibfield  {journal} {\bibinfo  {journal} {Appl. Phys. Lett.}\ }\textbf
  {\bibinfo {volume} {96}},\ \bibinfo {pages} {203110} (\bibinfo {year}
  {2010})}\BibitemShut {NoStop}%
\bibitem [{\citenamefont {Opremcak}\ \emph {et~al.}(2018)\citenamefont
  {Opremcak}, \citenamefont {Pechenezhskiy}, \citenamefont {Howington},
  \citenamefont {Christensen}, \citenamefont {Beck}, \citenamefont {Leonard},
  \citenamefont {Suttle}, \citenamefont {Wilen}, \citenamefont {Nesterov},
  \citenamefont {Ribeill}, \citenamefont {Thorbeck}, \citenamefont {Schlenker},
  \citenamefont {Vavilov}, \citenamefont {Plourde},\ and\ \citenamefont
  {McDermott}}]{Opremcak2018}%
  \BibitemOpen
  \bibfield  {author} {\bibinfo {author} {\bibfnamefont {A.}~\bibnamefont
  {Opremcak}}, \bibinfo {author} {\bibfnamefont {I.~V.}\ \bibnamefont
  {Pechenezhskiy}}, \bibinfo {author} {\bibfnamefont {C.}~\bibnamefont
  {Howington}}, \bibinfo {author} {\bibfnamefont {B.~G.}\ \bibnamefont
  {Christensen}}, \bibinfo {author} {\bibfnamefont {M.~A.}\ \bibnamefont
  {Beck}}, \bibinfo {author} {\bibfnamefont {E.}~\bibnamefont {Leonard}},
  \bibinfo {author} {\bibfnamefont {J.}~\bibnamefont {Suttle}}, \bibinfo
  {author} {\bibfnamefont {C.}~\bibnamefont {Wilen}}, \bibinfo {author}
  {\bibfnamefont {K.~N.}\ \bibnamefont {Nesterov}}, \bibinfo {author}
  {\bibfnamefont {G.~J.}\ \bibnamefont {Ribeill}}, \bibinfo {author}
  {\bibfnamefont {T.}~\bibnamefont {Thorbeck}}, \bibinfo {author}
  {\bibfnamefont {F.}~\bibnamefont {Schlenker}}, \bibinfo {author}
  {\bibfnamefont {M.~G.}\ \bibnamefont {Vavilov}}, \bibinfo {author}
  {\bibfnamefont {B.~L.~T.}\ \bibnamefont {Plourde}},\ and\ \bibinfo {author}
  {\bibfnamefont {R.}~\bibnamefont {McDermott}},\ }\bibfield  {title} {\bibinfo
  {title} {{Measurement of a superconducting qubit with a microwave photon
  counter}},\ }\href {https://doi.org/10.1126/science.aat4625} {\bibfield
  {journal} {\bibinfo  {journal} {Science}\ }\textbf {\bibinfo {volume}
  {361}},\ \bibinfo {pages} {1239} (\bibinfo {year} {2018})}\BibitemShut
  {NoStop}%
\bibitem [{\citenamefont {Opremcak}\ \emph {et~al.}(2020)\citenamefont
  {Opremcak}, \citenamefont {Liu}, \citenamefont {Wilen}, \citenamefont
  {Okubo}, \citenamefont {Christensen}, \citenamefont {Sank}, \citenamefont
  {White}, \citenamefont {Vainsencher}, \citenamefont {Giustina}, \citenamefont
  {Megrant} \emph {et~al.}}]{opremcak2020high}%
  \BibitemOpen
  \bibfield  {author} {\bibinfo {author} {\bibfnamefont {A.}~\bibnamefont
  {Opremcak}}, \bibinfo {author} {\bibfnamefont {C.}~\bibnamefont {Liu}},
  \bibinfo {author} {\bibfnamefont {C.}~\bibnamefont {Wilen}}, \bibinfo
  {author} {\bibfnamefont {K.}~\bibnamefont {Okubo}}, \bibinfo {author}
  {\bibfnamefont {B.}~\bibnamefont {Christensen}}, \bibinfo {author}
  {\bibfnamefont {D.}~\bibnamefont {Sank}}, \bibinfo {author} {\bibfnamefont
  {T.}~\bibnamefont {White}}, \bibinfo {author} {\bibfnamefont
  {A.}~\bibnamefont {Vainsencher}}, \bibinfo {author} {\bibfnamefont
  {M.}~\bibnamefont {Giustina}}, \bibinfo {author} {\bibfnamefont
  {A.}~\bibnamefont {Megrant}}, \emph {et~al.},\ }\bibfield  {title} {\bibinfo
  {title} {High-fidelity measurement of a superconducting qubit using an
  on-chip microwave photon counter},\ }\href@noop {} {\bibfield  {journal}
  {\bibinfo  {journal} {arXiv:2008.02346}\ } (\bibinfo {year}
  {2020})}\BibitemShut {NoStop}%
\bibitem [{\citenamefont {Dehmelt}(1982)}]{Dehmelt1982}%
  \BibitemOpen
  \bibfield  {author} {\bibinfo {author} {\bibfnamefont {H.~G.}\ \bibnamefont
  {Dehmelt}},\ }\bibfield  {title} {\bibinfo {title} {{Monoion oscillator as
  potential ultimate laser frequency standard}},\ }\href
  {https://doi.org/10.1109/TIM.1982.6312526} {\bibfield  {journal} {\bibinfo
  {journal} {IEEE Trans. Instrum. Meas.}\ }\textbf {\bibinfo {volume}
  {IM-31}},\ \bibinfo {pages} {83} (\bibinfo {year} {1982})}\BibitemShut
  {NoStop}%
\bibitem [{\citenamefont {Myerson}\ \emph {et~al.}(2008)\citenamefont
  {Myerson}, \citenamefont {Szwer}, \citenamefont {Webster}, \citenamefont
  {Allcock}, \citenamefont {Curtis}, \citenamefont {Imreh}, \citenamefont
  {Sherman}, \citenamefont {Stacey}, \citenamefont {Steane},\ and\
  \citenamefont {Lucas}}]{Myerson2008}%
  \BibitemOpen
  \bibfield  {author} {\bibinfo {author} {\bibfnamefont {A.~H.}\ \bibnamefont
  {Myerson}}, \bibinfo {author} {\bibfnamefont {D.~J.}\ \bibnamefont {Szwer}},
  \bibinfo {author} {\bibfnamefont {S.~C.}\ \bibnamefont {Webster}}, \bibinfo
  {author} {\bibfnamefont {D.~T.~C.}\ \bibnamefont {Allcock}}, \bibinfo
  {author} {\bibfnamefont {M.~J.}\ \bibnamefont {Curtis}}, \bibinfo {author}
  {\bibfnamefont {G.}~\bibnamefont {Imreh}}, \bibinfo {author} {\bibfnamefont
  {J.~A.}\ \bibnamefont {Sherman}}, \bibinfo {author} {\bibfnamefont {D.~N.}\
  \bibnamefont {Stacey}}, \bibinfo {author} {\bibfnamefont {A.~M.}\
  \bibnamefont {Steane}},\ and\ \bibinfo {author} {\bibfnamefont {D.~M.}\
  \bibnamefont {Lucas}},\ }\bibfield  {title} {\bibinfo {title} {{High-Fidelity
  Readout of Trapped-Ion Qubits}},\ }\href
  {https://doi.org/10.1103/PhysRevLett.100.200502} {\bibfield  {journal}
  {\bibinfo  {journal} {Phys. Rev. Lett.}\ }\textbf {\bibinfo {volume} {100}},\
  \bibinfo {pages} {200502} (\bibinfo {year} {2008})}\BibitemShut {NoStop}%
\bibitem [{\citenamefont {Britton}\ \emph {et~al.}(2012)\citenamefont
  {Britton}, \citenamefont {Sawyer}, \citenamefont {Keith}, \citenamefont
  {Wang}, \citenamefont {Freericks}, \citenamefont {Uys}, \citenamefont
  {Biercuk},\ and\ \citenamefont {Bollinger}}]{Britton2012}%
  \BibitemOpen
  \bibfield  {author} {\bibinfo {author} {\bibfnamefont {J.~W.}\ \bibnamefont
  {Britton}}, \bibinfo {author} {\bibfnamefont {B.~C.}\ \bibnamefont {Sawyer}},
  \bibinfo {author} {\bibfnamefont {A.~C.}\ \bibnamefont {Keith}}, \bibinfo
  {author} {\bibfnamefont {C.-C.~J.}\ \bibnamefont {Wang}}, \bibinfo {author}
  {\bibfnamefont {J.~K.}\ \bibnamefont {Freericks}}, \bibinfo {author}
  {\bibfnamefont {H.}~\bibnamefont {Uys}}, \bibinfo {author} {\bibfnamefont
  {M.~J.}\ \bibnamefont {Biercuk}},\ and\ \bibinfo {author} {\bibfnamefont
  {J.~J.}\ \bibnamefont {Bollinger}},\ }\bibfield  {title} {\bibinfo {title}
  {{Engineered two-dimensional Ising interactions in a trapped-ion quantum
  simulator with hundreds of spins}},\ }\href
  {https://doi.org/10.1038/nature10981} {\bibfield  {journal} {\bibinfo
  {journal} {Nature}\ }\textbf {\bibinfo {volume} {484}},\ \bibinfo {pages}
  {489} (\bibinfo {year} {2012})}\BibitemShut {NoStop}%
\bibitem [{\citenamefont {Dilling}\ \emph {et~al.}(2018)\citenamefont
  {Dilling}, \citenamefont {Blaum}, \citenamefont {Brodeur},\ and\
  \citenamefont {Eliseev}}]{Dilling2018}%
  \BibitemOpen
  \bibfield  {author} {\bibinfo {author} {\bibfnamefont {J.}~\bibnamefont
  {Dilling}}, \bibinfo {author} {\bibfnamefont {K.}~\bibnamefont {Blaum}},
  \bibinfo {author} {\bibfnamefont {M.}~\bibnamefont {Brodeur}},\ and\ \bibinfo
  {author} {\bibfnamefont {S.}~\bibnamefont {Eliseev}},\ }\bibfield  {title}
  {\bibinfo {title} {{Penning-Trap Mass Measurements in Atomic and Nuclear
  Physics}},\ }\href {https://doi.org/10.1146/annurev-nucl-102711-094939}
  {\bibfield  {journal} {\bibinfo  {journal} {Annu. Rev. Nucl. Part. Sci.}\
  }\textbf {\bibinfo {volume} {68}},\ \bibinfo {pages} {45} (\bibinfo {year}
  {2018})}\BibitemShut {NoStop}%
\bibitem [{\citenamefont {Jain}\ \emph {et~al.}(2020)\citenamefont {Jain},
  \citenamefont {Alonso}, \citenamefont {Grau},\ and\ \citenamefont
  {Home}}]{Jain2020}%
  \BibitemOpen
  \bibfield  {author} {\bibinfo {author} {\bibfnamefont {S.}~\bibnamefont
  {Jain}}, \bibinfo {author} {\bibfnamefont {J.}~\bibnamefont {Alonso}},
  \bibinfo {author} {\bibfnamefont {M.}~\bibnamefont {Grau}},\ and\ \bibinfo
  {author} {\bibfnamefont {J.~P.}\ \bibnamefont {Home}},\ }\bibfield  {title}
  {\bibinfo {title} {{Scalable Arrays of Micro-Penning Traps for Quantum
  Computing and Simulation}},\ }\href
  {https://doi.org/10.1103/PhysRevX.10.031027} {\bibfield  {journal} {\bibinfo
  {journal} {Phys. Rev. X}\ }\textbf {\bibinfo {volume} {10}},\ \bibinfo
  {pages} {031027} (\bibinfo {year} {2020})}\BibitemShut {NoStop}%
\bibitem [{\citenamefont {Macalpine}\ and\ \citenamefont
  {Schildknecht}(1959)}]{Macalpine1959}%
  \BibitemOpen
  \bibfield  {author} {\bibinfo {author} {\bibfnamefont {W.}~\bibnamefont
  {Macalpine}}\ and\ \bibinfo {author} {\bibfnamefont {R.}~\bibnamefont
  {Schildknecht}},\ }\bibfield  {title} {\bibinfo {title} {{Coaxial Resonators
  with Helical Inner Conductor}},\ }\href
  {https://doi.org/10.1109/JRPROC.1959.287128} {\bibfield  {journal} {\bibinfo
  {journal} {Proc. IRE}\ }\textbf {\bibinfo {volume} {47}},\ \bibinfo {pages}
  {2099} (\bibinfo {year} {1959})}\BibitemShut {NoStop}%
\bibitem [{\citenamefont {Siverns}\ \emph {et~al.}(2012)\citenamefont
  {Siverns}, \citenamefont {Simkins}, \citenamefont {Weidt},\ and\
  \citenamefont {Hensinger}}]{Siverns2012}%
  \BibitemOpen
  \bibfield  {author} {\bibinfo {author} {\bibfnamefont {J.~D.}\ \bibnamefont
  {Siverns}}, \bibinfo {author} {\bibfnamefont {L.~R.}\ \bibnamefont
  {Simkins}}, \bibinfo {author} {\bibfnamefont {S.}~\bibnamefont {Weidt}},\
  and\ \bibinfo {author} {\bibfnamefont {W.~K.}\ \bibnamefont {Hensinger}},\
  }\bibfield  {title} {\bibinfo {title} {{On the application of radio frequency
  voltages to ion traps via helical resonators}},\ }\href
  {https://doi.org/10.1007/s00340-011-4837-0} {\bibfield  {journal} {\bibinfo
  {journal} {Appl. Phys. B}\ }\textbf {\bibinfo {volume} {107}},\ \bibinfo
  {pages} {921} (\bibinfo {year} {2012})}\BibitemShut {NoStop}%
\bibitem [{\citenamefont {Jefferts}\ \emph {et~al.}(1995)\citenamefont
  {Jefferts}, \citenamefont {Monroe}, \citenamefont {Bell},\ and\ \citenamefont
  {Wineland}}]{Jefferts1995}%
  \BibitemOpen
  \bibfield  {author} {\bibinfo {author} {\bibfnamefont {S.~R.}\ \bibnamefont
  {Jefferts}}, \bibinfo {author} {\bibfnamefont {C.}~\bibnamefont {Monroe}},
  \bibinfo {author} {\bibfnamefont {E.~W.}\ \bibnamefont {Bell}},\ and\
  \bibinfo {author} {\bibfnamefont {D.~J.}\ \bibnamefont {Wineland}},\
  }\bibfield  {title} {\bibinfo {title} {{Coaxial-resonator-driven rf (Paul)
  trap for strong confinement}},\ }\href
  {https://doi.org/10.1103/PhysRevA.51.3112} {\bibfield  {journal} {\bibinfo
  {journal} {Phys. Rev. A}\ }\textbf {\bibinfo {volume} {51}},\ \bibinfo
  {pages} {3112} (\bibinfo {year} {1995})}\BibitemShut {NoStop}%
\bibitem [{\citenamefont {Gandolfi}\ \emph {et~al.}(2012)\citenamefont
  {Gandolfi}, \citenamefont {Niedermayr}, \citenamefont {Kumph}, \citenamefont
  {Brownnutt},\ and\ \citenamefont {Blatt}}]{Gandolfi2012}%
  \BibitemOpen
  \bibfield  {author} {\bibinfo {author} {\bibfnamefont {D.}~\bibnamefont
  {Gandolfi}}, \bibinfo {author} {\bibfnamefont {M.}~\bibnamefont
  {Niedermayr}}, \bibinfo {author} {\bibfnamefont {M.}~\bibnamefont {Kumph}},
  \bibinfo {author} {\bibfnamefont {M.}~\bibnamefont {Brownnutt}},\ and\
  \bibinfo {author} {\bibfnamefont {R.}~\bibnamefont {Blatt}},\ }\bibfield
  {title} {\bibinfo {title} {{Compact radio-frequency resonator for cryogenic
  ion traps}},\ }\href {https://doi.org/10.1063/1.4737889} {\bibfield
  {journal} {\bibinfo  {journal} {Rev. Sci. Instrum.}\ }\textbf {\bibinfo
  {volume} {83}},\ \bibinfo {pages} {084705} (\bibinfo {year}
  {2012})}\BibitemShut {NoStop}%
\bibitem [{\citenamefont {Brandl}\ \emph {et~al.}(2016)\citenamefont {Brandl},
  \citenamefont {Schindler}, \citenamefont {Monz},\ and\ \citenamefont
  {Blatt}}]{Brandl2016}%
  \BibitemOpen
  \bibfield  {author} {\bibinfo {author} {\bibfnamefont {M.~F.}\ \bibnamefont
  {Brandl}}, \bibinfo {author} {\bibfnamefont {P.}~\bibnamefont {Schindler}},
  \bibinfo {author} {\bibfnamefont {T.}~\bibnamefont {Monz}},\ and\ \bibinfo
  {author} {\bibfnamefont {R.}~\bibnamefont {Blatt}},\ }\bibfield  {title}
  {\bibinfo {title} {{Cryogenic resonator design for trapped ion experiments in
  Paul traps}},\ }\href {https://doi.org/10.1007/s00340-016-6430-z} {\bibfield
  {journal} {\bibinfo  {journal} {Appl. Phys. B}\ }\textbf {\bibinfo {volume}
  {122}},\ \bibinfo {pages} {157} (\bibinfo {year} {2016})}\BibitemShut
  {NoStop}%
\bibitem [{\citenamefont {Allcock}(2011)}]{Allcock2011}%
  \BibitemOpen
  \bibfield  {author} {\bibinfo {author} {\bibfnamefont {D.~T.~C.}\
  \bibnamefont {Allcock}},\ }\emph {\bibinfo {title} {{Surface-Electrode Ion
  Traps for Scalable Quantum Computing}}},\ \href@noop {} {Ph.D. thesis},\
  \bibinfo  {school} {University of Oxford} (\bibinfo {year}
  {2011})\BibitemShut {NoStop}%
\bibitem [{\citenamefont {Harty}(2013)}]{Harty2013}%
  \BibitemOpen
  \bibfield  {author} {\bibinfo {author} {\bibfnamefont {T.~P.}\ \bibnamefont
  {Harty}},\ }\emph {\bibinfo {title} {{High-Fidelity Quantum Logic in
  Intermediate-Field $^{43}\mathrm{Ca}^+$}}},\ \href@noop {} {Ph.D. thesis},\
  \bibinfo  {school} {Univeristy of Oxford} (\bibinfo {year}
  {2013})\BibitemShut {NoStop}%
\bibitem [{\citenamefont {Johnson}\ \emph {et~al.}(2016)\citenamefont
  {Johnson}, \citenamefont {Wong-Campos}, \citenamefont {Restelli},
  \citenamefont {Landsman}, \citenamefont {Neyenhuis}, \citenamefont
  {Mizrahi},\ and\ \citenamefont {Monroe}}]{Johnson2016}%
  \BibitemOpen
  \bibfield  {author} {\bibinfo {author} {\bibfnamefont {K.~G.}\ \bibnamefont
  {Johnson}}, \bibinfo {author} {\bibfnamefont {J.~D.}\ \bibnamefont
  {Wong-Campos}}, \bibinfo {author} {\bibfnamefont {A.}~\bibnamefont
  {Restelli}}, \bibinfo {author} {\bibfnamefont {K.~A.}\ \bibnamefont
  {Landsman}}, \bibinfo {author} {\bibfnamefont {B.}~\bibnamefont {Neyenhuis}},
  \bibinfo {author} {\bibfnamefont {J.}~\bibnamefont {Mizrahi}},\ and\ \bibinfo
  {author} {\bibfnamefont {C.}~\bibnamefont {Monroe}},\ }\bibfield  {title}
  {\bibinfo {title} {{Active stabilization of ion trap radiofrequency
  potentials}},\ }\href {https://doi.org/10.1063/1.4948734} {\bibfield
  {journal} {\bibinfo  {journal} {Rev. Sci. Instrum.}\ }\textbf {\bibinfo
  {volume} {87}},\ \bibinfo {pages} {053110} (\bibinfo {year}
  {2016})}\BibitemShut {NoStop}%
\bibitem [{\citenamefont {Saffman}(2016)}]{Saffman2016}%
  \BibitemOpen
  \bibfield  {author} {\bibinfo {author} {\bibfnamefont {M.}~\bibnamefont
  {Saffman}},\ }\bibfield  {title} {\bibinfo {title} {{Quantum computing with
  atomic qubits and Rydberg interactions: progress and challenges}},\ }\href
  {https://doi.org/10.1088/0953-4075/49/20/202001} {\bibfield  {journal}
  {\bibinfo  {journal} {J. Phys. B At. Mol. Opt. Phys.}\ }\textbf {\bibinfo
  {volume} {49}},\ \bibinfo {pages} {202001} (\bibinfo {year}
  {2016})}\BibitemShut {NoStop}%
\bibitem [{\citenamefont {Henriet}\ \emph {et~al.}(2020)\citenamefont
  {Henriet}, \citenamefont {Beguin}, \citenamefont {Signoles}, \citenamefont
  {Lahaye}, \citenamefont {Browaeys}, \citenamefont {Reymond},\ and\
  \citenamefont {Jurczak}}]{Henriet2020}%
  \BibitemOpen
  \bibfield  {author} {\bibinfo {author} {\bibfnamefont {L.}~\bibnamefont
  {Henriet}}, \bibinfo {author} {\bibfnamefont {L.}~\bibnamefont {Beguin}},
  \bibinfo {author} {\bibfnamefont {A.}~\bibnamefont {Signoles}}, \bibinfo
  {author} {\bibfnamefont {T.}~\bibnamefont {Lahaye}}, \bibinfo {author}
  {\bibfnamefont {A.}~\bibnamefont {Browaeys}}, \bibinfo {author}
  {\bibfnamefont {G.-O.}\ \bibnamefont {Reymond}},\ and\ \bibinfo {author}
  {\bibfnamefont {C.}~\bibnamefont {Jurczak}},\ }\bibfield  {title} {\bibinfo
  {title} {{Quantum computing with neutral atoms}},\ }\href
  {https://doi.org/10.22331/q-2020-09-21-327} {\bibfield  {journal} {\bibinfo
  {journal} {Quantum}\ }\textbf {\bibinfo {volume} {4}},\ \bibinfo {pages}
  {327} (\bibinfo {year} {2020})}\BibitemShut {NoStop}%
\bibitem [{\citenamefont {Dobrovitski}\ \emph {et~al.}(2013)\citenamefont
  {Dobrovitski}, \citenamefont {Fuchs}, \citenamefont {Falk}, \citenamefont
  {Santori},\ and\ \citenamefont {Awschalom}}]{Dobrovitski2013}%
  \BibitemOpen
  \bibfield  {author} {\bibinfo {author} {\bibfnamefont {V.~V.}\ \bibnamefont
  {Dobrovitski}}, \bibinfo {author} {\bibfnamefont {G.~D.}\ \bibnamefont
  {Fuchs}}, \bibinfo {author} {\bibfnamefont {A.~L.}\ \bibnamefont {Falk}},
  \bibinfo {author} {\bibfnamefont {C.}~\bibnamefont {Santori}},\ and\ \bibinfo
  {author} {\bibfnamefont {D.~D.}\ \bibnamefont {Awschalom}},\ }\bibfield
  {title} {\bibinfo {title} {{Quantum Control over Single Spins in Diamond}},\
  }\href {https://doi.org/10.1146/annurev-conmatphys-030212-184238} {\bibfield
  {journal} {\bibinfo  {journal} {Annu. Rev. Condens. Matter Phys.}\ }\textbf
  {\bibinfo {volume} {4}},\ \bibinfo {pages} {23} (\bibinfo {year}
  {2013})}\BibitemShut {NoStop}%
\bibitem [{\citenamefont {Doherty}\ \emph {et~al.}(2013)\citenamefont
  {Doherty}, \citenamefont {Manson}, \citenamefont {Delaney}, \citenamefont
  {Jelezko}, \citenamefont {Wrachtrup},\ and\ \citenamefont
  {Hollenberg}}]{Doherty2013}%
  \BibitemOpen
  \bibfield  {author} {\bibinfo {author} {\bibfnamefont {M.~W.}\ \bibnamefont
  {Doherty}}, \bibinfo {author} {\bibfnamefont {N.~B.}\ \bibnamefont {Manson}},
  \bibinfo {author} {\bibfnamefont {P.}~\bibnamefont {Delaney}}, \bibinfo
  {author} {\bibfnamefont {F.}~\bibnamefont {Jelezko}}, \bibinfo {author}
  {\bibfnamefont {J.}~\bibnamefont {Wrachtrup}},\ and\ \bibinfo {author}
  {\bibfnamefont {L.~C.}\ \bibnamefont {Hollenberg}},\ }\bibfield  {title}
  {\bibinfo {title} {{The nitrogen-vacancy colour centre in diamond}},\ }\href
  {https://doi.org/10.1016/j.physrep.2013.02.001} {\bibfield  {journal}
  {\bibinfo  {journal} {Phys. Rep.}\ }\textbf {\bibinfo {volume} {528}},\
  \bibinfo {pages} {1} (\bibinfo {year} {2013})}\BibitemShut {NoStop}%
\bibitem [{\citenamefont {Korpel}(1981)}]{Korpel1981}%
  \BibitemOpen
  \bibfield  {author} {\bibinfo {author} {\bibfnamefont {A.}~\bibnamefont
  {Korpel}},\ }\bibfield  {title} {\bibinfo {title} {{Acousto-optics—A review
  of fundamentals}},\ }\href {https://doi.org/10.1109/PROC.1981.11919}
  {\bibfield  {journal} {\bibinfo  {journal} {Proc. IEEE}\ }\textbf {\bibinfo
  {volume} {69}},\ \bibinfo {pages} {48} (\bibinfo {year} {1981})}\BibitemShut
  {NoStop}%
\bibitem [{\citenamefont {Desmarais}(1997)}]{Desmarais1997}%
  \BibitemOpen
  \bibfield  {author} {\bibinfo {author} {\bibfnamefont {L.}~\bibnamefont
  {Desmarais}},\ }\href@noop {} {\emph {\bibinfo {title} {{Applied Electro
  Optics}}}}\ (\bibinfo  {publisher} {Pearson Education},\ \bibinfo {year}
  {1997})\BibitemShut {NoStop}%
\bibitem [{\citenamefont {Pound}(1946)}]{Pound1946}%
  \BibitemOpen
  \bibfield  {author} {\bibinfo {author} {\bibfnamefont {R.~V.}\ \bibnamefont
  {Pound}},\ }\bibfield  {title} {\bibinfo {title} {{Electronic Frequency
  Stabilization of Microwave Oscillators}},\ }\href
  {https://doi.org/10.1063/1.1770414} {\bibfield  {journal} {\bibinfo
  {journal} {Rev. Sci. Instrum.}\ }\textbf {\bibinfo {volume} {17}},\ \bibinfo
  {pages} {490} (\bibinfo {year} {1946})}\BibitemShut {NoStop}%
\bibitem [{\citenamefont {Drever}\ \emph {et~al.}(1983)\citenamefont {Drever},
  \citenamefont {Hall}, \citenamefont {Kowalski}, \citenamefont {Hough},
  \citenamefont {Ford}, \citenamefont {Munley},\ and\ \citenamefont
  {Ward}}]{Drever1983}%
  \BibitemOpen
  \bibfield  {author} {\bibinfo {author} {\bibfnamefont {R.~W.~P.}\
  \bibnamefont {Drever}}, \bibinfo {author} {\bibfnamefont {J.~L.}\
  \bibnamefont {Hall}}, \bibinfo {author} {\bibfnamefont {F.~V.}\ \bibnamefont
  {Kowalski}}, \bibinfo {author} {\bibfnamefont {J.}~\bibnamefont {Hough}},
  \bibinfo {author} {\bibfnamefont {G.~M.}\ \bibnamefont {Ford}}, \bibinfo
  {author} {\bibfnamefont {A.~J.}\ \bibnamefont {Munley}},\ and\ \bibinfo
  {author} {\bibfnamefont {H.}~\bibnamefont {Ward}},\ }\bibfield  {title}
  {\bibinfo {title} {{Laser phase and frequency stabilization using an optical
  resonator}},\ }\href {https://doi.org/10.1007/BF00702605} {\bibfield
  {journal} {\bibinfo  {journal} {Appl. Phys. B Photophysics Laser Chem.}\
  }\textbf {\bibinfo {volume} {31}},\ \bibinfo {pages} {97} (\bibinfo {year}
  {1983})}\BibitemShut {NoStop}%
\bibitem [{\citenamefont {Mizrahi}\ \emph {et~al.}(2014)\citenamefont
  {Mizrahi}, \citenamefont {Neyenhuis}, \citenamefont {Johnson}, \citenamefont
  {Campbell}, \citenamefont {Senko}, \citenamefont {Hayes},\ and\ \citenamefont
  {Monroe}}]{Mizrahi2014}%
  \BibitemOpen
  \bibfield  {author} {\bibinfo {author} {\bibfnamefont {J.}~\bibnamefont
  {Mizrahi}}, \bibinfo {author} {\bibfnamefont {B.}~\bibnamefont {Neyenhuis}},
  \bibinfo {author} {\bibfnamefont {K.~G.}\ \bibnamefont {Johnson}}, \bibinfo
  {author} {\bibfnamefont {W.~C.}\ \bibnamefont {Campbell}}, \bibinfo {author}
  {\bibfnamefont {C.}~\bibnamefont {Senko}}, \bibinfo {author} {\bibfnamefont
  {D.}~\bibnamefont {Hayes}},\ and\ \bibinfo {author} {\bibfnamefont
  {C.}~\bibnamefont {Monroe}},\ }\bibfield  {title} {\bibinfo {title} {{Quantum
  control of qubits and atomic motion using ultrafast laser pulses}},\ }\href
  {https://doi.org/10.1007/s00340-013-5717-6} {\bibfield  {journal} {\bibinfo
  {journal} {Appl. Phys. B}\ }\textbf {\bibinfo {volume} {114}},\ \bibinfo
  {pages} {45} (\bibinfo {year} {2014})}\BibitemShut {NoStop}%
\bibitem [{\citenamefont {Louisell}(1960)}]{Louisell1960}%
  \BibitemOpen
  \bibfield  {author} {\bibinfo {author} {\bibfnamefont {W.~H.}\ \bibnamefont
  {Louisell}},\ }\href@noop {} {\emph {\bibinfo {title} {{Coupled Mode and
  Parametric Electronics}}}}\ (\bibinfo  {publisher} {John Wiley},\ \bibinfo
  {address} {New York},\ \bibinfo {year} {1960})\BibitemShut {NoStop}%
\bibitem [{\citenamefont {Mumford}(1960)}]{Mumford1960}%
  \BibitemOpen
  \bibfield  {author} {\bibinfo {author} {\bibfnamefont {W.~W.}\ \bibnamefont
  {Mumford}},\ }\bibfield  {title} {\bibinfo {title} {{Some Notes on the
  History of Parametric Transducers}},\ }\href
  {https://doi.org/10.1109/JRPROC.1960.287620} {\bibfield  {journal} {\bibinfo
  {journal} {Proc. IRE}\ }\textbf {\bibinfo {volume} {48}},\ \bibinfo {pages}
  {848} (\bibinfo {year} {1960})}\BibitemShut {NoStop}%
\bibitem [{\citenamefont {Castellanos-Beltran}\ and\ \citenamefont
  {Lehnert}(2007)}]{Castellanos-Beltran2007}%
  \BibitemOpen
  \bibfield  {author} {\bibinfo {author} {\bibfnamefont {M.~A.}\ \bibnamefont
  {Castellanos-Beltran}}\ and\ \bibinfo {author} {\bibfnamefont {K.~W.}\
  \bibnamefont {Lehnert}},\ }\bibfield  {title} {\bibinfo {title} {{Widely
  tunable parametric amplifier based on a superconducting quantum interference
  device array resonator}},\ }\href {https://doi.org/10.1063/1.2773988}
  {\bibfield  {journal} {\bibinfo  {journal} {Appl. Phys. Lett.}\ }\textbf
  {\bibinfo {volume} {91}},\ \bibinfo {pages} {083509} (\bibinfo {year}
  {2007})}\BibitemShut {NoStop}%
\bibitem [{\citenamefont {Yamamoto}\ \emph {et~al.}(2008)\citenamefont
  {Yamamoto}, \citenamefont {Inomata}, \citenamefont {Watanabe}, \citenamefont
  {Matsuba}, \citenamefont {Miyazaki}, \citenamefont {Oliver}, \citenamefont
  {Nakamura},\ and\ \citenamefont {Tsai}}]{Yamamoto2008}%
  \BibitemOpen
  \bibfield  {author} {\bibinfo {author} {\bibfnamefont {T.}~\bibnamefont
  {Yamamoto}}, \bibinfo {author} {\bibfnamefont {K.}~\bibnamefont {Inomata}},
  \bibinfo {author} {\bibfnamefont {M.}~\bibnamefont {Watanabe}}, \bibinfo
  {author} {\bibfnamefont {K.}~\bibnamefont {Matsuba}}, \bibinfo {author}
  {\bibfnamefont {T.}~\bibnamefont {Miyazaki}}, \bibinfo {author}
  {\bibfnamefont {W.~D.}\ \bibnamefont {Oliver}}, \bibinfo {author}
  {\bibfnamefont {Y.}~\bibnamefont {Nakamura}},\ and\ \bibinfo {author}
  {\bibfnamefont {J.~S.}\ \bibnamefont {Tsai}},\ }\bibfield  {title} {\bibinfo
  {title} {{Flux-driven Josephson parametric amplifier}},\ }\href
  {https://doi.org/10.1063/1.2964182} {\bibfield  {journal} {\bibinfo
  {journal} {Appl. Phys. Lett.}\ }\textbf {\bibinfo {volume} {93}},\ \bibinfo
  {pages} {042510} (\bibinfo {year} {2008})}\BibitemShut {NoStop}%
\bibitem [{\citenamefont {Vijay}(2008)}]{Vijay2008}%
  \BibitemOpen
  \bibfield  {author} {\bibinfo {author} {\bibfnamefont {R.}~\bibnamefont
  {Vijay}},\ }\emph {\bibinfo {title} {{Josephson Bifurcation Amplifier:
  Amplifying quantum signals using a dynamical bifurcation}}},\ \href@noop {}
  {Ph.D. thesis},\ \bibinfo  {school} {Yale University} (\bibinfo {year}
  {2008})\BibitemShut {NoStop}%
\bibitem [{\citenamefont {Bergeal}\ \emph {et~al.}(2010)\citenamefont
  {Bergeal}, \citenamefont {Schackert}, \citenamefont {Metcalfe}, \citenamefont
  {Vijay}, \citenamefont {Manucharyan}, \citenamefont {Frunzio}, \citenamefont
  {Prober}, \citenamefont {Schoelkopf}, \citenamefont {Girvin},\ and\
  \citenamefont {Devoret}}]{Bergeal2010a}%
  \BibitemOpen
  \bibfield  {author} {\bibinfo {author} {\bibfnamefont {N.}~\bibnamefont
  {Bergeal}}, \bibinfo {author} {\bibfnamefont {F.}~\bibnamefont {Schackert}},
  \bibinfo {author} {\bibfnamefont {M.}~\bibnamefont {Metcalfe}}, \bibinfo
  {author} {\bibfnamefont {R.}~\bibnamefont {Vijay}}, \bibinfo {author}
  {\bibfnamefont {V.~E.}\ \bibnamefont {Manucharyan}}, \bibinfo {author}
  {\bibfnamefont {L.}~\bibnamefont {Frunzio}}, \bibinfo {author} {\bibfnamefont
  {D.~E.}\ \bibnamefont {Prober}}, \bibinfo {author} {\bibfnamefont {R.~J.}\
  \bibnamefont {Schoelkopf}}, \bibinfo {author} {\bibfnamefont {S.~M.}\
  \bibnamefont {Girvin}},\ and\ \bibinfo {author} {\bibfnamefont {M.~H.}\
  \bibnamefont {Devoret}},\ }\bibfield  {title} {\bibinfo {title}
  {{Phase-preserving amplification near the quantum limit with a Josephson ring
  modulator.}},\ }\href {https://doi.org/10.1038/nature09035} {\bibfield
  {journal} {\bibinfo  {journal} {Nature}\ }\textbf {\bibinfo {volume} {465}},\
  \bibinfo {pages} {64} (\bibinfo {year} {2010})}\BibitemShut {NoStop}%
\bibitem [{\citenamefont {Hatridge}\ \emph {et~al.}(2013)\citenamefont
  {Hatridge}, \citenamefont {Shankar}, \citenamefont {Mirrahimi}, \citenamefont
  {Schackert}, \citenamefont {Geerlings}, \citenamefont {Brecht}, \citenamefont
  {Sliwa}, \citenamefont {Abdo}, \citenamefont {Frunzio}, \citenamefont
  {Girvin}, \citenamefont {Schoelkopf},\ and\ \citenamefont
  {Devoret}}]{Hatridge2013}%
  \BibitemOpen
  \bibfield  {author} {\bibinfo {author} {\bibfnamefont {M.}~\bibnamefont
  {Hatridge}}, \bibinfo {author} {\bibfnamefont {S.}~\bibnamefont {Shankar}},
  \bibinfo {author} {\bibfnamefont {M.}~\bibnamefont {Mirrahimi}}, \bibinfo
  {author} {\bibfnamefont {F.}~\bibnamefont {Schackert}}, \bibinfo {author}
  {\bibfnamefont {K.}~\bibnamefont {Geerlings}}, \bibinfo {author}
  {\bibfnamefont {T.}~\bibnamefont {Brecht}}, \bibinfo {author} {\bibfnamefont
  {K.~M.}\ \bibnamefont {Sliwa}}, \bibinfo {author} {\bibfnamefont
  {B.}~\bibnamefont {Abdo}}, \bibinfo {author} {\bibfnamefont {L.}~\bibnamefont
  {Frunzio}}, \bibinfo {author} {\bibfnamefont {S.~M.}\ \bibnamefont {Girvin}},
  \bibinfo {author} {\bibfnamefont {R.~J.}\ \bibnamefont {Schoelkopf}},\ and\
  \bibinfo {author} {\bibfnamefont {M.~H.}\ \bibnamefont {Devoret}},\
  }\bibfield  {title} {\bibinfo {title} {{Quantum Back-Action of an Individual
  Variable-Strength Measurement}},\ }\href
  {https://doi.org/10.1126/science.1226897} {\bibfield  {journal} {\bibinfo
  {journal} {Science}\ }\textbf {\bibinfo {volume} {339}},\ \bibinfo {pages}
  {178} (\bibinfo {year} {2013})}\BibitemShut {NoStop}%
\bibitem [{\citenamefont {Mutus}\ \emph {et~al.}(2014)\citenamefont {Mutus},
  \citenamefont {White}, \citenamefont {Barends}, \citenamefont {Chen},
  \citenamefont {Chen}, \citenamefont {Chiaro}, \citenamefont {Dunsworth},
  \citenamefont {Jeffrey}, \citenamefont {Kelly}, \citenamefont {Megrant},
  \citenamefont {Neill}, \citenamefont {O'Malley}, \citenamefont {Roushan},
  \citenamefont {Sank}, \citenamefont {Vainsencher}, \citenamefont {Wenner},
  \citenamefont {Sundqvist}, \citenamefont {Cleland},\ and\ \citenamefont
  {Martinis}}]{Mutus2014}%
  \BibitemOpen
  \bibfield  {author} {\bibinfo {author} {\bibfnamefont {J.~Y.}\ \bibnamefont
  {Mutus}}, \bibinfo {author} {\bibfnamefont {T.~C.}\ \bibnamefont {White}},
  \bibinfo {author} {\bibfnamefont {R.}~\bibnamefont {Barends}}, \bibinfo
  {author} {\bibfnamefont {Y.}~\bibnamefont {Chen}}, \bibinfo {author}
  {\bibfnamefont {Z.}~\bibnamefont {Chen}}, \bibinfo {author} {\bibfnamefont
  {B.}~\bibnamefont {Chiaro}}, \bibinfo {author} {\bibfnamefont
  {A.}~\bibnamefont {Dunsworth}}, \bibinfo {author} {\bibfnamefont
  {E.}~\bibnamefont {Jeffrey}}, \bibinfo {author} {\bibfnamefont
  {J.}~\bibnamefont {Kelly}}, \bibinfo {author} {\bibfnamefont
  {A.}~\bibnamefont {Megrant}}, \bibinfo {author} {\bibfnamefont
  {C.}~\bibnamefont {Neill}}, \bibinfo {author} {\bibfnamefont {P.~J.}\
  \bibnamefont {O'Malley}}, \bibinfo {author} {\bibfnamefont {P.}~\bibnamefont
  {Roushan}}, \bibinfo {author} {\bibfnamefont {D.}~\bibnamefont {Sank}},
  \bibinfo {author} {\bibfnamefont {A.}~\bibnamefont {Vainsencher}}, \bibinfo
  {author} {\bibfnamefont {J.}~\bibnamefont {Wenner}}, \bibinfo {author}
  {\bibfnamefont {K.~M.}\ \bibnamefont {Sundqvist}}, \bibinfo {author}
  {\bibfnamefont {A.~N.}\ \bibnamefont {Cleland}},\ and\ \bibinfo {author}
  {\bibfnamefont {J.~M.}\ \bibnamefont {Martinis}},\ }\bibfield  {title}
  {\bibinfo {title} {{Strong environmental coupling in a Josephson parametric
  amplifier}},\ }\bibfield  {journal} {\bibinfo  {journal} {Appl. Phys. Lett.}\
  }\textbf {\bibinfo {volume} {104}},\ \href
  {https://doi.org/10.1063/1.4886408} {10.1063/1.4886408} (\bibinfo {year}
  {2014})\BibitemShut {NoStop}%
\bibitem [{\citenamefont {Roy}\ \emph {et~al.}(2015)\citenamefont {Roy},
  \citenamefont {Kundu}, \citenamefont {Chand}, \citenamefont {Vadiraj},
  \citenamefont {Ranadive}, \citenamefont {Nehra}, \citenamefont {Patankar},
  \citenamefont {Aumentado}, \citenamefont {Clerk},\ and\ \citenamefont
  {Vijay}}]{Roy2015}%
  \BibitemOpen
  \bibfield  {author} {\bibinfo {author} {\bibfnamefont {T.}~\bibnamefont
  {Roy}}, \bibinfo {author} {\bibfnamefont {S.}~\bibnamefont {Kundu}}, \bibinfo
  {author} {\bibfnamefont {M.}~\bibnamefont {Chand}}, \bibinfo {author}
  {\bibfnamefont {A.~M.}\ \bibnamefont {Vadiraj}}, \bibinfo {author}
  {\bibfnamefont {A.}~\bibnamefont {Ranadive}}, \bibinfo {author}
  {\bibfnamefont {N.}~\bibnamefont {Nehra}}, \bibinfo {author} {\bibfnamefont
  {M.~P.}\ \bibnamefont {Patankar}}, \bibinfo {author} {\bibfnamefont
  {J.}~\bibnamefont {Aumentado}}, \bibinfo {author} {\bibfnamefont {A.~A.}\
  \bibnamefont {Clerk}},\ and\ \bibinfo {author} {\bibfnamefont
  {R.}~\bibnamefont {Vijay}},\ }\bibfield  {title} {\bibinfo {title}
  {{Broadband parametric amplification with impedance engineering: Beyond the
  gain-bandwidth product}},\ }\href {https://doi.org/10.1063/1.4939148}
  {\bibfield  {journal} {\bibinfo  {journal} {Appl. Phys. Lett.}\ }\textbf
  {\bibinfo {volume} {107}},\ \bibinfo {pages} {262601} (\bibinfo {year}
  {2015})}\BibitemShut {NoStop}%
\bibitem [{\citenamefont {Naaman}\ \emph {et~al.}(2019)\citenamefont {Naaman},
  \citenamefont {Ferguson}, \citenamefont {Marakov}, \citenamefont {Khalil},
  \citenamefont {Koehl},\ and\ \citenamefont {Epstein}}]{Naaman2019}%
  \BibitemOpen
  \bibfield  {author} {\bibinfo {author} {\bibfnamefont {O.}~\bibnamefont
  {Naaman}}, \bibinfo {author} {\bibfnamefont {D.~G.}\ \bibnamefont
  {Ferguson}}, \bibinfo {author} {\bibfnamefont {A.}~\bibnamefont {Marakov}},
  \bibinfo {author} {\bibfnamefont {M.}~\bibnamefont {Khalil}}, \bibinfo
  {author} {\bibfnamefont {W.~F.}\ \bibnamefont {Koehl}},\ and\ \bibinfo
  {author} {\bibfnamefont {R.~J.}\ \bibnamefont {Epstein}},\ }\bibfield
  {title} {\bibinfo {title} {{High Saturation Power Josephson Parametric
  Amplifier with GHz Bandwidth}},\ }in\ \href
  {https://doi.org/10.1109/MWSYM.2019.8701068} {\emph {\bibinfo {booktitle}
  {2019 IEEE MTT-S Int. Microw. Symp.}}},\ Vol.\ \bibinfo {volume} {2019-June}\
  (\bibinfo  {publisher} {IEEE},\ \bibinfo {year} {2019})\ pp.\ \bibinfo
  {pages} {259--262}\BibitemShut {NoStop}%
\bibitem [{\citenamefont {Macklin}\ \emph {et~al.}(2015)\citenamefont
  {Macklin}, \citenamefont {O'Brien}, \citenamefont {Hover}, \citenamefont
  {Schwartz}, \citenamefont {Bolkhovsky}, \citenamefont {Zhang}, \citenamefont
  {Oliver},\ and\ \citenamefont {Siddiqi}}]{Macklin2015}%
  \BibitemOpen
  \bibfield  {author} {\bibinfo {author} {\bibfnamefont {C.}~\bibnamefont
  {Macklin}}, \bibinfo {author} {\bibfnamefont {K.}~\bibnamefont {O'Brien}},
  \bibinfo {author} {\bibfnamefont {D.}~\bibnamefont {Hover}}, \bibinfo
  {author} {\bibfnamefont {M.~E.}\ \bibnamefont {Schwartz}}, \bibinfo {author}
  {\bibfnamefont {V.}~\bibnamefont {Bolkhovsky}}, \bibinfo {author}
  {\bibfnamefont {X.}~\bibnamefont {Zhang}}, \bibinfo {author} {\bibfnamefont
  {W.~D.}\ \bibnamefont {Oliver}},\ and\ \bibinfo {author} {\bibfnamefont
  {I.}~\bibnamefont {Siddiqi}},\ }\bibfield  {title} {\bibinfo {title} {{A
  near-quantum-limited Josephson traveling-wave parametric amplifier}},\ }\href
  {https://doi.org/10.1126/science.aaa8525} {\bibfield  {journal} {\bibinfo
  {journal} {Science}\ }\textbf {\bibinfo {volume} {350}},\ \bibinfo {pages}
  {307} (\bibinfo {year} {2015})}\BibitemShut {NoStop}%
\bibitem [{\citenamefont {Abdo}\ \emph {et~al.}(2013)\citenamefont {Abdo},
  \citenamefont {Sliwa}, \citenamefont {Frunzio},\ and\ \citenamefont
  {Devoret}}]{Abdo2013}%
  \BibitemOpen
  \bibfield  {author} {\bibinfo {author} {\bibfnamefont {B.}~\bibnamefont
  {Abdo}}, \bibinfo {author} {\bibfnamefont {K.}~\bibnamefont {Sliwa}},
  \bibinfo {author} {\bibfnamefont {L.}~\bibnamefont {Frunzio}},\ and\ \bibinfo
  {author} {\bibfnamefont {M.}~\bibnamefont {Devoret}},\ }\bibfield  {title}
  {\bibinfo {title} {{Directional Amplification with a Josephson Circuit}},\
  }\href {https://doi.org/10.1103/PhysRevX.3.031001} {\bibfield  {journal}
  {\bibinfo  {journal} {Phys. Rev. X}\ }\textbf {\bibinfo {volume} {3}},\
  \bibinfo {pages} {031001} (\bibinfo {year} {2013})}\BibitemShut {NoStop}%
\bibitem [{\citenamefont {Lecocq}\ \emph {et~al.}(2017)\citenamefont {Lecocq},
  \citenamefont {Ranzani}, \citenamefont {Peterson}, \citenamefont {Cicak},
  \citenamefont {Simmonds}, \citenamefont {Teufel},\ and\ \citenamefont
  {Aumentado}}]{Lecocq2017}%
  \BibitemOpen
  \bibfield  {author} {\bibinfo {author} {\bibfnamefont {F.}~\bibnamefont
  {Lecocq}}, \bibinfo {author} {\bibfnamefont {L.}~\bibnamefont {Ranzani}},
  \bibinfo {author} {\bibfnamefont {G.~A.}\ \bibnamefont {Peterson}}, \bibinfo
  {author} {\bibfnamefont {K.}~\bibnamefont {Cicak}}, \bibinfo {author}
  {\bibfnamefont {R.~W.}\ \bibnamefont {Simmonds}}, \bibinfo {author}
  {\bibfnamefont {J.~D.}\ \bibnamefont {Teufel}},\ and\ \bibinfo {author}
  {\bibfnamefont {J.}~\bibnamefont {Aumentado}},\ }\bibfield  {title} {\bibinfo
  {title} {{Nonreciprocal Microwave Signal Processing with a Field-Programmable
  Josephson Amplifier}},\ }\href
  {https://doi.org/10.1103/PhysRevApplied.7.024028} {\bibfield  {journal}
  {\bibinfo  {journal} {Phys. Rev. Appl.}\ }\textbf {\bibinfo {volume} {7}},\
  \bibinfo {pages} {024028} (\bibinfo {year} {2017})}\BibitemShut {NoStop}%
\bibitem [{\citenamefont {Ranzani}\ and\ \citenamefont
  {Aumentado}(2019)}]{Ranzani2019}%
  \BibitemOpen
  \bibfield  {author} {\bibinfo {author} {\bibfnamefont {L.}~\bibnamefont
  {Ranzani}}\ and\ \bibinfo {author} {\bibfnamefont {J.}~\bibnamefont
  {Aumentado}},\ }\bibfield  {title} {\bibinfo {title} {{Circulators at the
  Quantum Limit: Recent Realizations of Quantum-Limited Superconducting
  Circulators and Related Approaches}},\ }\href
  {https://doi.org/10.1109/MMM.2019.2891381} {\bibfield  {journal} {\bibinfo
  {journal} {IEEE Microw. Mag.}\ }\textbf {\bibinfo {volume} {20}},\ \bibinfo
  {pages} {112} (\bibinfo {year} {2019})}\BibitemShut {NoStop}%
\bibitem [{\citenamefont {Mahoney}\ \emph
  {et~al.}(2017{\natexlab{a}})\citenamefont {Mahoney}, \citenamefont {Colless},
  \citenamefont {Pauka}, \citenamefont {Hornibrook}, \citenamefont {Watson},
  \citenamefont {Gardner}, \citenamefont {Manfra}, \citenamefont {Doherty},\
  and\ \citenamefont {Reilly}}]{Mahoney2017}%
  \BibitemOpen
  \bibfield  {author} {\bibinfo {author} {\bibfnamefont {A.~C.}\ \bibnamefont
  {Mahoney}}, \bibinfo {author} {\bibfnamefont {J.~I.}\ \bibnamefont
  {Colless}}, \bibinfo {author} {\bibfnamefont {S.~J.}\ \bibnamefont {Pauka}},
  \bibinfo {author} {\bibfnamefont {J.~M.}\ \bibnamefont {Hornibrook}},
  \bibinfo {author} {\bibfnamefont {J.~D.}\ \bibnamefont {Watson}}, \bibinfo
  {author} {\bibfnamefont {G.~C.}\ \bibnamefont {Gardner}}, \bibinfo {author}
  {\bibfnamefont {M.~J.}\ \bibnamefont {Manfra}}, \bibinfo {author}
  {\bibfnamefont {A.~C.}\ \bibnamefont {Doherty}},\ and\ \bibinfo {author}
  {\bibfnamefont {D.~J.}\ \bibnamefont {Reilly}},\ }\bibfield  {title}
  {\bibinfo {title} {{On-Chip Microwave Quantum Hall Circulator}},\ }\href
  {https://doi.org/10.1103/PhysRevX.7.011007} {\bibfield  {journal} {\bibinfo
  {journal} {Phys. Rev. X}\ }\textbf {\bibinfo {volume} {7}},\ \bibinfo {pages}
  {011007} (\bibinfo {year} {2017}{\natexlab{a}})}\BibitemShut {NoStop}%
\bibitem [{\citenamefont {Sliwa}\ \emph {et~al.}(2015)\citenamefont {Sliwa},
  \citenamefont {Hatridge}, \citenamefont {Narla}, \citenamefont {Shankar},
  \citenamefont {Frunzio}, \citenamefont {Schoelkopf},\ and\ \citenamefont
  {Devoret}}]{Sliwa2015}%
  \BibitemOpen
  \bibfield  {author} {\bibinfo {author} {\bibfnamefont {K.~M.}\ \bibnamefont
  {Sliwa}}, \bibinfo {author} {\bibfnamefont {M.}~\bibnamefont {Hatridge}},
  \bibinfo {author} {\bibfnamefont {A.}~\bibnamefont {Narla}}, \bibinfo
  {author} {\bibfnamefont {S.}~\bibnamefont {Shankar}}, \bibinfo {author}
  {\bibfnamefont {L.}~\bibnamefont {Frunzio}}, \bibinfo {author} {\bibfnamefont
  {R.~J.}\ \bibnamefont {Schoelkopf}},\ and\ \bibinfo {author} {\bibfnamefont
  {M.~H.}\ \bibnamefont {Devoret}},\ }\bibfield  {title} {\bibinfo {title}
  {{Reconfigurable Josephson Circulator / Directional Amplifier}},\ }\href
  {https://doi.org/10.1103/PhysRevX.5.041020} {\bibfield  {journal} {\bibinfo
  {journal} {Phys. Rev. X}\ }\textbf {\bibinfo {volume} {5}},\ \bibinfo {pages}
  {041020} (\bibinfo {year} {2015})}\BibitemShut {NoStop}%
\bibitem [{\citenamefont {Lecocq}\ \emph
  {et~al.}(2020{\natexlab{a}})\citenamefont {Lecocq}, \citenamefont {Ranzani},
  \citenamefont {Peterson}, \citenamefont {Cicak}, \citenamefont {Metelmann},
  \citenamefont {Kotler}, \citenamefont {Simmonds}, \citenamefont {Teufel},\
  and\ \citenamefont {Aumentado}}]{Lecocq2020b}%
  \BibitemOpen
  \bibfield  {author} {\bibinfo {author} {\bibfnamefont {F.}~\bibnamefont
  {Lecocq}}, \bibinfo {author} {\bibfnamefont {L.}~\bibnamefont {Ranzani}},
  \bibinfo {author} {\bibfnamefont {G.}~\bibnamefont {Peterson}}, \bibinfo
  {author} {\bibfnamefont {K.}~\bibnamefont {Cicak}}, \bibinfo {author}
  {\bibfnamefont {A.}~\bibnamefont {Metelmann}}, \bibinfo {author}
  {\bibfnamefont {S.}~\bibnamefont {Kotler}}, \bibinfo {author} {\bibfnamefont
  {R.}~\bibnamefont {Simmonds}}, \bibinfo {author} {\bibfnamefont
  {J.}~\bibnamefont {Teufel}},\ and\ \bibinfo {author} {\bibfnamefont
  {J.}~\bibnamefont {Aumentado}},\ }\bibfield  {title} {\bibinfo {title}
  {{Microwave Measurement beyond the Quantum Limit with a Nonreciprocal
  Amplifier}},\ }\href {https://doi.org/10.1103/PhysRevApplied.13.044005}
  {\bibfield  {journal} {\bibinfo  {journal} {Phys. Rev. Appl.}\ }\textbf
  {\bibinfo {volume} {13}},\ \bibinfo {pages} {044005} (\bibinfo {year}
  {2020}{\natexlab{a}})}\BibitemShut {NoStop}%
\bibitem [{\citenamefont {Abdo}\ \emph {et~al.}(2020)\citenamefont {Abdo},
  \citenamefont {Jinka}, \citenamefont {Bronn}, \citenamefont {Olivadese},\
  and\ \citenamefont {Brink}}]{Abdo2020}%
  \BibitemOpen
  \bibfield  {author} {\bibinfo {author} {\bibfnamefont {B.}~\bibnamefont
  {Abdo}}, \bibinfo {author} {\bibfnamefont {O.}~\bibnamefont {Jinka}},
  \bibinfo {author} {\bibfnamefont {N.~T.}\ \bibnamefont {Bronn}}, \bibinfo
  {author} {\bibfnamefont {S.}~\bibnamefont {Olivadese}},\ and\ \bibinfo
  {author} {\bibfnamefont {M.}~\bibnamefont {Brink}},\ }\bibfield  {title}
  {\bibinfo {title} {{On-chip single-pump interferometric Josephson isolator
  for quantum measurements}},\ }\href@noop {} {\bibfield  {journal} {\bibinfo
  {journal} {arXiv:2006.01918}\ } (\bibinfo {year} {2020})}\BibitemShut
  {NoStop}%
\bibitem [{\citenamefont {Rosenthal}\ \emph {et~al.}(2020)\citenamefont
  {Rosenthal}, \citenamefont {Schneider}, \citenamefont {Malnou}, \citenamefont
  {Zhao}, \citenamefont {Leditzky}, \citenamefont {Chapman}, \citenamefont
  {Wustmann}, \citenamefont {Ma}, \citenamefont {Palken}, \citenamefont
  {Zanner}, \citenamefont {Vale}, \citenamefont {Hilton}, \citenamefont {Gao},
  \citenamefont {Smith}, \citenamefont {Kirchmair},\ and\ \citenamefont
  {Lehnert}}]{Rosenthal2020}%
  \BibitemOpen
  \bibfield  {author} {\bibinfo {author} {\bibfnamefont {E.~I.}\ \bibnamefont
  {Rosenthal}}, \bibinfo {author} {\bibfnamefont {C.~M.~F.}\ \bibnamefont
  {Schneider}}, \bibinfo {author} {\bibfnamefont {M.}~\bibnamefont {Malnou}},
  \bibinfo {author} {\bibfnamefont {Z.}~\bibnamefont {Zhao}}, \bibinfo {author}
  {\bibfnamefont {F.}~\bibnamefont {Leditzky}}, \bibinfo {author}
  {\bibfnamefont {B.~J.}\ \bibnamefont {Chapman}}, \bibinfo {author}
  {\bibfnamefont {W.}~\bibnamefont {Wustmann}}, \bibinfo {author}
  {\bibfnamefont {X.}~\bibnamefont {Ma}}, \bibinfo {author} {\bibfnamefont
  {D.~A.}\ \bibnamefont {Palken}}, \bibinfo {author} {\bibfnamefont {M.~F.}\
  \bibnamefont {Zanner}}, \bibinfo {author} {\bibfnamefont {L.~R.}\
  \bibnamefont {Vale}}, \bibinfo {author} {\bibfnamefont {G.~C.}\ \bibnamefont
  {Hilton}}, \bibinfo {author} {\bibfnamefont {J.}~\bibnamefont {Gao}},
  \bibinfo {author} {\bibfnamefont {G.}~\bibnamefont {Smith}}, \bibinfo
  {author} {\bibfnamefont {G.}~\bibnamefont {Kirchmair}},\ and\ \bibinfo
  {author} {\bibfnamefont {K.~W.}\ \bibnamefont {Lehnert}},\ }\bibfield
  {title} {\bibinfo {title} {{Efficient and low-backaction quantum measurement
  using a chip-scale detector}},\ }\href@noop {} {\bibfield  {journal}
  {\bibinfo  {journal} {arXiv:2008.03805}\ } (\bibinfo {year}
  {2020})}\BibitemShut {NoStop}%
\bibitem [{\citenamefont {Lecocq}\ \emph
  {et~al.}(2020{\natexlab{b}})\citenamefont {Lecocq}, \citenamefont {Ranzani},
  \citenamefont {Peterson}, \citenamefont {Cicak}, \citenamefont {Jin},
  \citenamefont {Simmonds}, \citenamefont {Teufel},\ and\ \citenamefont
  {Aumentado}}]{Lecocq2020a}%
  \BibitemOpen
  \bibfield  {author} {\bibinfo {author} {\bibfnamefont {F.}~\bibnamefont
  {Lecocq}}, \bibinfo {author} {\bibfnamefont {L.}~\bibnamefont {Ranzani}},
  \bibinfo {author} {\bibfnamefont {G.~A.}\ \bibnamefont {Peterson}}, \bibinfo
  {author} {\bibfnamefont {K.}~\bibnamefont {Cicak}}, \bibinfo {author}
  {\bibfnamefont {X.~Y.}\ \bibnamefont {Jin}}, \bibinfo {author} {\bibfnamefont
  {R.~W.}\ \bibnamefont {Simmonds}}, \bibinfo {author} {\bibfnamefont {J.~D.}\
  \bibnamefont {Teufel}},\ and\ \bibinfo {author} {\bibfnamefont
  {J.}~\bibnamefont {Aumentado}},\ }\bibfield  {title} {\bibinfo {title}
  {{Efficient qubit measurement with a nonreciprocal microwave amplifier}},\
  }\href@noop {} {\bibfield  {journal} {\bibinfo  {journal} {arXiv:2009.08863}\
  } (\bibinfo {year} {2020}{\natexlab{b}})}\BibitemShut {NoStop}%
\bibitem [{\citenamefont {Bosco}\ \emph {et~al.}(2017)\citenamefont {Bosco},
  \citenamefont {Haupt},\ and\ \citenamefont {DiVincenzo}}]{Bosco2017}%
  \BibitemOpen
  \bibfield  {author} {\bibinfo {author} {\bibfnamefont {S.}~\bibnamefont
  {Bosco}}, \bibinfo {author} {\bibfnamefont {F.}~\bibnamefont {Haupt}},\ and\
  \bibinfo {author} {\bibfnamefont {D.~P.}\ \bibnamefont {DiVincenzo}},\
  }\bibfield  {title} {\bibinfo {title} {Self-impedance-matched hall-effect
  gyrators and circulators},\ }\href
  {https://doi.org/10.1103/PhysRevApplied.7.024030} {\bibfield  {journal}
  {\bibinfo  {journal} {Phys. Rev. Applied}\ }\textbf {\bibinfo {volume} {7}},\
  \bibinfo {pages} {024030} (\bibinfo {year} {2017})}\BibitemShut {NoStop}%
\bibitem [{\citenamefont {Mahoney}\ \emph
  {et~al.}(2017{\natexlab{b}})\citenamefont {Mahoney}, \citenamefont {Colless},
  \citenamefont {Peeters}, \citenamefont {Pauka}, \citenamefont {Fox},
  \citenamefont {Kou}, \citenamefont {Pan}, \citenamefont {Wang}, \citenamefont
  {Goldhaber-Gordon},\ and\ \citenamefont {Reilly}}]{MahoneyTI17}%
  \BibitemOpen
  \bibfield  {author} {\bibinfo {author} {\bibfnamefont {A.~C.}\ \bibnamefont
  {Mahoney}}, \bibinfo {author} {\bibfnamefont {J.~I.}\ \bibnamefont
  {Colless}}, \bibinfo {author} {\bibfnamefont {L.}~\bibnamefont {Peeters}},
  \bibinfo {author} {\bibfnamefont {S.~J.}\ \bibnamefont {Pauka}}, \bibinfo
  {author} {\bibfnamefont {E.~J.}\ \bibnamefont {Fox}}, \bibinfo {author}
  {\bibfnamefont {X.}~\bibnamefont {Kou}}, \bibinfo {author} {\bibfnamefont
  {L.}~\bibnamefont {Pan}}, \bibinfo {author} {\bibfnamefont {K.~L.}\
  \bibnamefont {Wang}}, \bibinfo {author} {\bibfnamefont {D.}~\bibnamefont
  {Goldhaber-Gordon}},\ and\ \bibinfo {author} {\bibfnamefont {D.~J.}\
  \bibnamefont {Reilly}},\ }\bibfield  {title} {\bibinfo {title} {Zero-field
  edge plasmons in a magnetic topological insulator},\ }\href
  {https://doi.org/10.1038/s41467-017-01984-5} {\bibfield  {journal} {\bibinfo
  {journal} {Nat. Commun.}\ }\textbf {\bibinfo {volume} {8}},\ \bibinfo {pages}
  {1836} (\bibinfo {year} {2017}{\natexlab{b}})}\BibitemShut {NoStop}%
\bibitem [{\citenamefont {Bosco}\ \emph {et~al.}(2019)\citenamefont {Bosco},
  \citenamefont {DiVincenzo},\ and\ \citenamefont {Reilly}}]{Bosco}%
  \BibitemOpen
  \bibfield  {author} {\bibinfo {author} {\bibfnamefont {S.}~\bibnamefont
  {Bosco}}, \bibinfo {author} {\bibfnamefont {D.}~\bibnamefont {DiVincenzo}},\
  and\ \bibinfo {author} {\bibfnamefont {D.}~\bibnamefont {Reilly}},\
  }\bibfield  {title} {\bibinfo {title} {Transmission lines and metamaterials
  based on quantum hall plasmonics},\ }\href
  {https://doi.org/10.1103/PhysRevApplied.12.014030} {\bibfield  {journal}
  {\bibinfo  {journal} {Phys. Rev. Applied}\ }\textbf {\bibinfo {volume}
  {12}},\ \bibinfo {pages} {014030} (\bibinfo {year} {2019})}\BibitemShut
  {NoStop}%
\bibitem [{\citenamefont {Herr}\ \emph {et~al.}(2002)\citenamefont {Herr},
  \citenamefont {Smith},\ and\ \citenamefont {Wire}}]{herr2002high}%
  \BibitemOpen
  \bibfield  {author} {\bibinfo {author} {\bibfnamefont {Q.~P.}\ \bibnamefont
  {Herr}}, \bibinfo {author} {\bibfnamefont {A.~D.}\ \bibnamefont {Smith}},\
  and\ \bibinfo {author} {\bibfnamefont {M.~S.}\ \bibnamefont {Wire}},\
  }\bibfield  {title} {\bibinfo {title} {High speed data link between digital
  superconductor chips},\ }\href@noop {} {\bibfield  {journal} {\bibinfo
  {journal} {Applied physics letters}\ }\textbf {\bibinfo {volume} {80}},\
  \bibinfo {pages} {3210} (\bibinfo {year} {2002})}\BibitemShut {NoStop}%
\bibitem [{\citenamefont {Franke}\ \emph {et~al.}(2019)\citenamefont {Franke},
  \citenamefont {Clarke}, \citenamefont {Vandersypen},\ and\ \citenamefont
  {Veldhorst}}]{franke2019rent}%
  \BibitemOpen
  \bibfield  {author} {\bibinfo {author} {\bibfnamefont {D.~P.}\ \bibnamefont
  {Franke}}, \bibinfo {author} {\bibfnamefont {J.~S.}\ \bibnamefont {Clarke}},
  \bibinfo {author} {\bibfnamefont {L.~M.~K.}\ \bibnamefont {Vandersypen}},\
  and\ \bibinfo {author} {\bibfnamefont {M.}~\bibnamefont {Veldhorst}},\
  }\bibfield  {title} {\bibinfo {title} {Rent’s rule and extensibility in
  quantum computing},\ }\href@noop {} {\bibfield  {journal} {\bibinfo
  {journal} {Microprocessors and Microsystems}\ }\textbf {\bibinfo {volume}
  {67}},\ \bibinfo {pages} {1} (\bibinfo {year} {2019})}\BibitemShut {NoStop}%
\bibitem [{\citenamefont {Hornibrook}\ \emph {et~al.}(2015)\citenamefont
  {Hornibrook}, \citenamefont {Colless}, \citenamefont {Lamb}, \citenamefont
  {Pauka}, \citenamefont {Lu}, \citenamefont {Gossard}, \citenamefont {Watson},
  \citenamefont {Gardner}, \citenamefont {Fallahi}, \citenamefont {Manfra}
  \emph {et~al.}}]{Hornibrook2015}%
  \BibitemOpen
  \bibfield  {author} {\bibinfo {author} {\bibfnamefont {J.}~\bibnamefont
  {Hornibrook}}, \bibinfo {author} {\bibfnamefont {J.}~\bibnamefont {Colless}},
  \bibinfo {author} {\bibfnamefont {I.~C.}\ \bibnamefont {Lamb}}, \bibinfo
  {author} {\bibfnamefont {S.}~\bibnamefont {Pauka}}, \bibinfo {author}
  {\bibfnamefont {H.}~\bibnamefont {Lu}}, \bibinfo {author} {\bibfnamefont
  {A.}~\bibnamefont {Gossard}}, \bibinfo {author} {\bibfnamefont
  {J.}~\bibnamefont {Watson}}, \bibinfo {author} {\bibfnamefont
  {G.}~\bibnamefont {Gardner}}, \bibinfo {author} {\bibfnamefont
  {S.}~\bibnamefont {Fallahi}}, \bibinfo {author} {\bibfnamefont
  {M.}~\bibnamefont {Manfra}}, \emph {et~al.},\ }\bibfield  {title} {\bibinfo
  {title} {Cryogenic control architecture for large-scale quantum computing},\
  }\href@noop {} {\bibfield  {journal} {\bibinfo  {journal} {Phys. Rev. Appl.}\
  }\textbf {\bibinfo {volume} {3}},\ \bibinfo {pages} {024010} (\bibinfo {year}
  {2015})}\BibitemShut {NoStop}%
\bibitem [{\citenamefont {Pauka}\ \emph {et~al.}(2020)\citenamefont {Pauka},
  \citenamefont {Das}, \citenamefont {Hornibrook}, \citenamefont {Gardner},
  \citenamefont {Manfra}, \citenamefont {Cassidy},\ and\ \citenamefont
  {Reilly}}]{PaukaPRApp}%
  \BibitemOpen
  \bibfield  {author} {\bibinfo {author} {\bibfnamefont {S.}~\bibnamefont
  {Pauka}}, \bibinfo {author} {\bibfnamefont {K.}~\bibnamefont {Das}}, \bibinfo
  {author} {\bibfnamefont {J.}~\bibnamefont {Hornibrook}}, \bibinfo {author}
  {\bibfnamefont {G.}~\bibnamefont {Gardner}}, \bibinfo {author} {\bibfnamefont
  {M.}~\bibnamefont {Manfra}}, \bibinfo {author} {\bibfnamefont
  {M.}~\bibnamefont {Cassidy}},\ and\ \bibinfo {author} {\bibfnamefont
  {D.}~\bibnamefont {Reilly}},\ }\bibfield  {title} {\bibinfo {title}
  {Characterizing quantum devices at scale with custom cryo-cmos},\ }\href@noop
  {} {\bibfield  {journal} {\bibinfo  {journal} {Phys. Rev. Appl.}\ }\textbf
  {\bibinfo {volume} {13}},\ \bibinfo {pages} {054072} (\bibinfo {year}
  {2020})}\BibitemShut {NoStop}%
\bibitem [{\citenamefont {Reilly}(2019)}]{Reilly_IEDM}%
  \BibitemOpen
  \bibfield  {author} {\bibinfo {author} {\bibfnamefont {D.}~\bibnamefont
  {Reilly}},\ }\bibfield  {title} {\bibinfo {title} {Challenges in scaling-up
  the control interface of a quantum computer},\ }in\ \href@noop {} {\emph
  {\bibinfo {booktitle} {2019 IEEE International Electron Devices Meeting
  (IEDM)}}}\ (\bibinfo {organization} {IEEE},\ \bibinfo {year} {2019})\ pp.\
  \bibinfo {pages} {31--7}\BibitemShut {NoStop}%
\bibitem [{\citenamefont {Pauka}\ \emph {et~al.}(2019)\citenamefont {Pauka},
  \citenamefont {Das}, \citenamefont {Kalra}, \citenamefont {Moini},
  \citenamefont {Yang}, \citenamefont {Trainer}, \citenamefont {Bousquet},
  \citenamefont {Cantaloube}, \citenamefont {Dick}, \citenamefont {Gardner}
  \emph {et~al.}}]{Pauka_arxiv}%
  \BibitemOpen
  \bibfield  {author} {\bibinfo {author} {\bibfnamefont {S.}~\bibnamefont
  {Pauka}}, \bibinfo {author} {\bibfnamefont {K.}~\bibnamefont {Das}}, \bibinfo
  {author} {\bibfnamefont {R.}~\bibnamefont {Kalra}}, \bibinfo {author}
  {\bibfnamefont {A.}~\bibnamefont {Moini}}, \bibinfo {author} {\bibfnamefont
  {Y.}~\bibnamefont {Yang}}, \bibinfo {author} {\bibfnamefont {M.}~\bibnamefont
  {Trainer}}, \bibinfo {author} {\bibfnamefont {A.}~\bibnamefont {Bousquet}},
  \bibinfo {author} {\bibfnamefont {C.}~\bibnamefont {Cantaloube}}, \bibinfo
  {author} {\bibfnamefont {N.}~\bibnamefont {Dick}}, \bibinfo {author}
  {\bibfnamefont {G.}~\bibnamefont {Gardner}}, \emph {et~al.},\ }\bibfield
  {title} {\bibinfo {title} {A cryogenic interface for controlling many
  qubits},\ }\href@noop {} {\bibfield  {journal} {\bibinfo  {journal}
  {arXiv:1912.01299}\ } (\bibinfo {year} {2019})}\BibitemShut {NoStop}%
\bibitem [{\citenamefont {Stuart}\ \emph {et~al.}(2019)\citenamefont {Stuart},
  \citenamefont {Panock}, \citenamefont {Bruzewicz}, \citenamefont {Sedlacek},
  \citenamefont {McConnell}, \citenamefont {Chuang}, \citenamefont {Sage},\
  and\ \citenamefont {Chiaverini}}]{Stuart2019}%
  \BibitemOpen
  \bibfield  {author} {\bibinfo {author} {\bibfnamefont {J.}~\bibnamefont
  {Stuart}}, \bibinfo {author} {\bibfnamefont {R.}~\bibnamefont {Panock}},
  \bibinfo {author} {\bibfnamefont {C.}~\bibnamefont {Bruzewicz}}, \bibinfo
  {author} {\bibfnamefont {J.}~\bibnamefont {Sedlacek}}, \bibinfo {author}
  {\bibfnamefont {R.}~\bibnamefont {McConnell}}, \bibinfo {author}
  {\bibfnamefont {I.}~\bibnamefont {Chuang}}, \bibinfo {author} {\bibfnamefont
  {J.}~\bibnamefont {Sage}},\ and\ \bibinfo {author} {\bibfnamefont
  {J.}~\bibnamefont {Chiaverini}},\ }\bibfield  {title} {\bibinfo {title}
  {{Chip-Integrated Voltage Sources for Control of Trapped Ions}},\ }\href
  {https://doi.org/10.1103/PhysRevApplied.11.024010} {\bibfield  {journal}
  {\bibinfo  {journal} {Phys. Rev. Appl.}\ }\textbf {\bibinfo {volume} {11}},\
  \bibinfo {pages} {024010} (\bibinfo {year} {2019})}\BibitemShut {NoStop}%
\bibitem [{\citenamefont {Patra}\ \emph {et~al.}(2017)\citenamefont {Patra},
  \citenamefont {Incandela}, \citenamefont {Van~Dijk}, \citenamefont {Homulle},
  \citenamefont {Song}, \citenamefont {Shahmohammadi}, \citenamefont
  {Staszewski}, \citenamefont {Vladimirescu}, \citenamefont {Babaie},
  \citenamefont {Sebastiano} \emph {et~al.}}]{patra2017cryo}%
  \BibitemOpen
  \bibfield  {author} {\bibinfo {author} {\bibfnamefont {B.}~\bibnamefont
  {Patra}}, \bibinfo {author} {\bibfnamefont {R.~M.}\ \bibnamefont
  {Incandela}}, \bibinfo {author} {\bibfnamefont {J.~P.}\ \bibnamefont
  {Van~Dijk}}, \bibinfo {author} {\bibfnamefont {H.~A.}\ \bibnamefont
  {Homulle}}, \bibinfo {author} {\bibfnamefont {L.}~\bibnamefont {Song}},
  \bibinfo {author} {\bibfnamefont {M.}~\bibnamefont {Shahmohammadi}}, \bibinfo
  {author} {\bibfnamefont {R.~B.}\ \bibnamefont {Staszewski}}, \bibinfo
  {author} {\bibfnamefont {A.}~\bibnamefont {Vladimirescu}}, \bibinfo {author}
  {\bibfnamefont {M.}~\bibnamefont {Babaie}}, \bibinfo {author} {\bibfnamefont
  {F.}~\bibnamefont {Sebastiano}}, \emph {et~al.},\ }\bibfield  {title}
  {\bibinfo {title} {Cryo-cmos circuits and systems for quantum computing
  applications},\ }\href@noop {} {\bibfield  {journal} {\bibinfo  {journal}
  {IEEE Journal of Solid-State Circuits}\ }\textbf {\bibinfo {volume} {53}},\
  \bibinfo {pages} {309} (\bibinfo {year} {2017})}\BibitemShut {NoStop}%
\bibitem [{\citenamefont {Patra}\ \emph {et~al.}(2020)\citenamefont {Patra},
  \citenamefont {Van~Dijk}, \citenamefont {Corna}, \citenamefont {Xue},
  \citenamefont {Samkharadze}, \citenamefont {Sammak}, \citenamefont
  {Scappucci}, \citenamefont {Veldhorst}, \citenamefont {Vandersypen},
  \citenamefont {Babaie} \emph {et~al.}}]{patra2020scalable}%
  \BibitemOpen
  \bibfield  {author} {\bibinfo {author} {\bibfnamefont {B.}~\bibnamefont
  {Patra}}, \bibinfo {author} {\bibfnamefont {J.~P.}\ \bibnamefont {Van~Dijk}},
  \bibinfo {author} {\bibfnamefont {A.}~\bibnamefont {Corna}}, \bibinfo
  {author} {\bibfnamefont {X.}~\bibnamefont {Xue}}, \bibinfo {author}
  {\bibfnamefont {N.}~\bibnamefont {Samkharadze}}, \bibinfo {author}
  {\bibfnamefont {A.}~\bibnamefont {Sammak}}, \bibinfo {author} {\bibfnamefont
  {G.}~\bibnamefont {Scappucci}}, \bibinfo {author} {\bibfnamefont
  {M.}~\bibnamefont {Veldhorst}}, \bibinfo {author} {\bibfnamefont {L.~M.}\
  \bibnamefont {Vandersypen}}, \bibinfo {author} {\bibfnamefont
  {M.}~\bibnamefont {Babaie}}, \emph {et~al.},\ }\bibfield  {title} {\bibinfo
  {title} {A scalable cryo-{CMOS} 2-to-20 {GHz} digitally-intensive controller
  for 4$\times$ 32 frequency multiplexed spin qubits/transmons in 22-nm
  {FinFET} technology for quantum computers},\ }in\ \href@noop {} {\emph
  {\bibinfo {booktitle} {2020 International Solid-State Circuits Conference}}}\
  (\bibinfo {year} {2020})\BibitemShut {NoStop}%
\bibitem [{\citenamefont {Bardin}\ \emph
  {et~al.}(2019{\natexlab{a}})\citenamefont {Bardin}, \citenamefont {Jeffrey},
  \citenamefont {Lucero}, \citenamefont {Huang}, \citenamefont {Naaman},
  \citenamefont {Barends}, \citenamefont {White}, \citenamefont {Giustina},
  \citenamefont {Sank}, \citenamefont {Roushan} \emph {et~al.}}]{bardin201929}%
  \BibitemOpen
  \bibfield  {author} {\bibinfo {author} {\bibfnamefont {J.~C.}\ \bibnamefont
  {Bardin}}, \bibinfo {author} {\bibfnamefont {E.}~\bibnamefont {Jeffrey}},
  \bibinfo {author} {\bibfnamefont {E.}~\bibnamefont {Lucero}}, \bibinfo
  {author} {\bibfnamefont {T.}~\bibnamefont {Huang}}, \bibinfo {author}
  {\bibfnamefont {O.}~\bibnamefont {Naaman}}, \bibinfo {author} {\bibfnamefont
  {R.}~\bibnamefont {Barends}}, \bibinfo {author} {\bibfnamefont
  {T.}~\bibnamefont {White}}, \bibinfo {author} {\bibfnamefont
  {M.}~\bibnamefont {Giustina}}, \bibinfo {author} {\bibfnamefont
  {D.}~\bibnamefont {Sank}}, \bibinfo {author} {\bibfnamefont {P.}~\bibnamefont
  {Roushan}}, \emph {et~al.},\ }\bibfield  {title} {\bibinfo {title} {{A 28 nm
  Bulk-CMOS 4-to-8 GHz 2 mW cryogenic pulse modulator for scalable quantum
  computing}},\ }in\ \href@noop {} {\emph {\bibinfo {booktitle} {2019 IEEE
  International Solid-State Circuits Conference-(ISSCC)}}}\ (\bibinfo
  {organization} {IEEE},\ \bibinfo {year} {2019})\ pp.\ \bibinfo {pages}
  {456--458}\BibitemShut {NoStop}%
\bibitem [{\citenamefont {Bardin}\ \emph
  {et~al.}(2019{\natexlab{b}})\citenamefont {Bardin}, \citenamefont {Jeffrey},
  \citenamefont {Lucero}, \citenamefont {Huang}, \citenamefont {Das},
  \citenamefont {Sank}, \citenamefont {Naaman}, \citenamefont {Megrant},
  \citenamefont {Barends}, \citenamefont {White} \emph
  {et~al.}}]{bardin2019design}%
  \BibitemOpen
  \bibfield  {author} {\bibinfo {author} {\bibfnamefont {J.~C.}\ \bibnamefont
  {Bardin}}, \bibinfo {author} {\bibfnamefont {E.}~\bibnamefont {Jeffrey}},
  \bibinfo {author} {\bibfnamefont {E.}~\bibnamefont {Lucero}}, \bibinfo
  {author} {\bibfnamefont {T.}~\bibnamefont {Huang}}, \bibinfo {author}
  {\bibfnamefont {S.}~\bibnamefont {Das}}, \bibinfo {author} {\bibfnamefont
  {D.~T.}\ \bibnamefont {Sank}}, \bibinfo {author} {\bibfnamefont
  {O.}~\bibnamefont {Naaman}}, \bibinfo {author} {\bibfnamefont {A.~E.}\
  \bibnamefont {Megrant}}, \bibinfo {author} {\bibfnamefont {R.}~\bibnamefont
  {Barends}}, \bibinfo {author} {\bibfnamefont {T.}~\bibnamefont {White}},
  \emph {et~al.},\ }\bibfield  {title} {\bibinfo {title} {{Design and
  characterization of a 28-nm bulk-CMOS cryogenic quantum controller
  dissipating less than 2 mW at 3 K}},\ }\href@noop {} {\bibfield  {journal}
  {\bibinfo  {journal} {IEEE Journal of Solid-State Circuits}\ }\textbf
  {\bibinfo {volume} {54}},\ \bibinfo {pages} {3043} (\bibinfo {year}
  {2019}{\natexlab{b}})}\BibitemShut {NoStop}%
\bibitem [{\citenamefont {Bashir}\ \emph {et~al.}(2019)\citenamefont {Bashir},
  \citenamefont {Asker}, \citenamefont {Cetintepe}, \citenamefont {Leipold},
  \citenamefont {Esmailiyan}, \citenamefont {Wang}, \citenamefont
  {Siriburanon}, \citenamefont {Giounanlis}, \citenamefont {Blokhina},
  \citenamefont {Pomorski} \emph {et~al.}}]{bashir2019mixed}%
  \BibitemOpen
  \bibfield  {author} {\bibinfo {author} {\bibfnamefont {I.}~\bibnamefont
  {Bashir}}, \bibinfo {author} {\bibfnamefont {M.}~\bibnamefont {Asker}},
  \bibinfo {author} {\bibfnamefont {C.}~\bibnamefont {Cetintepe}}, \bibinfo
  {author} {\bibfnamefont {D.}~\bibnamefont {Leipold}}, \bibinfo {author}
  {\bibfnamefont {A.}~\bibnamefont {Esmailiyan}}, \bibinfo {author}
  {\bibfnamefont {H.}~\bibnamefont {Wang}}, \bibinfo {author} {\bibfnamefont
  {T.}~\bibnamefont {Siriburanon}}, \bibinfo {author} {\bibfnamefont
  {P.}~\bibnamefont {Giounanlis}}, \bibinfo {author} {\bibfnamefont
  {E.}~\bibnamefont {Blokhina}}, \bibinfo {author} {\bibfnamefont
  {K.}~\bibnamefont {Pomorski}}, \emph {et~al.},\ }\bibfield  {title} {\bibinfo
  {title} {A mixed-signal control core for a fully integrated semiconductor
  quantum computer system-on-chip},\ }in\ \href@noop {} {\emph {\bibinfo
  {booktitle} {ESSCIRC 2019-IEEE 45th European Solid State Circuits Conference
  (ESSCIRC)}}}\ (\bibinfo {organization} {IEEE},\ \bibinfo {year} {2019})\ pp.\
  \bibinfo {pages} {125--128}\BibitemShut {NoStop}%
\bibitem [{\citenamefont {{Degenhardt}}\ \emph {et~al.}(2019)\citenamefont
  {{Degenhardt}}, \citenamefont {{Artanov}}, \citenamefont {{Geck}},
  \citenamefont {{Grewing}}, \citenamefont {{Kruth}}, \citenamefont
  {{Nielinger}}, \citenamefont {{Vliex}}, \citenamefont {{Zambanini}},\ and\
  \citenamefont {{van Waasen}}}]{8702442}%
  \BibitemOpen
  \bibfield  {author} {\bibinfo {author} {\bibfnamefont {C.}~\bibnamefont
  {{Degenhardt}}}, \bibinfo {author} {\bibfnamefont {A.}~\bibnamefont
  {{Artanov}}}, \bibinfo {author} {\bibfnamefont {L.}~\bibnamefont {{Geck}}},
  \bibinfo {author} {\bibfnamefont {C.}~\bibnamefont {{Grewing}}}, \bibinfo
  {author} {\bibfnamefont {A.}~\bibnamefont {{Kruth}}}, \bibinfo {author}
  {\bibfnamefont {D.}~\bibnamefont {{Nielinger}}}, \bibinfo {author}
  {\bibfnamefont {P.}~\bibnamefont {{Vliex}}}, \bibinfo {author} {\bibfnamefont
  {A.}~\bibnamefont {{Zambanini}}},\ and\ \bibinfo {author} {\bibfnamefont
  {S.}~\bibnamefont {{van Waasen}}},\ }\bibfield  {title} {\bibinfo {title}
  {Systems engineering of cryogenic cmos electronics for scalable quantum
  computers},\ }in\ \href@noop {} {\emph {\bibinfo {booktitle} {2019 IEEE
  International Symposium on Circuits and Systems (ISCAS)}}}\ (\bibinfo {year}
  {2019})\ pp.\ \bibinfo {pages} {1--5}\BibitemShut {NoStop}%
\bibitem [{\citenamefont {Langford}\ \emph {et~al.}(2017)\citenamefont
  {Langford}, \citenamefont {Sagastizabal}, \citenamefont {Kounalakis},
  \citenamefont {Dickel}, \citenamefont {Bruno}, \citenamefont {Luthi},
  \citenamefont {Thoen}, \citenamefont {Endo},\ and\ \citenamefont
  {DiCarlo}}]{langford2017experimentally}%
  \BibitemOpen
  \bibfield  {author} {\bibinfo {author} {\bibfnamefont {N.}~\bibnamefont
  {Langford}}, \bibinfo {author} {\bibfnamefont {R.}~\bibnamefont
  {Sagastizabal}}, \bibinfo {author} {\bibfnamefont {M.}~\bibnamefont
  {Kounalakis}}, \bibinfo {author} {\bibfnamefont {C.}~\bibnamefont {Dickel}},
  \bibinfo {author} {\bibfnamefont {A.}~\bibnamefont {Bruno}}, \bibinfo
  {author} {\bibfnamefont {F.}~\bibnamefont {Luthi}}, \bibinfo {author}
  {\bibfnamefont {D.}~\bibnamefont {Thoen}}, \bibinfo {author} {\bibfnamefont
  {A.}~\bibnamefont {Endo}},\ and\ \bibinfo {author} {\bibfnamefont
  {L.}~\bibnamefont {DiCarlo}},\ }\bibfield  {title} {\bibinfo {title}
  {Experimentally simulating the dynamics of quantum light and matter at
  deep-strong coupling},\ }\href@noop {} {\bibfield  {journal} {\bibinfo
  {journal} {Nat. Commun.}\ }\textbf {\bibinfo {volume} {8}},\ \bibinfo {pages}
  {1} (\bibinfo {year} {2017})}\BibitemShut {NoStop}%
\bibitem [{\citenamefont {Rol}\ \emph {et~al.}(2019)\citenamefont {Rol},
  \citenamefont {Battistel}, \citenamefont {Malinowski}, \citenamefont
  {Bultink}, \citenamefont {Tarasinski}, \citenamefont {Vollmer}, \citenamefont
  {Haider}, \citenamefont {Muthusubramanian}, \citenamefont {Bruno},
  \citenamefont {Terhal} \emph {et~al.}}]{rol2019fast}%
  \BibitemOpen
  \bibfield  {author} {\bibinfo {author} {\bibfnamefont {M.}~\bibnamefont
  {Rol}}, \bibinfo {author} {\bibfnamefont {F.}~\bibnamefont {Battistel}},
  \bibinfo {author} {\bibfnamefont {F.}~\bibnamefont {Malinowski}}, \bibinfo
  {author} {\bibfnamefont {C.}~\bibnamefont {Bultink}}, \bibinfo {author}
  {\bibfnamefont {B.}~\bibnamefont {Tarasinski}}, \bibinfo {author}
  {\bibfnamefont {R.}~\bibnamefont {Vollmer}}, \bibinfo {author} {\bibfnamefont
  {N.}~\bibnamefont {Haider}}, \bibinfo {author} {\bibfnamefont
  {N.}~\bibnamefont {Muthusubramanian}}, \bibinfo {author} {\bibfnamefont
  {A.}~\bibnamefont {Bruno}}, \bibinfo {author} {\bibfnamefont
  {B.}~\bibnamefont {Terhal}}, \emph {et~al.},\ }\bibfield  {title} {\bibinfo
  {title} {Fast, high-fidelity conditional-phase gate exploiting leakage
  interference in weakly anharmonic superconducting qubits},\ }\href@noop {}
  {\bibfield  {journal} {\bibinfo  {journal} {Phys. rev. lett.}\ }\textbf
  {\bibinfo {volume} {123}},\ \bibinfo {pages} {120502} (\bibinfo {year}
  {2019})}\BibitemShut {NoStop}%
\bibitem [{LNF()}]{LNF}%
  \BibitemOpen
  \href {https://www.lownoisefactory.com/files/7015/7825/6000/LNF-LNC4\_8C.pdf}
  {\emph {\bibinfo {title} {{LNF-LNC4\_8C} 4--8 {GHz} Cryogenic Low Noise
  Amplifier}}},\ \bibinfo {organization} {Low Noise Factory, G{\"{o}}teborg,
  Sweden}\BibitemShut {NoStop}%
\bibitem [{\citenamefont {Montazeri}\ \emph {et~al.}(2015)\citenamefont
  {Montazeri}, \citenamefont {Wong}, \citenamefont {Coskun},\ and\
  \citenamefont {Bardin}}]{montazeri2015ultra}%
  \BibitemOpen
  \bibfield  {author} {\bibinfo {author} {\bibfnamefont {S.}~\bibnamefont
  {Montazeri}}, \bibinfo {author} {\bibfnamefont {W.-T.}\ \bibnamefont {Wong}},
  \bibinfo {author} {\bibfnamefont {A.~H.}\ \bibnamefont {Coskun}},\ and\
  \bibinfo {author} {\bibfnamefont {J.~C.}\ \bibnamefont {Bardin}},\ }\bibfield
   {title} {\bibinfo {title} {Ultra-low-power cryogenic {SiGe} low-noise
  amplifiers: {Theory} and demonstration},\ }\href@noop {} {\bibfield
  {journal} {\bibinfo  {journal} {IEEE Trans. Microw. Theory and Techn.}\
  }\textbf {\bibinfo {volume} {64}},\ \bibinfo {pages} {178} (\bibinfo {year}
  {2015})}\BibitemShut {NoStop}%
\bibitem [{\citenamefont {Wong}\ \emph {et~al.}(2020)\citenamefont {Wong},
  \citenamefont {Hosseini}, \citenamefont {R\"ucker},\ and\ \citenamefont
  {Bardin}}]{wongIMS2020}%
  \BibitemOpen
  \bibfield  {author} {\bibinfo {author} {\bibfnamefont {W.-T.}\ \bibnamefont
  {Wong}}, \bibinfo {author} {\bibfnamefont {M.}~\bibnamefont {Hosseini}},
  \bibinfo {author} {\bibfnamefont {H.}~\bibnamefont {R\"ucker}},\ and\
  \bibinfo {author} {\bibfnamefont {J.~C.}\ \bibnamefont {Bardin}},\ }\bibfield
   {title} {\bibinfo {title} {A 1\,{mW} cryogenic {LNA} exploiting optimized
  {SiGe HBTs} to achieve an average noise temperature of 3.2\,{K} from 4–-8
  {GHz}},\ }in\ \href@noop {} {\emph {\bibinfo {booktitle} {Proc. IEEE IMS}}}\
  (\bibinfo {year} {2020})\ pp.\ \bibinfo {pages} {181--184}\BibitemShut
  {NoStop}%
\bibitem [{\citenamefont {v.~{Dijk}}\ \emph {et~al.}(2019)\citenamefont
  {v.~{Dijk}}, \citenamefont {{Vladimirescu}}, \citenamefont {{Babaie}},
  \citenamefont {{Charbon}},\ and\ \citenamefont {{Sebastiano}}}]{8791334}%
  \BibitemOpen
  \bibfield  {author} {\bibinfo {author} {\bibfnamefont {J.}~\bibnamefont
  {v.~{Dijk}}}, \bibinfo {author} {\bibfnamefont {A.}~\bibnamefont
  {{Vladimirescu}}}, \bibinfo {author} {\bibfnamefont {M.}~\bibnamefont
  {{Babaie}}}, \bibinfo {author} {\bibfnamefont {E.}~\bibnamefont
  {{Charbon}}},\ and\ \bibinfo {author} {\bibfnamefont {F.}~\bibnamefont
  {{Sebastiano}}},\ }\bibfield  {title} {\bibinfo {title} {Spine (spin
  emulator) - a quantum-electronics interface simulator},\ }in\ \href@noop {}
  {\emph {\bibinfo {booktitle} {2019 IEEE 8th International Workshop on
  Advances in Sensors and Interfaces (IWASI)}}}\ (\bibinfo {year} {2019})\ pp.\
  \bibinfo {pages} {23--28}\BibitemShut {NoStop}%
\end{thebibliography}
\end{document}